\documentclass[11pt]{article}
\usepackage[margin=1.4in,bottom=1in]{geometry}
\usepackage{amssymb, amsmath, amsthm}  

\usepackage{mathrsfs, amsfonts, bbm}        
\usepackage{graphics, subcaption}  
\usepackage{enumerate}  
\usepackage{extarrows}            
\usepackage{multirow}             
\usepackage{hyperref}
\usepackage{setspace}
\usepackage{xcolor}
\usepackage{epstopdf}
\usepackage{float}
\usepackage{verbatim}             
\usepackage{soul}
\usepackage{amsmath}
\allowdisplaybreaks[4]
\usepackage{perpage} 
\MakePerPage{footnote} 
\usepackage{adjustbox}  

\usepackage{natbib}
\usepackage{hyperref} 
\definecolor{darkblue}{rgb}{0,0,.4}
\hypersetup{
colorlinks=true,
citecolor=blue,
urlcolor=blue,
linkcolor=blue,
}

\usepackage[symbol*]{footmisc}
\DefineFNsymbolsTM{myfnsymbols}{
  \textasteriskcentered *
  {\textasteriskcentered\textasteriskcentered}{**}
  {\textasteriskcentered\textasteriskcentered\textasteriskcentered}{***}
  \textdagger    \dagger
  \textdaggerdbl \ddagger
  \textsection   \mathsection
  \textbardbl    \|%
  \textparagraph \mathparagraph
}%
\setfnsymbol{myfnsymbols}

\newtheorem{theorem}{Theorem}
\newtheorem{assumption}{Assumption}
\newtheorem{proposition}{Proposition}

\newtheorem{lemma}{Lemma}
\newtheorem{example}{Example}
\newtheorem{remark}{Remark}

\renewcommand{\thesection}{\arabic{section}}
\renewcommand{\theequation}{\arabic{section}.\arabic{equation}}

\newcommand{\lf}{\lfloor}
\newcommand{\rf}{\rfloor}
\newcommand{\oline}{\overline}
\newcommand{\mbf}{\mathbf}

\newcommand{\Ra}{\Rightarrow}   
\newcommand{\ra}{\rightarrow}   

\newcommand{\bb}{\mathbb}
\newcommand{\bbm}{\mathbbm}
\newcommand{\mcal}{\mathcal}

\newcommand{\tq}{\triangleq}

\DeclareMathOperator*{\argmax}{arg\,max}
\newcommand{\Var}{\mathrm{Var}}
\newcommand{\Cov}{\mathrm{Cov}}
\newcommand{\convas}{\stackrel{a.s.}{\rightarrow}}
\newcommand{\convp}{\stackrel{p}{\rightarrow}}

\soulregister\cite7
\soulregister\citep7
\soulregister\citet7
\soulregister\ref7
\soulregister\pageref7

\begin{document}
\title{Asymptotic Properties of the Maximum Likelihood Estimator in Regime-Switching Models with Time-Varying Transition Probabilities\thanks{We are grateful to Yoosoon Chang, Yoshihiko Nishiyama, and Joon Y. Park for their guidance in writing the paper. All errors are our own.}}
\author{Chaojun Li \footnote{Academy of Statistics and Interdisciplinary Sciences, Faculty of Economics and Management, East China Normal University, Shanghai 200062, China. Email: cjli@fem.ecnu.edu.cn}\\ East China Normal University\and
Yan Liu\footnote{Department of Economics, Boston University, 270 Bay State Road, Boston, MA 02215, USA. Email: yanliu@bu.edu}\\Boston University}
\date{\today}
\maketitle

\begin{abstract}
      We prove the asymptotic properties of the maximum likelihood estimator (MLE) in time-varying transition probability (TVTP) regime-switching models. This class of models extends the constant regime transition probability in Markov-switching models to a time-varying probability by including information from observations. An important feature in this proof is the mixing rate of the regime process conditional on the observations, which is time varying owing to the time-varying transition probabilities. Consistency and asymptotic normality follow from the almost deterministic geometrically decaying bound of the mixing rate. The assumptions are verified in regime-switching autoregressive models with widely-applied TVTP specifications. A simulation study examines the finite-sample distributions of the MLE and compares the estimates of the asymptotic variance constructed from the Hessian matrix and the outer product of the score. The simulation results favour the latter. As an empirical example, we compare three leading economic indicators in terms of describing U.S. industrial production.
\end{abstract}

\newpage

\section{Introduction}
\setcounter{equation}{0}
\setcounter{theorem}{0}
\setcounter{assumption}{0}
\setcounter{proposition}{0}
\setcounter{corollary}{0}
\setcounter{lemma}{0}
\setcounter{example}{0}
\setcounter{remark}{0}

Regime-switching models have been applied extensively since \citet{hamilton1989new} to study how time-series patterns change across different underlying economic states, such as boom and recession, high-volatility and low-volatility financial market environments, and active and passive monetary and fiscal policies.
This class of models features a bivariate process $(S_t,Y_t)$, where $(S_t)$ is an unobservable Markov chain determining the regime in each period, and $(Y_t)$ is an observable process whose conditional distribution is governed by the underlying regime $(S_t)$.

In basic Markov-switching models, like \citet{hamilton1989new}, the unobservable regime process is assumed to follow a homogeneous Markov chain. It implies the transition probability and expected duration of each regime are constant through all periods, regardless of the level of the observable and how long the regime has lasted.
The setup is restrictive in application and conflicts with empirical findings, such as those of \citet{burns1969progress} and \citet{diebold1992have}.
The basic Markov-switching model is then extended in different ways to include the information of observations into the regime transition probabilities. When the observations take different values over time, the transition probabilities change accordingly, or simply, the model possesses time-varying transition probabilities (TVTP).
\citet{diebold1994regime} modelled the transition probability as a function of predetermined variables $(X_t)$, which can be chosen as economic covariates helpful to  predict the regime change. The model is then applied widely; see, e.g., \citet{filardo1994business}, \citet{bekaert1995time}, \citet{gray1996modeling}, \citet{filardo1998business}, \citet{ang2008term}. 
\citet{chib2004non}, \citet{kim2008estimation}, and \citet{chang2017new} proposed another way to model the regime process. They introduced an autoregressive latent process, and the regime at each period is determined according to whether the latent process takes a value above or below a threshold. 
The information of the observable process $(Y_t)$ can be incorporated into the regime transition probabilities by allowing the innovations to $(Y_t)$ to be correlated with the innovations to the latent process.
The information of predetermined variables can also be incorporated by embedding them to the latent process. This approach is then applied and extended; see, e.g., \citet{cheng2018new}, \citet{song2018volatility}, and \citet{chang2021origins}. 

The literature on the asymptotic properties of the maximum likelihood estimator (MLE) of TVTP regime-switching models is sparse, which might be owing to its theoretical challenges. The general difficulty in the proof of asymptotic theories with regime-switching models is that the predictive densities of the observable given past realizations do not form a stationary sequence, and thus, the ergodic theorem does not directly apply.
The strategy, originated from \cite{baum1966statistical}, is to approximate the log-likelihood function by the partial sum of a stationary ergodic sequence.
The cornerstone of the approximation is the almost surely geometrically decaying bound of the mixing rate of the conditional chain $(S|Y,X)$ in probability and in $L^p$. \citet{douc2004asymptotic} and \citet{kasahara2019asymptotic} showed the approximation and the asymptotic properties of the MLE only in basic Markov-switching models.\footnote{\citet{baum1966statistical}, \citet{leroux1992maximum}, \citet{bickel1996inference}, \citet{bickel1998asymptotic}, \citet{jensen1999asymptotic}, \citet{le2000exponential} and \citet{douc2001asymptotics} contribute to the asymptotic theories with the basic Markov-switching models less general than \citet{douc2004asymptotic} and \citet{kasahara2019asymptotic}. Their models, usually referred to as hidden Markov models, do not allow autoregression and ($Y_t$) are conditionally independent given the current regime.} \citet{kasahara2019asymptotic} considered a more general model than \citet{douc2004asymptotic} in that they allowed the observations to depend on lagged regimes and some elements of the regime transition probabilities to be zero. 
\citet{kasahara2019asymptotic} showed only the approximation in probability and resorted to the dominated convergence theorem to show the approximation in $L^p$ by imposing some high-level assumptions on the moments of the period score and Hessian.
The time-varying regime transition probability makes it more complex to show the bound, because it enters the mixing rate, causes the rate to approach unity as the transition probability approaches zero, and makes the rate non-decaying.

The theoretical works on TVTP regime-switching models include \citet{ailliot2015consistency} and \citet{pouzo2018maximum}.  \citet{ailliot2015consistency} showed consistency of the MLE. \citet{pouzo2018maximum} showed consistency and asymptotic normality of the MLE under possible misspecification. Some of their assumptions are either too restrictive to hold in some widely-applied models or difficult to be verified in practice. First, they required the one-step transition probability of regimes to be positive. This assumption is key to showing the geometrically decaying bound of the mixing rate, because otherwise the bound would be unity and non-decaying. This assumption, however, excludes the models where the conditional distribution of $Y_t$ depends on lagged regimes. 
This assumption also precludes some elements of the transition probabilities being zero. Second, \citet{ailliot2015consistency} particularly assumed the transition probabilities to be bounded away from zero. 
This assumption does not hold in the models such as \citet{diebold1994regime}, where transition probability is a logistic function of $X_t$ and approaches zero as $X_t$ goes to infinity. 
Instead, \citet{pouzo2018maximum} assumed the time-varying transition probabilities not to be too close to zero. Their assumption, though less restrictive than the one in \citet{ailliot2015consistency}, involved sums of infinite terms, which might be difficult to verify in practice. Third, \cite{pouzo2018maximum} made assumptions on the support of the conditional distribution of $Y_t$, which rules out the widely applied models with normal distributions. 

This study shows consistency and asymptotic normality of the MLE in a wide range of TVTP regime-switching models. The main theoretical contribution of this study is that we show the mixing rate is decaying geometrically with a large probability and in $L^p$ by assuming there is a small probability that the observations take extreme values. 
Compared to \citet{kasahara2019asymptotic}, our results in the approximation in $L^p$ do not rely on any high-level assumptions. We show the approximation in $L^p$ with the same assumptions that are used to show the approximation in probability.
Compared to \citet{ailliot2015consistency} and \citet{pouzo2018maximum}, our model 
imposes weak conditions on the regime transition probabilities, which is verified to hold in models of \citet{diebold1994regime} and \citet{chang2017new}, and imposes weak conditions on the conditional density of $Y_t$, which allow for the density of a normal distribution. Then for some widely-applied models, such as \citet{filardo1994business}, which extended the mean-switching AR(4) model in \citet{hamilton1989new} to allow for time-varying transition probabilities,
our theory applies while the theories of \citet{ailliot2015consistency} and \citet{pouzo2018maximum} do not.

We use simulation to examine how well the asymptotic distribution approximates the distribution of the MLE in finite samples. From the theoretical results, we can construct consistent estimates of the asymptotic variance of the MLE either from the negative of the inverse of the Hessian matrix or from the outer product of the score. We compare their performance and find that it is preferable to make inference based on the outer product of the score instead of the Hessian matrix.

The rest of this paper is organized as follows. Section \ref{secmodel} lists the main assumptions and examples of TVTP regime-switching models. Section \ref{secapprox} shows the geometrically decaying bounds of the mixing rate of the conditional chain in probability and in $L^p$. Sections \ref{secconsistency} and \ref{secnormality} show consistency and asymptotic normality of the MLE, respectively. Section \ref{secdiscussion} verifies the assumptions hold in a regime-switching autoregressive process with logistic or probit transition probabilities in \citet{diebold1994regime} or transition probabilities in \citet{chang2017new}. Section \ref{secsim} reports the simulation results. Section \ref{sec:empirical} presents an empirical illustration and examines the contribution of including leading economic indicators into transition probabilities to describing U.S. industrial production. Section \ref{secconclusion} concludes.

\section{Models, notations, and assumptions}\label{secmodel}
\setcounter{equation}{0}
\setcounter{theorem}{0}
\setcounter{assumption}{0}
\setcounter{proposition}{0}
\setcounter{corollary}{0}
\setcounter{lemma}{0}
\setcounter{example}{0}
\setcounter{remark}{0}

\textbf{Notation:} Let $\tq$ denote ``equals by definition". Let ``i.o." stand for ``infinitely often". For a matrix or vector $M$, $\|M\|\tq\sum|M_{ij}|$. For two probability measures $\mu_1$ and $\mu_2$, define the total variation distance between $\mu_1$ and $\mu_2$ as $\|\mu_1-\mu_2\|_{TV}\tq\sup_A|\mu_1(A)-\mu_2(A)|$. Let $\bbm{1}\{A\}$ denote an indicator function that takes the value of one when $A$ is true and zero otherwise. Let $a\wedge b\tq\min\{a,b\}$. Let $\lf x\rf$ denote the largest integer less than or equal to $x$. For any $\{x_i\}$, define $\sum_{i=a}^b x_i\tq0$ and $\prod_{i=a}^b x_i\tq1$ when $b<a$. For a square matrix $A$, its spectral radius is denoted as $\rho(A)$. $\Phi(\cdot)$ and $\varphi(\cdot)$ are the distribution function and the density function of $\bb{N}(0,1)$, respectively.

\subsection{Models}\label{subsecmodel}
We are interested in regime-switching models where the transition probabilities are allowed to include information from observations. In a regime-switching model, the conditional distribution of the observable process $(Y_t)$ is governed by an unobservable regime process $(S_t)$:
\begin{align}
    Y_t=f_{\theta}(S_t,\oline{\mbf{Y}}_{t-1},X_t;U_t),\label{intromodel}
\end{align}
where $\oline{\mbf{Y}}_t\tq(Y_t,\ldots,Y_{t-p+1})'$, $X_t$ is a predetermined variable (vector), $U_t$ is an independent and identically distributed (i.i.d.) sequence of random variables, and $f_\theta$ is a family of functions indexed by $\theta$. We allow $S_t$ to be a vector
\begin{align}
    S_t=(\tilde{S}_t,\ldots,\tilde{S}_{t-d+1})' \label{defs}
\end{align}
so that the conditional density of $Y_t$ can depend on both the current and lagged regimes. 
One example is the autoregressive model with switching in mean and
variance:
\begin{align}\label{ar}
    \gamma(L)(Y_t-\mu(\tilde{S}_t))=\gamma_X'X_t+\sigma(\tilde{S}_t)U_t,
\end{align}
where $\gamma(z)=1-\gamma_1z-\ldots-\gamma_pz^p$. The seminal mean-switching AR(4) model of \citet{hamilton1989new} is a special case of (\ref{ar}) with $p=4,d=5$ and no switching in the variance. 

We model the regime transition probability as
\begin{align}
    q_{\theta}(S_t|S_{t-1},\oline{\mbf{Y}}_{t-1},X_t)\label{introtransition}
\end{align}
where $q_\theta$ is a family of probabilities indexed by $\theta$. We allow the regime transition probability  (\ref{introtransition}) to include information from observations, $\oline{\mbf{Y}}_{t-1}$ and $X_t$. In the literature, \citet{diebold1994regime} and \citet{chang2017new} are two examples of such transition probabilities. 
\begin{example}\label{egdiebold}
\textnormal{(\citealp{diebold1994regime}) Transition probabilities of $\tilde{S}_t$ are functions mapping $X_t$ to $[0,1]$.
In the special case with two regimes $\tilde{S}_t=1$ or 2,}
\begin{align}
    \tilde{q}_\theta(\tilde{S}_t=j|\tilde{S}_{t-1}=i,X_t)=
    \begin{pmatrix}
        p_{11}(X_t) & 1-p_{11}(X_t)\\
        p_{21}(X_t) & 1-p_{21}(X_t)
    \end{pmatrix}.
    \label{Diebold}
\end{align}
\textnormal{The functions used most often are logistic functions (\citealt{diebold1994regime}; \citealt{filardo1994business}) and probit functions (\citealt{gray1996modeling,filardo1998business}). The case in which $p_{11}(X_t)$ and $p_{21}(X_t)$ are constants reduces to the basic Markov-switching model.}
\end{example}

\begin{example}\label{egccp}
\textnormal{(\citealp{chang2017new})
There are two regimes $\tilde{S}_t=0$ or 1. The regime is decided using an autoregressive latent factor, depending on whether the factor takes a value above or below a threshold $\tau$. The latent factor follows an AR(1) process}
\begin{align}\label{CCPW}
    W_t = \alpha W_{t-1}+V_t
\end{align}
\textnormal{for $t=1,2,\dots$ with parameter $\alpha\in(-1,1)$ and i.i.d. standard normal innovations $(V_t)$. $(U_t)$ and $(V_t)$ are jointly i.i.d. and distributed as}
\begin{align}\label{CCPrho}
    \begin{pmatrix}
        U_t\\V_{t+1}
      \end{pmatrix}\sim\bb{N}
      \begin{pmatrix}
        \begin{pmatrix}
            0\\0
        \end{pmatrix},
        \begin{pmatrix}
            1&\rho\\ \rho&1
        \end{pmatrix}
    \end{pmatrix}.
\end{align}
\textnormal{$W_t$ is initialized as}
\begin{equation}\label{ccpw0}
    W_0\sim \bb{N}(0,1/(1-\alpha^2)).
\end{equation}
\textnormal{The regime is decided by}
\begin{align}\label{CCPS}
    \tilde{S}_t =
    \begin{cases}
        1\qquad &\textnormal{ if }W_t\geq \tau\\
        0\qquad &\textnormal{ if }W_t<\tau
    \end{cases}.
\end{align}
\textnormal{The correlation $\rho$ between innovations to the latent factor and lagged innovations to the observed process connects the regime transition probability to the observed process. 
A zero-correlation model reduces to the basic Markov-switching model.}

\textnormal{\citet[Theorem 3.1]{chang2017new} clarifies the transition probability. It states that if $Y_t=m(Y_{t-1},\ldots,Y_{t-p+1})+\sigma(\tilde{S}_t,\ldots,\tilde{S}_{t-p+1})U_t$, $U_t\sim\text{i.i.d.}\bb{N}(0,1)$, $|\alpha|<1$, and $|\rho|<1$, $(\tilde{S}_t,Y_t)$ together follow a $p$th-order Markov process, and the transition probability is}
\begin{align}
    \tilde{q}_\theta(\tilde{S}_t|\tilde{S}_{t-1},\dots,\tilde{S}_{t-p},Y_{t-1},\dots,Y_{t-p}) = (1-\tilde{S}_t)\omega_\rho+\tilde{S}_t(1-\omega_\rho)\label{CCP}
\end{align}
\textnormal{with $\omega_\rho = \omega_\rho(\tilde{S}_{t-1},\dots,\tilde{S}_{t-p},Y_{t-1},\dots,Y_{t-p})$ defined as}
\begin{align}
      \omega_\rho=\frac{\big[(1-\tilde{S}_{t-1})\int_{-\infty}^{\tau\sqrt{1-\alpha^2}}+\tilde{S}_{t-1}\int_{\tau\sqrt{1-\alpha^2}}^\infty\big]\Phi\Big(\frac{\tau-\rho U_{t-1}}{\sqrt{1-\rho^2}}-\frac{\alpha x}{\sqrt{1-\alpha^2}\sqrt{1-\rho^2}}\Big)\varphi(x)dx}
      {(1-\tilde{S}_{t-1})\Phi(\tau\sqrt{1-\alpha^2})+\tilde{S}_{t-1}\big(1-\Phi(\tau\sqrt{1-\alpha^2})\big)}\label{wrho}
\end{align}
\textnormal{where $U_t=\frac{Y_t-m(Y_{t-1},\ldots,Y_{t-p+1})}{\sigma(\tilde{S}_t,\ldots,\tilde{S}_{t-p+1})}$. }
\end{example}

The above two models specify the transition of $\tilde{S}_t$ in (\ref{Diebold}) and (\ref{CCP}). To obtain (\ref{introtransition}) when $d\geq 2$, we write
\begin{align*}
    &q_\theta(S_t=(\tilde{s}_{t},\ldots,\tilde{s}_{t-d+1})|S_{t-1}=(\tilde{s}_{t-1}',\ldots,\tilde{s}_{t-d}'),\oline{\mbf{Y}}_{t-1},X_t)\\
    =&\tilde{q}_\theta(\tilde{S}_t=\tilde{s}_t|\tilde{S}_{t-1}=\tilde{s}_{t-1}',\ldots,\tilde{S}_{t-d}=\tilde{s}_{t-d}',\oline{\mbf{Y}}_{t-1},X_t)
    \bbm{1}\{\tilde{s}_{t-1}=\tilde{s}_{t-1}',\ldots,\tilde{s}_{t-d+1}=\tilde{s}_{t-d+1}'\}.
\end{align*}

\subsection{Assumptions}
For short notation, we define
\begin{align*}
      \mbf{Y}_m^n \tq (Y_n,Y_{n-1},\dots,Y_m)',
      \quad \oline{\mbf{Y}}_m^n\tq(\oline{\mbf{Y}}_n,\dots,\oline{\mbf{Y}}_m)'
\end{align*}
for $n\geq m$. We similarly define $\tilde{\mbf{S}}_m^n$, $\mbf{S}_m^n$, and $\mbf{X}_m^n$.

We assume that $\{S_t\}_{t=0}^{\infty}$ takes a value in a discrete set $\bb{S}$ and use $\mcal{P}(\bb{S})$ to denote the power set of $\bb{S}$.
For each $t\geq 1$ and given $(\oline{\mbf{Y}}_{t-1},S_{t-1},X_t)$, $S_t$ is conditionally independent of $(\mbf{Y}_{-p+1}^{t-p-1},\allowbreak \mbf{S}_{0}^{t-2},\allowbreak \mbf{X}_1^{t-1},\mbf{X}_{t+1}^\infty)$. The transition probability is $q_\theta(s_t|S_{t-1},\oline{\mbf{Y}}_{t-1},X_t)$.
$\{Y_t\}_{t=-p+1}^{\infty}$ takes a value in a set $\bb{Y}$, which is separable and metrizable by a complete metric. Let $\oline{\bb{Y}}\tq \bb{Y}^p$.
For each $t\geq1$ and given $(\oline{\mbf{Y}}_{t-1},S_{t},X_t)$, $Y_t$ is independent of $(\mbf{Y}_{-p+1}^{t-p-1},\mbf{S}_{0}^{t-1},\mbf{X}_1^{t-1},\allowbreak \mbf{X}_{t+1}^\infty)$. The conditional law has a density $g_\theta(y_t|\oline{\mbf{Y}}_{t-1},S_t,X_t)$ with respect to some fixed $\sigma-$finite measure $\nu$ on the Borel $\sigma-$field $\mcal{B}(\bb{Y})$.
$\{X_t\}_{t=1}^\infty$ takes a value in a set $\bb{X}$. Under the setup, conditional on $\mbf{X}_{1}^{\infty}$, $(S_t,\oline{\mbf{Y}}_t)$ follows a Markov chain with transition density
\begin{align*}
    p_\theta(S_t,\oline{\mbf{Y}}_t|S_{t-1},\oline{\mbf{Y}}_{t-1},X_t)
    =g_\theta(Y_t|\oline{\mbf{Y}}_{t-1},S_{t},X_t)
    q_\theta(S_t|S_{t-1},\oline{\mbf{Y}}_{t-1},X_t).
\end{align*}
Moreover, we assume conditionally on $X_t$, $\{X_k\}_{k\geq t+1}$ is independent of $\{Y_k\}_{k\leq t}$ and $\{S_k\}_{k\leq t}$, and conditionally on $X_{t}$, $\{X_k\}_{k\leq t-1}$ is independent of $\{Y_k\}_{k\geq t}$ and $\{S_k\}_{k\geq t}$. Then
$p_\theta(\mbf{Y}_1^t|\oline{\mbf{Y}}_0,S_0=s_0,\mbf{X}_1^n)
      =p_\theta(\mbf{Y}_1^t|\oline{\mbf{Y}}_0,S_0=s_0,\mbf{X}_1^t),
      p_\theta(\mbf{Y}_1^t|\oline{\mbf{Y}}_0,\mbf{X}_1^n)=p_\theta(\mbf{Y}_1^t|\oline{\mbf{Y}}_0,\mbf{X}_1^t).
$
The following are the basic assumptions.
\begin{assumption}\label{compact}
    The parameter $\theta$ belongs to $\Theta$. $\Theta$ is compact. Let $\theta^*$ denote the true parameter. $\theta^*$ lies in the interior of $\Theta$.
\end{assumption}
\begin{assumption}\label{ergodic}
    $(S_t,\oline{\mbf{Y}}_t,X_t)$ is a strictly stationary ergodic process.
\end{assumption}
\begin{assumption}\label{bound1}
    (a) There exists $r\geq1$ such that
    $\sigma_-(\oline{\mbf{y}}_0^{r-1},\mbf{x}_1^r)
    \tq\inf_\theta \min_{s_r,s_0\in\bb{S}} \bb{P}_\theta(S_r=s_r| S_0=s_0,\allowbreak\oline{\mbf{y}}_0^{r-1},\mbf{x}_1^r)>0$, for all $\oline{\mbf{y}}_0^{r-1}\in\oline{\bb{Y}}^{r}$ and $\mbf{x}_1^r\in\bb{X}^r$;
    (b) Let $b_-(y_1,\oline{\mbf{y}}_0,\allowbreak x_1)
    \tq\inf_\theta\min_{s_1\in\bb{S}}g_\theta(y_1|\oline{\mbf{y}}_{0},\allowbreak s_1,x_1)>0$, for all $y_1\in\bb{Y}$, $\oline{\mbf{y}}_0\in\oline{\bb{Y}}$, and $x_1\in\bb{X}$, and $\bb{E}_{\theta^*}[|\log b_-(y_1,
    \oline{\mbf{y}}_0,\allowbreak x_1)|]<\infty$;
    (c) $b_+\tq\sup_{\theta}\sup_{y_1,\oline{\mbf{y}}_0,x_1} \max_{s_1} g_\theta(y_1|\oline{\mbf{y}}_0,\allowbreak s_1,x_1)<\infty$.
\end{assumption}
\begin{remark}\label{remass3}
\textnormal{Assumption \ref{bound1}(a) involves the regime transition probability. \citet{ailliot2015consistency} and \citet{pouzo2018maximum} imposed similar assumptions but restricted $r=1$, based on which they established their key result on the mixing rate of the conditional chain $(S|Y,X)$.
Their assumption of $r=1$, however, is too restrictive and precludes examples in use. On the one hand, it excludes the model where $S_t$ is a vector with $d\geq2$. For instance, when $d=2$, $\bb{P}_\theta(S_1=(\tilde{s}_1,\tilde{s}_{0})|S_0=(\tilde{s}_0',\tilde{s}_{-1}'),\oline{\mbf{Y}}_{0},X_1)=0$ if $\tilde{s}_{0}\neq \tilde{s}_{0}'$. Thus, their theory does not allow the conditional distribution of $Y_t$ to depend on lagged regimes and cannot be applied to some widely-applied empirical examples, such as the mean-switching AR(4) process with transition probability (\ref{Diebold}) in \citet{filardo1994business} and the one with transition probability (\ref{CCP}) in \citet{chang2017new}. 
On the other hand, $r=1$ also precludes some elements of the transition probabilities being zero. We relax their assumption to allow for $r>1$ so that $S_t$ can be a vector and some elements of the transition probabilities can be zero.}

\textnormal{Assumptions \ref{bound1}(b) and \ref{bound1}(c) involve $g_\theta(\cdot)$. \citet{ailliot2015consistency} assumed $g_\theta(\cdot)$ to be the density of a normal or a gamma distribution. Because a normal or a gamma distribution has positive and bounded densities and their first moments are finite, the assumption about $g_\theta(\cdot)$ in \citet{ailliot2015consistency} is a special case of ours. \citet{pouzo2018maximum} assumed that, under our notation, there exists a $\bb{P}_{\theta^*}$-almost surely (a.s.) finite constant $C$ such that $\sup_{\theta}\frac{\max_{s_1}g_{\theta}(Y_1|\oline{\mbf{Y}}_0,s_1,X_1)}{\min_{s_1}g_{\theta}(Y_1|\oline{\mbf{Y}}_0,s_1,X_1)}\leq C$. Their assumption is much stronger than ours and does not allow $g_\theta(\cdot)$ to be the density of a normal distribution. Consider $Y_t=\mu(\tilde{S}_t)+U_t$ with $\tilde{S}_t=1,2$, $\mu(1)=0,$ $\mu(2)=2$, and $U_t\sim\text{i.i.d.}\bb{N}(0,1)$. Then $\frac{\max_{s_1}g_{\theta}(Y_1|s_1)}{\min_{s_1}g_{\theta}(Y_1|s_1)}
=\exp(2|Y_1-1|)$, which increases to infinity as $Y_1$ increases. Thus, the assumption in \citet{pouzo2018maximum} is not satisfied in this widely-applied example.}
\end{remark}
\begin{assumption}\label{bound2}
    Constants $\alpha_1>0$, $C_1\in(0,1]$, $C_2\in(0,+\infty)$, and $\beta_1>1$ exist such that, for any $\xi>0$,
    \begin{align}\label{bound21}
        \bb{P}_{\theta^*}( \sigma_-(\oline{\mbf{Y}}_0^{r-1},X_1^r)\leq C_1 e^{-\alpha_1 \xi} )\leq C_2 \xi^{-\beta_1}.
    \end{align}
\end{assumption}
\begin{remark}
\textnormal{Assumption \ref{bound2} restricts the probability of a small $\sigma_-(\oline{\mbf{Y}}_0^{r-1},\mbf{X}_1^r)$, which is key to our proof of the geometrically decaying bound of the mixing rate in Lemma \ref{mixingbound}. We show in Lemma \ref{suffsigma} a sufficient condition of Assumption \ref{bound2} is $\bb{E}_{\theta^*}[|\log\sigma_-(\oline{\mbf{Y}}_0^{r-1},\mbf{X}_1^r)|^{1+\delta}]<\infty$ for some $\delta>0$. In Subsection \ref{subsecdiscuss4}, we show   that this assumption is satisfied in autoregressive models with transition probabilities in Examples \ref{egdiebold} and \ref{egccp}.}

\textnormal{\citet{ailliot2015consistency} and \citet{pouzo2018maximum} imposed similar assumptions. \citet{ailliot2015consistency} assumed that 
$\inf_{\oline{\mbf{y}}_0,x_1}\sigma_-(\oline{\mbf{y}}_0,x_1)>0$,
which excludes examples in use. For instance, the transition probability in \citet{diebold1994regime}, $\tilde{q}_{\theta^*}(\tilde{S}_t=2|\tilde{S}_{t-1}=1,X_t)=\frac{1}{1+\exp(X_t'\beta_1)}$ approaches zero as $X_t$ goes to infinity. In \citet{chang2017new}, when $\rho>0$ and $\tilde{S_t}=0$, the transition probability  (\ref{CCP}) approaches zero as $U_{t-1}$ goes to infinity. 
\citet{pouzo2018maximum} assumed that, under our notation, $\sum_{k=0}^\infty\bb{E}_{\theta^*}[\prod_{i=0}^k(1-\sigma_-(\oline{\mbf{Y}}_i,X_{i+1}))]<\infty$. Their assumption involves sums of infinite terms, which might be difficult to verify in practice.}
\end{remark}
\begin{assumption}\label{continuityass}
      For all $(s',s)\in\bb{S}^2$, $(\oline{\mbf{y}},y')\in\oline{\bb{Y}}\times\bb{Y}$, and $x\in\bb{X}$, $\theta\mapsto q_\theta(s'|s,\oline{\mbf{y}},x)$ and $\theta\mapsto g_\theta(y'|\oline{\mbf{y}},s,x)$ are continuous.
\end{assumption}
\begin{assumption}\label{identifiability}
      $\theta$ and $\theta^*$ are identical (up to a permutation of regime indexes) if and only if $\bb{P}_\theta(\mbf{Y}_1^n\in\cdot|\oline{\mbf{Y}}_0,\mbf{X}_{-p+1}^n)=\bb{P}_{\theta^*}(\mbf{Y}_1^n\in\cdot|\oline{\mbf{Y}}_0,\mbf{X}_{-p+1}^n)$ for all $n\geq1$.
\end{assumption}

\section{Approximation with stationary ergodic sequence}\label{secapprox}
\setcounter{equation}{0}
\setcounter{theorem}{0}
\setcounter{assumption}{0}
\setcounter{proposition}{0}
\setcounter{corollary}{0}
\setcounter{lemma}{0}
\setcounter{example}{0}
\setcounter{remark}{0}

This study works with the conditional likelihood function given initial observations $\oline{\mbf{Y}}_0=(Y_0,\dots,Y_{-p+1})$, (unobservable) initial regime $S_0$, and predetermined variables $\mbf{X}_1^t$, owing to the difficulties obtaining the closed-form expression of the unconditional stationary likelihood function. We can write the conditional log-likelihood function as
\begin{align}\label{conditionL}
    \ell_n(\theta,s_0)
    =\log p_\theta(Y_1,\dots,Y_n|\oline{\mbf{Y}}_0,S_0=s_0,\mbf{X}_1^n) 
    =\sum_{t=1}^n \log p_\theta(Y_t|\oline{\mbf{Y}}_0^{t-1},S_0=s_0,\mbf{X}_1^t)
\end{align}
with the predictive density
\begin{align}
    p_\theta(Y_t|\oline{\mbf{Y}}_0^{t-1},S_0=s_0,\mbf{X}_1^t)
    =\sum_{s_t,s_{t-1}} (g_\theta(Y_t|\oline{\mbf{Y}}_{t-1},s_t,X_t) q_\theta(s_t|s_{t-1},\oline{\mbf{Y}}_{t-1},X_t)\nonumber\\
    \times\bb{P}_\theta(S_{t-1}=s_{t-1}|S_0=s_0,\oline{\mbf{Y}}_0^{t-1},\mbf{X}_1^{t-1}))\label{predensity}
\end{align}
for $t\geq1$. 
When the number of observations is $n+p$, we condition on the first $p$ observations, and arbitrarily choose the initial regime $s_{0}$.
The aim of this study is to show consistency and asymptotic normality of the MLE $\hat{\theta}_{n,s_0}=\argmax_\theta\ell_n(\theta,s_0)$ with any choice of $s_0$, even when it is not the true underlying initial regime.

Consistency and asymptotic normality follow if we can show the following two results:
(a) the normalized log-likelihood $n^{-1}\ell_n(\theta,s_0)$ converges to a deterministic function $\ell(\theta)$ uniformly with respect to $\theta$, and $\theta^*$ is a well-separated point of the maximum of $\ell(\theta)$; and (b) a central limit theorem for the Fisher score function and a locally uniform law of large numbers for the observed Fisher information.

The updating distribution $\bb{P}_\theta(S_{t-1}=s_{t-1}|S_0=s_0,\oline{\mbf{Y}}_0^{t-1},\mbf{X}_1^{t-1})$ in (\ref{predensity}) is not stationary ergodic and thus, the predictive density is not stationary ergodic. Therefore, the ergodic theorem cannot be applied directly to conclude a law of large numbers or central limit theorems. The strategy, originated from \cite{baum1966statistical}, is to approximate the predictive density with a stationary ergodic sequence. This section gives the theoretical foundations of the approximation.

We choose the stationary ergodic sequence to be
\begin{align}
      &p_\theta(Y_t|\oline{\mbf{Y}}_{-\infty}^{t-1},\mbf{X}_{-\infty}^{t-1})\nonumber\\
      =&\sum_{s_t,s_{t-1}} g_\theta(Y_t|\oline{\mbf{Y}}_{t-1},s_t,X_t)
      q_\theta(s_t|s_{t-1},\oline{\mbf{Y}}_{t-1},X_t)
      \bb{P}_\theta(S_{t-1}=s_{t-1}|\oline{\mbf{Y}}^{t-1}_{-\infty},\mbf{X}^{t-1}_{-\infty}).
      \label{stationarydensity}
\end{align}
(\ref{stationarydensity}) differs from (\ref{predensity}) in that the updating distribution now depends on the whole history of $(Y_t,X_t)$ from past infinity and does not depend on the initial regime. We can extend $(Y_t,X_t)$ to doubly infinite time $\{Y_t,X_t\}_{t=-\infty}^{+\infty}$ because of Assumption \ref{ergodic}. Stationary ergodicity of (\ref{stationarydensity}) follows from Theorem 7.1.3 in \citet{durrett2013probability}. The approximation builds on the almost surely geometrically decaying bound of the mixing rate of the conditional chain $(S|Y,X)$. The bound guarantees that the influence of the observations and initial regime far in the past quickly vanishes and thus, the difference between the exact predictive and approximated predictive log densities becomes asymptotically negligible. Lemma \ref{mixing}
establishes an upper bound for the mixing rate.
\begin{lemma}[Uniform ergodicity]\label{mixing}
      Assume \ref{bound1}. Let $m,n\in\bb{Z}$, $-m\leq n$, and $\theta\in \Theta$. Then, for $-m\leq k\leq n$, for all probability measures $\mu_1$ and $\mu_2$ defined on $\mcal{P}(\bb{S})$ and for all $\oline{\mbf{Y}}_{-m}^n$ and $\oline{\mbf{X}}_{-m}^n$,
      \begin{align}
     &\Big\|\sum_{s\in\bb{S}}\bb{P}_\theta(S_k\in\cdot|S_{-m}=s,\oline{\mbf{Y}}_{-m}^n,\mbf{X}_{-m}^n)\mu_1(s)
            -\sum_{s\in\bb{S}}\bb{P}_\theta(S_k\in\cdot|S_{-m}=s,\oline{\mbf{Y}}_{-m}^n,\mbf{X}_{-m}^n)\mu_2(s)\Big\|_{TV}\nonumber\\
     &\hspace{1cm}\leq \prod_{i=1}^{\lf(k+m)/r\rf}\big( 1-\sigma_-( \oline{\mbf{Y}}_{-m+ri-r}^{-m+ri-1},\mbf{X}_{-m+ri-r+1}^{-m+ri} ) \big).\label{equamixing}
    \end{align}
\end{lemma}
$\sigma_-(\oline{\mbf{Y}}_{k-r}^{k-1},\mbf{X}_{k-r+1}^{k})$ might approach zero when the observations take extreme values, as in the examples of \citet{diebold1994regime} and \citet{chang2017new}. Then in such extreme cases, the upper bound in (\ref{equamixing}) is unity and does not decay over time. 
In Lemma \ref{mixingbound}, We show that under Assumption \ref{bound2}, which restricts the probability of such extreme cases, the upper bound in (\ref{equamixing}) is decaying geometrically with a large probability and in $L^p$. 

\begin{lemma}[Bound for mixing rate]
\label{mixingbound}
Assume \ref{ergodic}--\ref{bound2}. Then, $\varepsilon_0\in(0,\frac{1}{16r})$ and $\rho\in(0,1)$ exist such that for all $m,n\in\bb{Z}^+$,
\begin{align}
    &\bb{P}_{\theta^*}\Big(\prod_{k=1}^{n}(1-\sigma_-(\oline{\mbf{Y}}_{t_k-r}^{t_k-1},\mbf{X}_{t_k-r+1}^{t_k}))\geq \rho^{(1-2\varepsilon_0)n}\ \mathrm{i.o.}\Big)=0,
    \label{mix1}\\
    &\bb{E}_{\theta^*}\Big[\prod_{k=1}^{n}(1-\sigma_-(\oline{\mbf{Y}}_{t_k-r}^{t_k-1},\mbf{X}_{t_k-r+1}^{t_k}))^m\Big]\leq 2\rho^n,\label{mix3}
\end{align}
where $\{t_k\}_{1\leq k\leq n}$ is a sequence of integers such that $t_{k}\neq t_{k'}$ for $1\leq k,k'\leq n$, and $k\neq k'$.
It also holds that for $t\geq 2r$,
\begin{align}
    &\bb{P}_{\theta^*}(\sigma_-(\oline{\mbf{Y}}_{t-r}^{t-1},\mbf{X}_{t-r+1}^{t})\leq C_1 \rho^{\varepsilon_0\lf(t-r)/r\rf}\ \mathrm{i.o.}
    )=0.\label{mix2}
\end{align}
\end{lemma}
\begin{remark}
\textnormal{(\ref{mix3}) establishes geometrically decaying rates of the moments of the upper bound. The rates play a role when we establish the approximation in $L^2(\bb{P}_{\theta^*})$. \cite{kasahara2019asymptotic} did not derive such rates. Instead, they made additional high-level assumptions on the moment of $\sup_{m\geq0}\sup_{|\theta-\theta^*|<\delta}\nabla_{\theta}^i \log p_\theta(Y_1|\oline{\mbf{Y}}_{-m}^0,\mbf{X}_{-m}^1)$ with $i=1,2$ for some $\delta>0$ and resorted to the dominated convergence theorem to establish the approximation in $L^2(\bb{P}_{\theta^*})$. By contrast, this study does not need such additional assumptions. (\ref{mix3}) is derived from the same assumptions as (\ref{mix1}) and (\ref{mix2}).}
\end{remark}

\section{Consistency}\label{secconsistency}
\setcounter{equation}{0}
\setcounter{theorem}{0}
\setcounter{assumption}{0}
\setcounter{proposition}{0}
\setcounter{corollary}{0}
\setcounter{lemma}{0}
\setcounter{example}{0}
\setcounter{remark}{0}

This sections proves consistency of the MLE $\hat{\theta}_{n,s_0}=\argmax_\theta \ell_n(\theta,s_0)$. We first show that the normalized log-likelihood $n^{-1}\ell_n(\theta,s_0)$ converges to a deterministic function $\ell(\theta)$ uniformly with respect to $\theta$. 
The first step is to approximate $n^{-1}\ell_n(\theta,s_0)$ with the sample mean of a $\bb{P}_{\theta^*}$-stationary ergodic sequence of random variables.
The construction of the stationary ergodic sequence is as follows. Define
\begin{align*}
      &\Delta_{t,m,s}(\theta)\tq\log p_\theta(Y_t|\oline{\mbf{Y}}_{-m}^{t-1},S_{-m}=s,\mbf{X}_{-m+1}^t),\\
      &\Delta_{t,m}(\theta)\tq\log p_\theta(Y_t|\oline{\mbf{Y}}_{-m}^{t-1},\mbf{X}_{-m+1}^t),
\end{align*}
so that $n^{-1}\ell_n(\theta,s_0)=\frac{1}{n}\sum_{t=1}^n\Delta_{t,0,s_0}(\theta)$. We show that $\{\Delta_{t,m,s}(\theta)\}_{m\geq0}$ is a uniformly Cauchy sequence with respect to $\theta\in\Theta$ $\bb{P}_{\theta^*}$-a.s. in (\ref{1.1}) of Lemma \ref{perioddistance}. 

\begin{lemma}\label{perioddistance}
    Assume \ref{ergodic}--\ref{bound2}. Then, there exist constants $\rho\in(0,1)$ and $M<\infty$ and a random sequence $\{A_{t,m}\}_{t\geq1, m\geq 0}$ such that, for all $t\geq 1$  and $m'\geq m \geq0$,
    \begin{align}
        &\sup_{\theta\in\Theta}\max_{s_i,s_j\in\bb{S}}|\Delta_{t,m,s_i}(\theta)-\Delta_{t,m',s_j}(\theta)|\leq A_{t,m}\rho^{\lf(t+m)/3r\rf},\label{1.1}\\
        &\sup_{\theta\in\Theta}\max_{s\in\bb{S}}|\Delta_{t,m,s}(\theta)-\Delta_{t,m}(\theta)|\leq A_{t,m}\rho^{\lf(t+m)/3r\rf}
            ,\label{1.2}
    \end{align}
    where $\bb{P}_{\theta^*}(A_{t,m}\geq M\ \mathrm{i.o.})=0.$ Also, $\{\Delta_{t,m,s}(\theta)\}_{m\geq0}$ is uniformly bounded in $L^1(\bb{P}_{\theta^*})$.
\end{lemma}
(\ref{1.1}) is derived from the results in Section 3. To see this, we take $\mu_1(s)=\delta_{s_i}(s)$ and
$\mu_2(s)=\bb{P}_{\theta}(S_{-m}=s|S_{-m'}=s_j,\oline{\mbf{Y}}_{-m'}^n,\mbf{X}_{-m'}^n)$ in Lemma \ref{mixing}. Then (\ref{equamixing}) gives an upper bound for the distance between $\bb{P}_{\theta}(S_k\in\cdot|S_{-m}=s_i,\oline{\mbf{Y}}_{-m}^n,\mbf{X}_{-m}^n)$
and $\bb{P}_{\theta}(S_k\in\cdot|S_{-m'}=s_j,\oline{\mbf{Y}}_{-m'}^n,\mbf{X}_{-m'}^n)$. 
(\ref{1.1}) follows by combining Lemmas \ref{mixing} and \ref{mixingbound}. We give the details in the online appendix.

 (\ref{1.1}) indicates $\{\Delta_{t,m,s}(\theta)\}_{m\geq0}$ converges uniformly $\bb{P}_{\theta^*}$-a.s. and the limit does not depend on $s$.
Define $\Delta_{t,\infty}(\theta)\tq\lim_{m\ra\infty}\Delta_{t,m,s}(\theta)$,
which is well-defined in $L^1(\bb{P}_{\theta^*})$. From (\ref{1.2}), $\Delta_{t,\infty}(\theta)=\lim_{m\ra\infty}\Delta_{t,m}(\theta)$. From Assumption \ref{ergodic}, $\{\Delta_{t,\infty}\}_{t\geq0}$ is stationary ergodic, and the ergodic theorem applies:
\begin{align}\label{2}
      \frac{1}{n}\sum_{t=1}^n \Delta_{t,\infty}(\theta)\convas\ell(\theta)\tq\bb{E}_{\theta^*}[\Delta_{0,\infty}(\theta)].
\end{align}
Let $m=0$ and $m'\ra\infty$ in (\ref{1.1}), which yields
\begin{align}\label{1}
      \frac{1}{n}\sum_{t=1}^n \sup_{\theta\in\Theta}\max_{s_0\in\bb{S}}|\Delta_{t,0,s_0}(\theta)-\Delta_{t,\infty}(\theta)|\convas 0.
\end{align}
(\ref{1}) shows the difference between $n^{-1}\ell_n(\theta,s_0)$ and $\frac{1}{n}\sum_{t=1}^n \Delta_{t,\infty}(\theta)$ is asymptotically negligible. Then $n^{-1}\ell_n(\theta,s_0)$ also converges to $\ell(\theta)$ $\bb{P}_{\theta^*}$-a.s. From Assumption \ref{continuityass}, we can show the continuity of $\ell(\theta)$. Then $n^{-1}\ell_n(\theta,s_0)$ converges to $\ell(\theta)$ uniformly with respect to $\theta$, which is shown in Proposition \ref{uconvergence}.
\begin{proposition}[Uniform convergence]
      \label{uconvergence}
      Assume \ref{compact}--\ref{continuityass}. Then, for any $s_0\in\bb{S}$,
      \begin{align*}
        \sup_{\theta\in\Theta}|n^{-1}\ell_n(\theta,s_0)-\ell(\theta)|\convas0, \quad\text{as }n\ra\infty.
      \end{align*}
\end{proposition}
Consistency follows once we show that $\theta^*$ is a well-separated point of the maximum of $\ell(\theta)$. Theorem \ref{consistency} summarizes the finding and the details of proof are in the online appendix.

\begin{theorem}\label{consistency}
      Assume \ref{compact}--\ref{identifiability}. Then, for any $s_0\in\bb{S}$, $\hat{\theta}_{n,s_0}\convas\theta^*$, as $n\ra\infty$.
\end{theorem}

\section{Asymptotic normality}\label{secnormality}
\setcounter{equation}{0}
\setcounter{theorem}{0}
\setcounter{assumption}{0}
\setcounter{proposition}{0}
\setcounter{corollary}{0}
\setcounter{lemma}{0}
\setcounter{example}{0}
\setcounter{remark}{0}

This section establishes asymptotic normality of the MLE. We need additional regularity assumptions. Assume a positive real $\delta$ exists such that on $G\tq \{\theta\in\Theta: |\theta-\theta^*|<\delta\}$, the following conditions hold:
\begin{assumption}\label{differentiable}
    For all $(s',s)\in\bb{S}^2$, $(\oline{\mbf{y}},y')\in\oline{\bb{Y}}\times\bb{Y}$, and $x\in\bb{X}$, the functions $\theta\mapsto q_\theta(s'|s,\oline{\mbf{y}},x)$ and $\theta\mapsto g_\theta(y'|\oline{\mbf{y}},s,x)$ are twice continuously differentiable.
\end{assumption}
\begin{assumption}\label{finitemoment}
    $\bb{E}_{\theta^*}[\sup_{\theta\in G}\max_{s,s'}\|\nabla_\theta \log q_\theta(s'|s,\oline{\mbf{Y}}_0,X_0)\|^4]<\infty$,\\
    $\bb{E}_{\theta^*}[\sup_{\theta\in G}\max_{s,s'}\|\nabla_\theta^2 \log q_\theta(s'|s,\oline{\mbf{Y}}_0,X_0)\|^2]<\infty$;\\
    $\bb{E}_{\theta^*}[\sup_{\theta\in G}\max_{s}\|\nabla_\theta \log g_\theta(Y_0|\oline{\mbf{Y}}_{-1},s,X_0)\|^4]<\infty$,\\
    $\bb{E}_{\theta^*}[\sup_{\theta\in G}\max_{s}\|\nabla_\theta^2 \log g_\theta(Y_0|\oline{\mbf{Y}}_{-1},s,X_0)\|^2]<\infty$.
\end{assumption}
\begin{assumption}\label{dominatedbounded}
    (a) For almost all $(\oline{\mbf{y}},y',x)$, a finite function $f_{\oline{\mbf{y}},y',x}:\bb{S}\ra\bb{R}^+$ exists such that $\sup_{\theta\in G} g_\theta(y'|\oline{\mbf{y}},s,x)\leq f_{\oline{\mbf{y}},y',x}(s)$.
    (b) For almost all $(\oline{\mbf{y}},s,x)$, functions $f^1_{\oline{\mbf{y}},s,x}:\bb{Y}\ra\bb{R}^+$ and $f^2_{\oline{\mbf{y}},s,x}:\bb{Y}\ra\bb{R}^+$ exist in $L^1(\nu)$ such that $\|\nabla_\theta g_\theta(y'|\oline{\mbf{y}},s,x)\|\leq f^1_{\oline{\mbf{y}},s,x}(y')$ and $\|\nabla^2_\theta g_\theta(y'|\oline{\mbf{y}},s,x)\|\leq f^2_{\oline{\mbf{y}},s,x}(y')$ for all $\theta\in G$.
\end{assumption}

\subsection{A central limit theorem for the score function}\label{subsecclt}
This subsection shows asymptotic normality of the score function. First, we show that the score function can be approximated by a sequence of integrable martingale increments. We use the Fisher identity (\citealp[p.353]{cappe2005inference}) to write the score function as the expectation of the complete score conditional on the observed data:
 \begin{align*}
  \nabla_\theta \ell_n(\theta,s_0)
     =\bb{E}_\theta\Big[ \sum_{t=1}^n \phi_{\theta,t}\Big|\oline{\mbf{Y}}_0^n,S_0=s_0,\mbf{X}_1^n\Big],
 \end{align*}
where $\phi_{\theta,t}$ is the short-hand notation for
\begin{align*}
      \phi_{\theta}( (S_t,Y_t),(S_{t-1},\oline{\mbf{Y}}_{t-1}),X_t)
      &\tq\nabla_\theta\log p_\theta(Y_t,S_t|\oline{\mbf{Y}}_{t-1},S_{t-1},X_t).
\end{align*}
We write the period score function as
\begin{align*}
      &\dot{\Delta}_{t,m,s}(\theta)\tq \nabla_\theta \Delta_{t,m,s}(\theta)
      =\nabla_\theta \log p_\theta(Y_t|\oline{\mbf{Y}}_{-m}^{t-1},S_{-m}=s,\oline{\mbf{X}}_{-m+1}^t)\nonumber\\
      =&\bb{E}_\theta\Big[\sum_{k=-m+1}^t \phi_{\theta,k}|\oline{\mbf{Y}}_{-m}^t,S_{-m}=s,\mbf{X}_{-m}^t\Big]
      -\bb{E}_\theta\Big[\sum_{k=-m+1}^{t-1}  \phi_{\theta,k}|\oline{\mbf{Y}}_{-m}^{t-1},S_{-m}=s,\mbf{X}_{-m}^{t-1}\Big],
\end{align*}
so that $\nabla_\theta \ell_n(\theta,s_0)=\sum_{t=1}^n\dot{\Delta}_{t,0,s_0}(\theta)$.
We can similarly define
\begin{align}
      \dot{\Delta}_{t,m}(\theta)
      \tq\bb{E}_\theta\Big[\sum_{k=-m+1}^t \phi_{\theta,k}|\oline{\mbf{Y}}_{-m}^t,\mbf{X}_{-m}^t\Big]
      -\bb{E}_\theta\Big[\sum_{k=-m+1}^{t-1}  \phi_{\theta,k}|\oline{\mbf{Y}}_{-m}^{t-1},\mbf{X}_{-m}^{t-1}\Big].
      \label{deltadot}
\end{align}
The stationary conditional score is constructed by conditioning on the whole history of $(Y_t,X_t)$ starting from past infinity: $\dot{\Delta}_{t,\infty}(\theta)\tq\lim_{m\ra\infty} \dot{\Delta}_{t,m}(\theta)$.
Define the filtration $\mcal{F}_t=\sigma\big((\oline{\mbf{Y}}_k,X_{k+1}),-\infty<k\leq t\big)$ for $t\in\bb{Z}$. (\ref{appdefine3})-(\ref{martingale2}) show that $\{\dot{\Delta}_{t,\infty}(\theta^*)\}_{t=-\infty}^{\infty}$ is an $(\mcal{F},\bb{P}_{\theta^*})$-adapted stationary ergodic and square integrable martingale increment sequence.
The central limit theorem for the sums of such a sequence shows that
\begin{align*}
      n^{-\frac{1}{2}}\sum_{t=1}^n \dot{\Delta}_{t,\infty}(\theta^*) \Ra \bb{N}(0,I(\theta^*)),
\end{align*}
where $I(\theta^*)\tq \bb{E}_{\theta^*}[\dot{\Delta}_{0,\infty}(\theta^*)\dot{\Delta}_{0,\infty}(\theta^*)']$ is the asymptotic Fisher information matrix.

Lemma \ref{scoreapproximation} shows that $\dot{\Delta}_{t,0,s_0}(\theta^*)$ can be approximated by $\dot{\Delta}_{t,\infty}(\theta^*)$ in $L^2(\bb{P}_{\theta^*})$. The approximation depends on the distance between regime processes that start from different time points, whose upper bound decays geometrically in $L^2(\bb{P}_{\theta^*})$ according to (\ref{mix3}).  
\begin{lemma}\label{scoreapproximation}
      Assume \ref{ergodic}--\ref{bound2} and \ref{differentiable}-\ref{finitemoment}. Then, for all $s_0\in \bb{S}$,
      \begin{align*}
            \bb{E}_{\theta^*}\Big[\Big\| \frac{1}{\sqrt{n}} \sum_{t=1}^n \big(\dot{\Delta}_{t,0,s_0}(\theta^*)-\dot{\Delta}_{t,\infty}(\theta^*)\big)\Big\|^2\Big]\ra0.
      \end{align*}
\end{lemma}
\begin{remark}
\textnormal{\citet[Lemmas 8 and 10]{douc2004asymptotic} showed the result in basic Markov-switching models when the conditional distribution of $Y_t$ only depends on the current regime. The mixing rate is bounded by a constant  $1-\frac{\inf_{\theta}\inf_{\tilde{s}',\tilde{s}}\tilde{q}_\theta(\tilde{s}'|\tilde{s})}{\sup_{\theta}\sup_{\tilde{s}',\tilde{s}}\tilde{q}_\theta(\tilde{s}'|\tilde{s})}$. Then the mixing rate in $L^2(\bb{P}_{\theta^*})$ is bounded by the same constant, and the approximation in $L^2(\bb{P}_{\theta^*})$ follows. In \citet{kasahara2019asymptotic}, the conditional distribution of $Y_t$ is allowed to depend on lagged regimes. The upper bound of the mixing rate is no longer a constant but depends on observations, which adds difficulties to the proof of the approximation in $L^2(\bb{P}_{\theta^*})$. For this reason, \citet{kasahara2019asymptotic} imposed high-level assumptions that bound the moments of $\dot{\Delta}_{t,m}(\theta)$ and $\dot{\Delta}_{t,m,s}(\theta)$.
We contribute to the literature by showing the geometrically decaying bound of the mixing rate in $L^2(\bb{P}_{\theta^*})$ in (\ref{mix3}), so that we can show the approximation in Lemma \ref{scoreapproximation} without additional assumptions.}
\end{remark}

$n^{-\frac{1}{2}}\sum_{t=1}^n\dot{\Delta}_{t,0,s_0}(\theta^*)$ has the same limiting distribution as $n^{-\frac{1}{2}}\sum_{t=1}^n\dot{\Delta}_{t,\infty}(\theta^*)$ by Lemma \ref{scoreapproximation}.
\begin{theorem}\label{scorenormality}
      Assume \ref{ergodic}--\ref{bound2} and \ref{differentiable}--\ref{dominatedbounded}. Then, for any $s_0\in\bb{S}$,
      \begin{align*}
            n^{-\frac{1}{2}}\nabla_\theta \ell_n(\theta^*,s_0) \Ra \bb{N}(0,I(\theta^*)).
      \end{align*}
\end{theorem}

\subsection{Law of large numbers for the observed Fisher information}
This subsection presents the law of large numbers for the observed Fisher information.
We use the Louis missing information principle (\citealp{louis1982finding}) to express the observed Fisher information in terms of the Hessian of the complete log-likelihood function:
\begin{align}
      \nabla^2_\theta \ell_n(\theta,s_0)
      =&\bb{E}_\theta\Big[ \sum_{t=1}^n \dot{\phi}_{\theta,t}\Big|\oline{\mbf{Y}}_0^n,S_0=s_0,\mbf{X}_1^n\Big]
      +\Var_\theta\Big[\sum_{t=1}^n \phi_{\theta,t}\Big|\oline{\mbf{Y}}_0^n,S_0=s_0,\mbf{X}_1^n\Big],\label{eqhessian}
\end{align}
where $\dot{\phi}_{\theta,t}$ is the short-hand notation for
\begin{align*}
      \dot{\phi}_{\theta}( (S_t,Y_t),(S_{t-1},\oline{\mbf{Y}}_{t-1}),X_t)
      \tq \nabla_\theta \phi_{\theta,t}
      =\nabla_\theta^2 \log p_\theta(Y_t,S_t|\oline{\mbf{Y}}_{t-1},S_{t-1},X_t).
\end{align*}
We approximate the expressions in (\ref{eqhessian}) with stationary ergodic sequences $\bb{P}_{\theta^*}$-a.s. and in $L^1(\bb{P}_{\theta^*})$. 
Similar to Lemma \ref{scoreapproximation}, we use (\ref{mix3}) to show the approximation in $L^1(\bb{P}_{\theta^*})$, while 
\citet{kasahara2019asymptotic} imposed high-level assumptions on the moments of derivatives of predictive densities. 
The construction of the stationary ergodic sequences is similar to the one in Subsection \ref{subsecclt}, and thus we leave the details in the online appendix and directly give the theoretical finding.
\begin{theorem}\label{HessianLLN}
    Assume \ref{ergodic}--\ref{continuityass} and \ref{differentiable}--\ref{dominatedbounded}. Let $\{\theta_n^*\}$ be any, possibly stochastic, sequence in $\Theta$ such that $\theta_n^*\convas\theta^*$. Then, for all $s_0\in\bb{S}$,
    \begin{align}\label{convhessian}
        -n^{-1}\nabla^2_\theta \ell_n(\theta^*_n,s_0)\allowbreak \convas I(\theta^*)
    \end{align}
\end{theorem}
Theorems \ref{scorenormality} and \ref{HessianLLN} together yield the following theorem of asymptotic normality.
\begin{theorem}\label{normality}
    Assume \ref{compact}--\ref{identifiability} and \ref{differentiable}--\ref{dominatedbounded}. Then, for any $s_0\in\bb{S}$,
    $n^{1/2} (\hat{\theta}_{n,s_0}-\theta^*)\Ra
        \bb{N}(0,I(\theta^*)^{-1}).
    $
\end{theorem}
From Theorems \ref{consistency}, \ref{HessianLLN}, and \ref{normality}, the negative of the inverse of the Hessian matrix at the MLE value is a consistent estimate of the asymptotic variance. Next proposition gives another consistent estimate constructed from the outer product of the score.
\begin{proposition}\label{ops}
    Assume \ref{ergodic}--\ref{continuityass} and \ref{differentiable}--\ref{dominatedbounded}. Let $\{\theta_n^*\}$ be any, possibly stochastic, sequence in $\Theta$ such that $\theta_n^*\convas\theta^*$. Then, for all $s_0\in\bb{S}$,
    \begin{align}\label{convops}
        \frac{1}{n}\sum_{t=1}^n\dot{\Delta}_{t,0,s_0}(\theta_n^*)\dot{\Delta}_{t,0,s_0}(\theta_n^*)' \convas I(\theta^*).
    \end{align}
\end{proposition}
\noindent A third consistent estimate, provided in \citet[p. 466]{cappe2005inference} and also motivated from Proposition \ref{ops}, can be constructed from the outer product of the demeaned score
\begin{align}\label{convops2}
    \frac{1}{n}\sum_{t=1}^n\Big(\dot{\Delta}_{t,0,s_0}(\hat{\theta}_{n,s_0})-\frac{1}{n}\sum_{t=1}^n\dot{\Delta}_{t,0,s_0}(\hat{\theta}_{n,s_0})\Big)
    \Big(\dot{\Delta}_{t,0,s_0}(\hat{\theta}_{n,s_0})-\frac{1}{n}\sum_{t=1}^n\dot{\Delta}_{t,0,s_0}(\hat{\theta}_{n,s_0})\Big)'.
\end{align}
We compare their performance in finite samples through simulation in Section \ref{secsim}.

\section{Discussion of Assumptions}\label{secdiscussion}
\setcounter{equation}{0}
\setcounter{theorem}{0}
\setcounter{assumption}{0}
\setcounter{proposition}{0}
\setcounter{corollary}{0}
\setcounter{lemma}{0}
\setcounter{example}{0}
\setcounter{remark}{0}

This section verifies the assumptions hold in an autoregressive process with regime-dependent means, variances, and autoregressive coefficients
\begin{align}\label{ar2}
    Y_t=\mu(\tilde{S}_t)+\sum_{j=1}^{k}\gamma_j(\tilde{S}_{t})(Y_{t-j}-\mu(\tilde{S}_{t-j}))+\sigma(\tilde{S}_t)U_t
\end{align}
and transition probabilities of Examples \ref{egdiebold} and \ref{egccp}.

\subsection{Discussion of Assumption \ref{ergodic}}
\citet{francq2001stationarity} gave conditions for Markov-switching autoregressive models to be stationary ergodic.
We extend their result to TVTP regime-switching models.
\begin{theorem}\label{thmergodic}
Suppose the transition equation of the observed process is
\begin{align}\label{vectorial}
    \oline{\mbf{Y}}_t = A(\tilde{S}_t)\oline{\mbf{Y}}_{t-1}+B(S_t,X_t),
\end{align}
where $A(\tilde{S}_t)$ is a $p\times p$ matrix and $B(S_t,X_t)$ is a $p\times 1$ vector. 
The regime $\tilde{S}_t$ takes $J$ possible values, and its transition probability, $\tilde{q}_\theta(\tilde{S}_t=j|\tilde{S}_{t-1}=i,X_t)$, is denoted for brevity as $\tilde{q}_{ij,t}$. $S_t=(\tilde{S}_t,\ldots,\tilde{S}_{t-d+1})'$.
If it is satisfied that 
(a) $(S_t,X_t)$ is stationary ergodic, 
(b) $\bb{E}_{\theta^*}[\|B(S_t,X_t)\|^2]<\infty$, and (c) when the autoregressive coefficients do not depend on $\tilde{S}_t$, i.e., $A(\tilde{S}_t)=A$, $\rho(A)<1$, (c')  when the autoregressive coefficients depend on $\tilde{S}_t$, the $Jp\times Jp$ matrix $M (X_t)$ defined in (\ref{defm}) satisfies $\bb{E}_{\theta^*}[\|M(
X_t)\|]<1$. Then $(\oline{\mbf{Y}}_t,S_t,X_t)$ is strictly stationary ergodic.
\begin{align}
    M (X_t)\tq 
\begin{pmatrix}
    (A(1)\otimes A(1))\tilde{q}_{11,t} &
    (A(1)\otimes A(1))\tilde{q}_{21,t} &
    \dots &
    (A(1)\otimes A(1))\tilde{q}_{J1,t}\\
    (A(2)\otimes A(2))\tilde{q}_{12,t} &
    (A(2)\otimes A(2))\tilde{q}_{22,t} &
    \dots &
    (A(2)\otimes A(2))\tilde{q}_{J2,t}\\
    \vdots & \vdots &  & \vdots\\
    (A(J)\otimes A(J))\tilde{q}_{1J,t} &
    (A(J)\otimes A(J))\tilde{q}_{2J,t} &
    \dots &
    (A(J)\otimes A(J))\tilde{q}_{JJ,t}
\end{pmatrix}\label{defm}
\end{align}
\end{theorem}
(\ref{vectorial}) covers a wide range of regime-switching autoregressive models. For instance, in (\ref{ar2}), if $p=k$ and $d=k+1$, we can rewrite (\ref{ar2}) as
$\oline{\mbf{Y}}_t=A(\tilde{S}_t)\oline{\mbf{Y}}_{t-1}+B(S_t,X_t)$
with 
\begin{align*}
    A(\tilde{S}_t)=
    \begin{pmatrix}
    \gamma_1(\tilde{S}_{t}) & \gamma_2(\tilde{S}_{t}) &\dots & \gamma_{k}(\tilde{S}_{t})\\
    1 & 0 & 0 & 0\\
    0 & \ddots & 0 & 0\\
    0 & 0 & 1 & 0
    \end{pmatrix}
\end{align*}
and $B(S_t,X_t)=(\mu(\tilde{S}_t)-\sum_{j=1}^k\gamma_j(\tilde{S}_t)\mu(\tilde{S}_{t-j})+\sigma(\tilde{S}_t)U_t,0,\ldots,0)'.$

If the autoregressive coefficient does not depend on $\tilde{S}_t$, i.e., $A(\tilde{S}_t)=A$, then (c) is the usual stationarity condition of autoregressive processes. For instance, for the seminal \citet{hamilton1989new} model, we only need to check whether $\rho(A)<1$. In the following analysis of this subsection, we consider the case where the autoregressive coefficient does depend on $\tilde{S}_t$.

In Example \ref{egdiebold}, (a) in Theorem \ref{thmergodic} holds if, for instance, $X_t$ follows $X_t=\eta X_{t-1}+\xi_t$, $\xi_t\sim $ i.i.d. $\bb{N}(0,\Sigma)$ with $\|\eta\|<1$, and the transition probabilities are strictly positive and continuous with respect to $X_t$. For (c') in Theorem \ref{thmergodic}, the expectation of the random matrix might not be easily computed analytically, even for the widely applied logistic and probit functions. Instead, we can verify (c') numerically for given parameter values.

For Example \ref{egccp},  note that the transition probability is essentially a function of $(\tilde{S}_t,\tilde{S}_{t-1},U_{t-1})$. Then we express the transition probability as $\tilde{q}_\theta(\tilde{S}_t|\tilde{S}_{t-1},U_{t-1})$, so that we can apply Theorem \ref{thmergodic}. 
From (\ref{CCPW})--(\ref{CCPS}), $(\tilde{S}_t,U_{t-1})$ is strictly stationary ergodic.
Next, we derive an expression of (c') in Theorem \ref{thmergodic}. Using $\tilde{q}_{ij}\tq\bb{E}_{\theta^*}[\tilde{q}_\theta(\tilde{S}_t=j|\tilde{S}_{t-1}=i,U_{t-1})]=(1-j)w+j(1-w)$, where $i,j=0,1$ and
\begin{align*}
    w=\frac{[(1-i)\int_{-\infty}^{\tau\sqrt{1-\alpha^2}}+i\int_{\tau\sqrt{1-\alpha^2}}^\infty]\Phi\big(\tau-\frac{\alpha x}{\sqrt{1-\alpha^2}}\big)\varphi(x)dx}
    {(1-i)\Phi(\tau\sqrt{1-\alpha^2})+i(1-\Phi(\tau\sqrt{1-\alpha^2}))},
\end{align*}
(c') is equivalent to $\|M\|<1$ with
\begin{align*}
    M\tq 
    \begin{pmatrix}
        A(0)\otimes A(0)\tilde{q}_{00} & A(0)\otimes A(0)\tilde{q}_{10}\\
        A(1)\otimes A(1)\tilde{q}_{01} & A(1)\otimes A(1)\tilde{q}_{11}
    \end{pmatrix}.
\end{align*}

\subsection{Discussion of Assumption \ref{bound2}}\label{subsecdiscuss4}
We give a sufficient condition for Assumption \ref{bound2} that is easier to verify in practice.
\begin{lemma}\label{suffsigma}
    A sufficient condition for Assumption \ref{bound2} is that for some $\delta>0$, 
    \begin{align}\label{logsigma}
        \bb{E}_{\theta^*}[|\log \sigma_-(\oline{\mbf{Y}}_0^{r-1},\mbf{X}_1^r)|^{1+\delta}]<\infty.
    \end{align}
\end{lemma}


First consider Example \ref{egdiebold} with $\tilde{q}_\theta(\tilde{S}_t=1|\tilde{S}_{t-1}=1,X_t)=\frac{\exp(X'_t\beta_1)}{1+\exp(X'_t\beta_1)}$, $\tilde{q}_\theta(\tilde{S}_t=1|\tilde{S}_{t-1}=2,X_t)=\frac{\exp(X'_t\beta_2)}{1+\exp(X'_t\beta_2)}$, $\bb{E}_{\theta^*}[\|X_t\|^2]<\infty$, and  $U_t\sim \text{i.i.d. }\bb{N}(0,1)$ in (\ref{ar2}). In the definition of $\oline{\mbf{Y}}_t$, $p=k$. In the definition of $S_t$, $d=k+1$. We choose $r=d$ and $\delta=1$. Then
\begin{align}
    &\sigma_-(\oline{\mbf{Y}}_0^{d-1},\mbf{X}_1^d)
    =\inf_\theta\min_{s_d,s_0}\bb{P}_\theta(S_d=s_d|S_0=s_0,\oline{\mbf{Y}}_{0}^{d-1},\mbf{X}_1^d)\nonumber\\
    =&\inf_\theta\min_{s_d,s_0}
    \frac{\sum_{\mbf{s}_{1}^{d-1}}(\prod_{\ell=1}^d q_{\theta}(s_\ell|s_{\ell-1},\oline{\mbf{Y}}_{\ell-1},X_\ell)\prod_{\ell=1}^{d-1}g_{\theta}(Y_\ell|\oline{\mbf{Y}}_{\ell-1},s_\ell,X_\ell))}
    {\sum_{\mbf{s}_{1}^{d}}(\prod_{\ell=1}^d q_{\theta}(s_\ell|s_{\ell-1},\oline{\mbf{Y}}_{\ell-1},X_\ell)\prod_{\ell=1}^{d-1}g_{\theta}(Y_\ell|\oline{\mbf{Y}}_{\ell-1},s_\ell,X_\ell))}\nonumber\\
    \geq &  \frac{\prod_{\ell=1}^{d-1}b_-(Y_\ell,\oline{\mbf{Y}}_{\ell-1},X_\ell)}{b_+^{d-1}}\inf_\theta\min_{\tilde{\mbf{s}}_0^d}\prod_{\ell=1}^d \tilde{q}_\theta(\tilde{s}_\ell|\tilde{s}_{\ell-1},X_\ell).\label{ineqsigma}
\end{align}
It follows that
\begin{align}
\begin{split}
&\bb{E}_{\theta^*}[|\log \sigma_-(\oline{\mbf{Y}}_0^{d-1},X_1^d)|^2]\\
\leq &\bb{E}_{\theta^*}\Big[\Big|\sum_{\ell=1}^{d-1}\log \frac{b_-(Y_\ell,\oline{\mbf{Y}}_{\ell-1},X_\ell)}{b_+}+\sum_{\ell=1}^d  \inf_\theta\min_{\tilde{s}_\ell,\tilde{s}_{\ell-1}}\log ( \tilde{q}_\theta(\tilde{s}_\ell|\tilde{s}_{\ell-1},X_\ell))\Big|^2\Big].
\end{split}\label{elogsigma}
\end{align}
Then it suffices to show that 
\begin{align}
    &\bb{E}_{\theta^*}[|\log b_-(Y_1,\oline{\mbf{Y}}_0,X_1)|^2]<\infty,\label{logb-}\\
    &\bb{E}_{\theta^*}[|\log (\inf_\theta\min_{\tilde{s}_0,\tilde{s}_1}\tilde{q}_\theta(\tilde{s}_1|\tilde{s}_0,X_1))|^2]<\infty.\label{logq}
\end{align}
(\ref{logb-}) follows from the normal distribution of $U_t$ and $\bb{E}_{\theta^*}[U_t^4]<\infty$. 
The following lemma shows that (\ref{logq}) is satisfied.
\begin{lemma}\label{lemlogistic}
    Suppose the regime transition probability is $\tilde{q}_\theta(\tilde{S}_t=1|\tilde{S}_{t-1}=1,X_t)=\frac{\exp(X'_t\beta_1)}{1+\exp(X'_t\beta_1)}$, $\tilde{q}_\theta(\tilde{S}_t=1|\tilde{S}_{t-1}=2,X_t)=\frac{\exp(X'_t\beta_2)}{1+\exp(X'_t\beta_2)}$, $\bb{E}_{\theta^*}[\|X_t\|^2]<\infty$, and $\Theta$ is compact. Then, 
    $\bb{E}_{\theta^*}[|\log (\inf_\theta\min_{\tilde{s}_0,\tilde{s}_1}\tilde{q}_\theta(\tilde{s}_1|\tilde{s}_0,X_1))|^2]<\infty$.
\end{lemma}

Next consider Example \ref{egdiebold} with $\tilde{q}_\theta(\tilde{S}_t=1|\tilde{S}_{t-1}=1,X_t)=\Phi(X'_t\beta_1)$, $\tilde{q}_\theta(\tilde{S}_t=1|\tilde{S}_{t-1}=2,X_t)=\Phi(X'_t\beta_2)$, $\bb{E}_{\theta^*}[\|X_t\|^4]<\infty$, and  $U_t\sim \text{i.i.d. }\bb{N}(0,1)$ in (\ref{ar2}). In this example, in the definition of $S_t$, $d=k+1$. We choose $r=d$ and $\delta=1$. Similar to the case of logistic functions, it suffices to show (\ref{logb-}) and (\ref{logq}). (\ref{logb-}) follows from the normal distribution of $U_t$ and $\bb{E}_{\theta^*}[U_t^4]<\infty$. The following lemma shows that (\ref{logq}) is satisfied.
\begin{lemma}\label{lemprobit}
    Suppose the regime transition probability is $\tilde{q}_\theta(\tilde{S}_t=1|\tilde{S}_{t-1}=1,X_t)=\Phi(X'_t\beta_1)$, $\tilde{q}_\theta(\tilde{S}_t=1|\tilde{S}_{t-1}=2,X_t)=\Phi(X'_t\beta_2)$, $\bb{E}_{\theta^*}[\|X_t\|^4]<\infty$, and $\Theta$ is compact. Then, 
    $\bb{E}_{\theta^*}[|\log (\inf_\theta\min_{\tilde{s}_0,\tilde{s}_1}\tilde{q}_\theta(\tilde{s}_1|\tilde{s}_0,X_1))|^2]<\infty$.
\end{lemma}

Last, consider Example \ref{egccp}. Because $(Y_t,\tilde{S}_t)$ is a Markov process of order $k+1$, in the definition of $S_t$, $d=k+1$, and in the definition of $\oline{\mbf{Y}}_t$, $p=k+1$. We choose $r=d$ and $\delta=1$. Then 
\begin{align*}
    &\sigma_-(\oline{\mbf{Y}}_0^{d-1})
    =\inf_\theta\min_{s_{d},s_0}\bb{P}_\theta(S_{d}=s_{d}|S_0=s_0,\oline{\mbf{Y}}_{0}^{d-1})\\
    =&\inf_\theta\min_{\tilde{\mbf{s}}_{-d+1}^d}\bb{P}_\theta(\tilde{S}_d=\tilde{s}_d,\ldots,\tilde{S}_1=\tilde{s}_1|\tilde{S}_0=\tilde{s}_0,\ldots,\tilde{S}_{-d+1}=\tilde{s}_{-d+1},\oline{\mbf{Y}}_0^{d-1})\\
    =&\inf_\theta\min_{\tilde{\mbf{s}}_{-d+1}^d}\prod_{\ell=1}^d\tilde{q}_\theta(\tilde{s}_\ell|\tilde{s}_{\ell-1},\ldots,\tilde{s}_{\ell-d},\oline{\mbf{Y}}_{\ell-1}).
\end{align*}
From 
$
\bb{E}_{\theta^*}[|\log\sigma_-(\oline{\mbf{Y}}_0^{d-1})|^2]\leq\bb{E}_{\theta^*}[|\sum_{\ell=1}^d\inf_{\theta}\min_{\tilde{\mbf{s}}_{\ell-d}^\ell}\log\tilde{q}_\theta(\tilde{s}_\ell|\tilde{s}_{\ell-1},\ldots,\tilde{s}_{\ell-d},\oline{\mbf{Y}}_{\ell-1})|^2],
$
it suffices to show 
$\bb{E}_{\theta^*}[|\inf_{\theta}\min_{\tilde{\mbf{s}}_{\ell-d}^\ell}\log\tilde{q}_\theta(\tilde{s}_\ell|\tilde{s}_{\ell-1},\ldots,\tilde{s}_{\ell-d},\oline{\mbf{Y}}_{\ell-1})|^2]<\infty$, which is shown in the following lemma.

\begin{lemma}\label{lemccp}
    Suppose the model is (\ref{ar2}) with the regime transition probability (\ref{CCP}), and $\Theta$ is compact. Then, 
    $\bb{E}_{\theta^*}[|\inf_{\theta}\min_{\tilde{\mbf{s}}_{-d+1}^1}\log\tilde{q}_\theta(\tilde{s}_1|\tilde{s}_{0},\ldots,\tilde{s}_{-d+1},\oline{\mbf{Y}}_{0})|^2]<\infty$.
\end{lemma}

\subsection{Discussion of Assumption \ref{identifiability}}
Write the conditional density as
\begin{align*}
    &p_\theta(Y_1,\dots,Y_n|\oline{\mbf{Y}}_0,\mbf{X}_{-p+1}^n)\\
    = &\sum_{\mbf{s}_{1}^n} \Big(\prod_{t=1}^n g_\theta(Y_t|\oline{\mbf{Y}}_{t-1},s_t,X_t) \prod_{t=2}^n q_\theta(s_t|s_{t-1},\oline{\mbf{Y}}_{t-1},X_t)\bb{P}_\theta(S_1=s_1|\oline{\mbf{Y}}_0,\mbf{X}_{-p+1}^1)\Big).
\end{align*}
It is a class of finite mixtures of the $n$-fold product densities $\prod_{t=1}^n g_\theta(Y_t|\oline{\mbf{Y}}_{t-1},s_t,X_t)$ with mixing distributions $\prod_{t=2}^n q_\theta(s_t|s_{t-1},\oline{\mbf{Y}}_{t-1},X_t)\bb{P}_\theta(S_1=s_1|\oline{\mbf{Y}}_0,\mbf{X}_{-p+1}^1)$. 
Because the class of finite mixtures of the normal family $g_\theta(\cdot)$ is identifiable (\citealp[Section 12.4.2]{cappe2005inference}), Assumption \ref{identifiability} holds for (\ref{ar2}) if the following two conditions hold. First, $\mu(\tilde{s}_t)+\sum_{j=1}^{k}\gamma_j(\tilde{s}_{t})(Y_{t-j}-\mu(\tilde{s}_{t-j}))\neq \mu(\tilde{s}_t')+\sum_{j=1}^{k}\gamma_j(\tilde{s}_{t}')(Y_{t-j}-\mu(\tilde{s}_{t-j}'))$ or $\sigma(\tilde{s}_t)\neq \sigma(\tilde{s}_t')$, for $\tilde{\mbf{s}}_{t-k}^t\neq \tilde{\mbf{s}}^{\prime t}_{t-k}$. That is, the means and variances in different regimes are distinct up to renumbering of regimes. Second, $q_\theta(\cdot)=q_{\theta^*}(\cdot)$ if and only if $\theta=\theta^*$.


\section{Simulation}\label{secsim}
\setcounter{equation}{0}
\setcounter{theorem}{0}
\setcounter{assumption}{0}
\setcounter{proposition}{0}
\setcounter{corollary}{0}
\setcounter{lemma}{0}
\setcounter{example}{0}
\setcounter{remark}{0}

We use simulation to examine how well the asymptotic distribution approximates the distribution of the MLE in finite samples. We consider the \cite{hamilton1989new} regime-switching model with transition probabilities $\tilde{q}_\theta(\tilde{S}_t=1|\tilde{S}_{t-1}=1,X_t)=\frac{\exp(\beta_{10}+\beta_{11}{X}_t)}{1+\exp(\beta_{10}+\beta_{11}{X}_t)}$ and $\tilde{q}_\theta(\tilde{S}_t=1|\tilde{S}_{t-1}=2,X_t)=\frac{\exp(\beta_{20}+\beta_{21}{X}_t)}{1+\exp(\beta_{20}+\beta_{21}{X}_t)}$.
The data generating process of $X_t$ is $X_t=aX_{t-1}+\xi_t$ with $a=0.4$ and $\xi_t\sim$ i.i.d.$\bb{N}(0,1)$. The true parameter is taken from the estimation result in Section \ref{sec:empirical}, $(\mu(1)^*,\mu(2)^*,\gamma_1^*,\gamma_2^*,\gamma_3^*,\gamma_4^*,\sigma^*,\beta_{10}^*,\allowbreak \beta_{11}^*,\beta_{20}^*,\beta_{21}^*)=(-2.33, 0.16, 0.08, 0.17, 0.15,0.005,0.50,\allowbreak -1.70,\allowbreak -1.61,-5.66,-4.85)$.
As commented in Remark \ref{remass3}, this model is not covered by the theory of \citet{ailliot2015consistency} and \citet{pouzo2018maximum} 
 but by our theory, and we have shown the assumptions hold in Section \ref{secdiscussion}.

The sample size ranges from 200 to 1000. For each sample size, we generate 1000 data sets. For each data, we use the Broyden--Fletcher--Goldfarb--Shanno algorithm to find the MLE. We construct the confidence interval from the Hessian matrix in (\ref{convhessian}), the outer product of the score in (\ref{convops}), and the outer product of the demeaned score in (\ref{convops2}).\footnote{The Hessian matrix is computed with the method in \citet[Chapter 10.3.4]{cappe2005inference} and is generalized to allow for TVTP specifications. The score is computed following  \citet{hamilton1996specification}.}

Table \ref{table:sim} reports the frequency at which the 95\% confidence interval contains the true parameter. It is worth mentioning that in some replications, the diagonal elements of the inverse of (\ref{convhessian}) are negative, and we fail to construct confidence intervals. We count it as the intervals not including the true parameters.\footnote{The numbers of replications with negative diagonal elements in the inverse of (\ref{convhessian}) are respectively 7, 7, and 0 when $n= 200$, 600, and 1000.} The coverage frequency of the interval constructed from (\ref{convops2}) is the same as the one constructed from (\ref{convops}), so we put their results together in Panel B of Table \ref{table:sim} to save space.\footnote{The reason why the converge frequencies based on (\ref{convops2}) and (\ref{convops}) are the same might be that the score mean $\frac{1}{n}\sum_{t=1}^n\dot{\Delta}_{t,0,s}(\hat{\theta}_{n,s})$ is of small scale. The average of the absolute value of the score mean across 1000 simulation experiments is no more than $1\times 10^{-7}$ when $n=200, 600,$ and 1000.}
In Panel A, the coverage frequency of $\sigma$ at $n=1000$, though improves slightly from $n=200$, is still much lower than 95\%.
In Panel B, the coverage frequencies of all parameters improve as $n$ increases from 200 to 1000 and are generally close to 95\%.  In this simulation experiment, the confidence interval constructed from the outer product of the (demeaned) score outperforms the one constructed from the Hessian matrix. Considering also the huge efforts in the computation of the Hessian matrix, it might be preferable to make inference based on the outer product of the score instead of the Hessian matrix.
\begin{table}
\caption{Number of coverage of 95\% confidence interval in 1000 replications}
\centering
\begin{tabular}{cccccccccccc}
\hline
\multicolumn{12}{l}{Panel A: confidence interval constructed from ($\ref{convhessian}$)}                                                                                   \\ \hline
\multicolumn{1}{c|}{$n$}  & $\mu(1)$ & $\mu(2)$ & $\gamma_1$ & $\gamma_2$ & $\gamma_3$ & $\gamma_4$ & $\sigma$ & $\beta_{10}$ & $\beta_{11}$ & $\beta_{20}$ & $\beta_{21}$ \\ \hline
\multicolumn{1}{c|}{200}  & 930      & 893      & 922        & 882        & 894        & 895        & 806      & 972          & 966          & 945          & 944          \\
\multicolumn{1}{c|}{600}  & 940      & 921      & 946        & 919        & 918        & 919        & 827      & 961          & 952          & 943          & 954          \\
\multicolumn{1}{c|}{1000} & 950      & 942      & 946        & 925        & 919        & 933        & 828      & 952          & 956          & 964          & 950          \\ \hline
\multicolumn{12}{l}{Panel B: confidence interval constructed from ($\ref{convops}$) and ($\ref{convops2}$)}                                                                \\ \hline
\multicolumn{1}{c|}{$n$}  & $\mu(1)$ & $\mu(2)$ & $\gamma_1$ & $\gamma_2$ & $\gamma_3$ & $\gamma_4$ & $\sigma$ & $\beta_{10}$ & $\beta_{11}$ & $\beta_{20}$ & $\beta_{21}$ \\ \hline
\multicolumn{1}{c|}{200}  & 937      & 911      & 937        & 898        & 914        & 913        & 953      & 982          & 986          & 959          & 964          \\
\multicolumn{1}{c|}{600}  & 949      & 928      & 952        & 931        & 920        & 921        & 949      & 965          & 966          & 953          & 958          \\
\multicolumn{1}{c|}{1000} & 950      & 940      & 950        & 927        & 919        & 931        & 952      & 951          & 956          & 962          & 945          \\ \hline
\end{tabular}
\label{table:sim}
\end{table}

\section{Empirical Example}\label{sec:empirical}
\setcounter{equation}{0}
\setcounter{theorem}{0}
\setcounter{assumption}{0}
\setcounter{proposition}{0}
\setcounter{corollary}{0}
\setcounter{lemma}{0}
\setcounter{example}{0}
\setcounter{remark}{0}

This section examines the contribution of including leading economic indicators into regime transition probabilities to describing the expansion and contraction of U.S. industrial production. In the study of business cycles with TVTP regime-switching models, the predetermined variables are usually those considered to be useful as business-cycle predictors (\citealp{filardo1994business}), and leading economic indicators are possible candidates. 
The Organisation for Economic Co-operation and Development (OECD) database provides three leading indicators: the composite leading indicator (CLI), the business confidence index (BCI), and the consumer confidence index (CCI). As far as we know, there have not been studies on the performance of the three indicators in  identifying the turning points of business cycles, and thus we fill the gap.

We model the monthly U.S. industrial production log growth rate as a mean-switching AR(4) process with two regimes as in \cite{hamilton1989new}, and transition probabilities are 
$\tilde{q}_\theta(\tilde{S}_t=1|\tilde{S}_{t-1}=1,X_t)=\frac{\exp(\beta_{10}+\beta_{11}{X}_t)}{1+\exp(\beta_{10}+\beta_{11}{X}_t)}$ and 
$\tilde{q}_\theta(\tilde{S}_t=1|\tilde{S}_{t-1}=2,X_t)=\frac{\exp(\beta_{20}+\beta_{21}{X}_t)}{1+\exp(\beta_{20}+\beta_{21}{X}_t)}.$
$X_t$ is taken to be the lagged demeaned log growth rate of montly CLI, BCI, or CCI. The data of U.S. industrial production, CLI, BCI, and CCI is downloaded from the OECD website.
The sample period is Jan 1984--Dec 2019. We choose the sample period because there seem structural breaks in the variance of industrial production respectively in 1984 and 2020.
We proceed to compare the performance of the three indicators from three perspectives: hypothesis tests of including the indicators, the response of transition probabilities to changes in the indicators, and the inferred probability of recessions.

\begin{table}[t]
\caption{Maximum Likelihood Estimation Result}
\begin{center}
\begin{tabular}{cccccc}
\hline
               & Markov                  & CLI                                                                    & BCI                                                      & CCI                                                      & CLI,BCI,CCI                                                             \\ \hline
$\mu(1)$       & \begin{tabular}[c]{@{}c@{}}-2.170\\ (0.196)\end{tabular}& \begin{tabular}[c]{@{}c@{}}-1.296\\ (0.099)\end{tabular}               & \begin{tabular}[c]{@{}c@{}}-2.327\\ (0.185)\end{tabular} & \begin{tabular}[c]{@{}c@{}}-2.139\\ (0.301)\end{tabular} & \begin{tabular}[c]{@{}c@{}}-1.757\\ (0.118)\end{tabular}                \\
$\mu(2)$       & \begin{tabular}[c]{@{}c@{}}0.164\\ (0.041)\end{tabular} & \begin{tabular}[c]{@{}c@{}}0.189\\ (0.046)\end{tabular}                & \begin{tabular}[c]{@{}c@{}}0.161\\ (0.042)\end{tabular}  & \begin{tabular}[c]{@{}c@{}}0.164\\ (0.044)\end{tabular}  & \begin{tabular}[c]{@{}c@{}}0.184\\ (0.041)\end{tabular}                 \\
$\beta_{10}$   &\begin{tabular}[c]{@{}c@{}}-0.733\\ (1.247)\end{tabular}               & \begin{tabular}[c]{@{}c@{}}0.962\\ (0.800)\end{tabular}                & \begin{tabular}[c]{@{}c@{}}-1.702\\ (2.574)\end{tabular} & \begin{tabular}[c]{@{}c@{}}-1.002\\ (1.191)\end{tabular} & \begin{tabular}[c]{@{}c@{}}-272.889\\ ($9.354\times10^7$)\end{tabular}  \\
$\beta_{11}$   &                      & \begin{tabular}[c]{@{}c@{}}-1.025\\ (1.425)\end{tabular}               & \begin{tabular}[c]{@{}c@{}}-1.610\\ (4.941)\end{tabular} & \begin{tabular}[c]{@{}c@{}}9.294\\ (24.429)\end{tabular} & \begin{tabular}[c]{@{}c@{}}-1242.074\\ ($3.734\times10^8$)\end{tabular} \\
$\beta_{12}$   & \multicolumn{1}{l}{} & \multicolumn{1}{l}{}                                                   & \multicolumn{1}{l}{}                                     & \multicolumn{1}{l}{}                                     & \begin{tabular}[c]{@{}c@{}}954.478\\ ($3.170\times10^8$)\end{tabular}   \\
$\beta_{13}$   & \multicolumn{1}{l}{} & \multicolumn{1}{l}{}                                                   & \multicolumn{1}{l}{}                                     & \multicolumn{1}{l}{}                                     & \begin{tabular}[c]{@{}c@{}}-88.191\\ ($5.591\times10^7$)\end{tabular}   \\
$\beta_{20}$   & \begin{tabular}[c]{@{}c@{}}-4.921\\ (0.693)\end{tabular}  & \begin{tabular}[c]{@{}c@{}}-754.443\\ ($4.733\times10^4$)\end{tabular}  & \begin{tabular}[c]{@{}c@{}}-5.662\\ (0.905)\end{tabular} & \begin{tabular}[c]{@{}c@{}}-5.330\\ (1.576)\end{tabular} & \begin{tabular}[c]{@{}c@{}}-6.692\\ (4.052)\end{tabular}                \\
$\beta_{21}$   &                      & \begin{tabular}[c]{@{}c@{}}-1463.306\\ ($9.198\times10^4$)\end{tabular} & \begin{tabular}[c]{@{}c@{}}-4.849\\ (1.979)\end{tabular} & \begin{tabular}[c]{@{}c@{}}-3.628\\ (3.175)\end{tabular} & \begin{tabular}[c]{@{}c@{}}-12.361\\ (9.912)\end{tabular}               \\
$\beta_{22}$   &                      &                                                                        &                                                          &                                                          & \begin{tabular}[c]{@{}c@{}}4.837\\ (10.309)\end{tabular}                \\
$\beta_{23}$   & \multicolumn{1}{l}{} & \multicolumn{1}{l}{}                                                   & \multicolumn{1}{l}{}                                     & \multicolumn{1}{l}{}                                     & \begin{tabular}[c]{@{}c@{}}-1.177\\ (4.383)\end{tabular}                \\
log-likelihood & -327.563              & -318.012                                                               & -325.201                                                 & -325.035                                                 & -313.165                                                                \\
$p$-value      & \multicolumn{1}{l}{} & $7.113\times10^{-5}$                                                       & 0.094                                                    & 0.080                                                    &   $6.649\times 10^{-5}$                                                         \\ \hline
\end{tabular}
\end{center}
\label{table:empirical}
\footnotesize
\renewcommand{\baselineskip}{11pt}
\textbf{Note: }Standard errors are in parentheses. The last row reports the $p$-value for the test of no predetermined variables in the transition probabilities. The complete result is in Table \ref{table:empirical2} in the online appendix.
\end{table}
Table \ref{table:empirical} reports the estimation result. Based on the simulation results in Section \ref{secsim}, we compute the standard errors according to (\ref{convops}).
Because regime-switching models are identified up to renumbering of regimes, we restrict $\mu(1)<\mu(2)$. Then regime 1 and 2 can be interpreted as recession and boom, respectively. The last row reports the $p$-value of the likelihood ratio test of not including predetermined variables. At the 1\% level of significance, we can reject the null hypothesis in the case of CLI but fail to reject the null in the case of BCI and CCI. 

\begin{figure}[h]
\centering
\includegraphics[trim=3.8cm 0.15cm 3cm 0.15cm,clip=true,width=13.3cm]{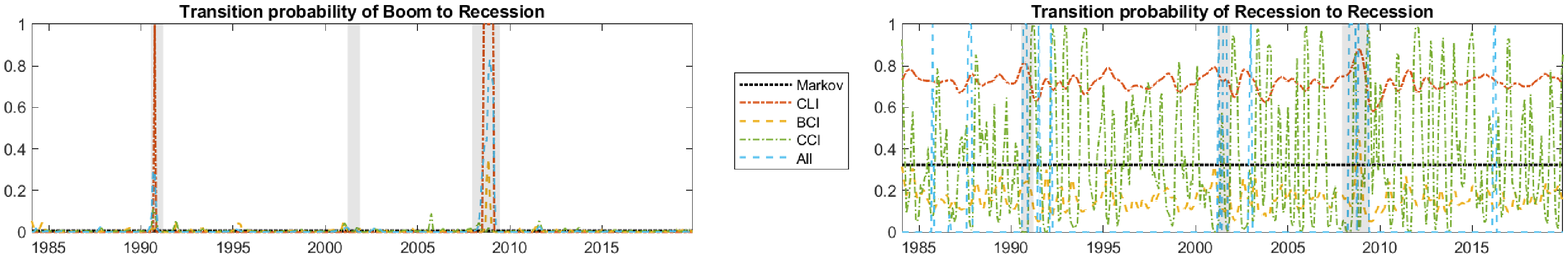}
\caption{Transition probabilities}
\label{fig:trans}
\end{figure}

Figure \ref{fig:trans} plots the 
inferred transition probabilities. 
It is worth noting that the transition probability from boom to recession responds strongly to changes in CLI. In fact, the transition probability to recession from boom is close to one during the early 1990s recession and the Great Recession.

\begin{figure}
\centering
\includegraphics[trim=2.2cm 0.8cm 2cm 0.5cm,clip=true,width=13cm]{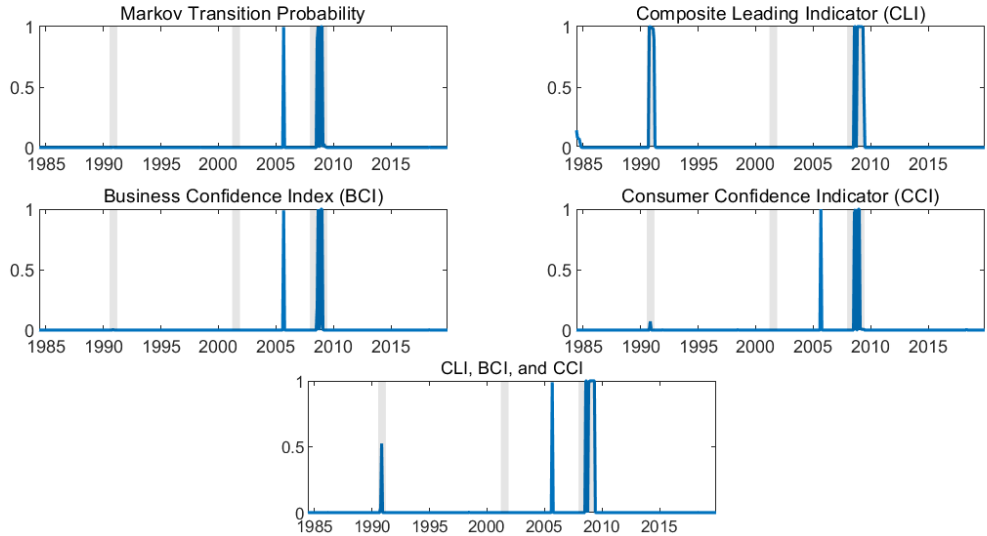}
\caption{Smoothed transition probability of the recession regime}
\label{fig:smooth}
\end{figure}

Figure \ref{fig:smooth} plots the smoothed probability of the recession regime and the NBER recession periods. All of the models produce high probabilities of recession during the Great Recession. Only the case of CLI produces high probabilities of recession in the early 1990s recession. All except the case of CLI gives false signals of recession in early 2005. All of the models, however, fail to identify the early 2000s recession.

The above comparison shows that CLI might be preferable from the perspective of hypothesis testing and its success at identifying the early 1990s recession. We also consider the model that includes all three leading indicators with $\tilde{q}_\theta(\tilde{S}_t=1|\tilde{S}_{t-1}=1,X_t)=\frac{\exp(\beta_{10}+\beta_{11}CLI_{t-1}+\beta_{12}BCI_{t-1}+\beta_{13}CCI_{t-1})}{1+\exp(\beta_{10}+\beta_{11}CLI_{t-1}+\beta_{12}BCI_{t-1}+\beta_{13}CCI_{t-1})}$, and 
$\tilde{q}_\theta(\tilde{S}_t=1|\tilde{S}_{t-1}=2,X_t)=\frac{\exp(\beta_{20}+\beta_{21}CLI_{t-1}+\beta_{22}BCI_{t-1}+\beta_{23}CCI_{t-1})}{1+\exp(\beta_{20}+\beta_{21}CLI_{t-1}+\beta_{22}BCI_{t-1}+\beta_{23}CCI_{t-1})}.$
The $p$-value of the likelihood ratio test of $\beta_{12}=\beta_{13}=\beta_{22}=\beta_{23}=0$ is 0.0459. 
From Figure \ref{fig:smooth}, the smoothed probability of recession is nearly 0.5 during the 1990s recession. Thus, the case of all three leading economic indicators hardly beats the case of CLI only.

\section{Conclusion}\label{secconclusion}
\setcounter{equation}{0}
\setcounter{theorem}{0}
This study shows consistency and asymptotic normality of the MLE in TVTP regime-switching models. The proof follows from the geometrically decaying bound of the mixing rate of $(S|Y,X)$ with a large probability and in $L^p$, which is shown based on the assumption that there is a small probability for the observations to take extreme values. We verify the assumptions to hold in TVTP regime-switching autoregressive models with transition probabilities in \citet{diebold1994regime} and \citet{chang2017new}. We examine how well the asymptotic distribution approximates the distribution of the MLE in finite samples. We compare the estimates of the asymptotic variance constructed from the negative of the inverse Hessian matrix and the inverse of the outer product of the score. The simulation results favour the latter. We apply the TVTP regime-switching model to U.S. industrial production data and investigate the performance of three leading economic indicators in identifying business-cycle turning points.

\section*{Acknowledgements}
We are grateful to Yoosoon Chang, Yoshihiko Nishiyama, and Joon Y. Park for their guidance in writing the paper. All errors are our own.

\bibliographystyle{asa}
\bibliography{AsymptoticsTVTPRS_ArXiv}

\section*{Appendix A: Proofs}
\renewcommand{\theequation}{A.\arabic{equation}}
\renewcommand{\thesection}{A}
\setcounter{equation}{0}

{\bf Notation: }Let $a\vee b\tq\max\{a,b\}$. Let $I_n$ denote the $n\times n$ identity matrix. For short notation, define $\mbf{V}_t\tq(\oline{\mbf{Y}}_{t-r}^{t-1},\mbf{X}_{t-r+1}^{t})$ and $\|\phi_{\theta^*,t}\|_\infty=\max_{s_t,s_{t-1}}\|\phi_{\theta^*}( (s_t,Y_t),(s_{t-1},\oline{\mbf{Y}}_{t-1}),X_t)\|$.

\subsection{Proof of lemmas and corollaries in Section 3}
\noindent\textbf{Proof of Lemma \ref{mixing}: }
The lemma is a consequence of Lemma \ref{minorization} and \citet[Lemma 7]{kasahara2019asymptotic}.\hfill$\square$

\noindent\textbf{Proof of Lemma \ref{mixingbound}: }
First, we show (\ref{mix3}). By Assumption \ref{bound2}, $C_1\in(0,1]$ and $C_2>0$ exist such that for any $\xi>0$,
\begin{align}
      \bb{P}_{\theta^*}\Big( 1-\sigma_-(\mbf{V}_t) \geq 1-C_1 e^{-\alpha_1\xi} \Big)\leq  C_2\xi^{-\beta_1}.\label{lemma20}
\end{align}
We choose $\xi_0$ such that it satisfies
\begin{align}
      \frac{1}{16r}(1-C_1 e^{-\alpha_1 \xi_0})^{mn}
      = C_2\xi_0^{-\beta_1}.\label{lemma21}
\end{align}
The existence of such $\xi_0>0$ is guaranteed by monotone increasing of $\frac{1}{16r}(1-C_1 e^{-\alpha_1\xi})^{mn}$ in $\xi\in(0,+\infty)$ from $\frac{1}{16r}(1-C_1)^{mn}$ to $\frac{1}{16r}$ and monotone decreasing of $C_2\xi^{-\beta_1}$ in $\xi$ from $+\infty$ to 0.
We define
\begin{align}
      \rho=1-C_1 e^{-\alpha_1\xi_0}\label{rho}
\end{align}
and
\begin{align}
      \bbm{1}_{t_k}\tq\bbm{1}\{1-\sigma_-(\mbf{V}_{t_k})\geq \rho\}\label{Itk}
\end{align}
Notice that $\rho\in (0,1)$. Using $(1-\sigma_-(\mbf{V}_{t_k}))^m\leq \rho^{(1-\bbm{1}_{t_k})m}$ and $mn\geq n$,
\begin{align*}
      \bb{E}_{\theta^*}\Big[\prod_{k=1}^{n}(1-\sigma_-(\mbf{V}_{t_k}))^m\Big]
      \leq \bb{E}_{\theta^*}\Big[\prod_{k=1}^{n}\rho^{(1-\bbm{1}_{t_k})m}\Big]
      =\rho^{mn}\bb{E}_{\theta^*}\Big[\prod_{k=1}^n\rho^{-m \bbm{1}_{t_k}}\Big]
      \leq\rho^{n}\bb{E}_{\theta^*}\Big[\prod_{k=1}^n\rho^{-m \bbm{1}_{t_k}}\Big].
\end{align*}
Next, we find an upper bound for $\bb{E}_{\theta^*}[\prod_{k=1}^n\rho^{-m \bbm{1}_{t_k}}]$. Using the generalized H\"{o}lder's inequality,
\begin{align}
      \bb{E}_{\theta^*}\Big[\prod_{k=1}^n\rho^{-m \bbm{1}_{t_k}}\Big]\leq \prod_{k=1}^n \big(\bb{E}_{\theta^*}[\rho^{-mn\bbm{1}_{t_k}}]\big)^{\frac{1}{n}}.\label{lemma22}
\end{align}
Given (\ref{lemma20}) and (\ref{lemma21}),
\begin{align}
    &\bb{E}_{\theta^*}[\rho^{-mn \bbm{1}_{t_k}}]
    = \rho^{-mn} \bb{P}_{\theta^*}(\bbm{1}_{t_k}=1)+\bb{P}_{\theta^*}(\bbm{1}_{t_k}=0)
    \leq \rho^{-mn}\bb{P}_{\theta^*}(1-\sigma_-(\mbf{V}_{t_k})\geq \rho)+1\nonumber\\
    &\leq \rho^{-mn}( C_2\xi_0^{-\beta_1}+(r-1)C_4\xi_0^{-\beta_2})+1
    =\rho^{-mn}\times \frac{1}{16r} \rho^{mn}+1\leq 2.\label{lemma23}
\end{align}
The proof is completed by plugging (\ref{lemma23}) back into (\ref{lemma22}).

Next, we show (\ref{mix1}). Let $\rho$ and $\bbm{1}_{t_k}$ be defined as in (\ref{rho}) and (\ref{Itk}), respectively. Define
\begin{align}
      \varepsilon_0= C_2\xi_0^{-\beta_1}.\label{epsilon0}
\end{align}
Then, $\varepsilon_0\in (0,\frac{1}{16r})$, and
\begin{align*}
      \bb{P}_{\theta^*}(1-\sigma_-(\mbf{V}_t) \geq \rho)\leq \varepsilon_0.
\end{align*}
Using $1-\sigma_-(\mbf{V}_{t_k})\leq \rho^{1-\bbm{1}_{t_k}}$,
\begin{align*}
      \prod_{k=1}^n \big( 1-\sigma_-(\mbf{V}_{t_k}) \big)
      \leq \rho^{n-\sum_{k=1}^n \bbm{1}_{t_k}}=\rho^n a_n
\end{align*}
with $a_n\tq \rho^{-\sum_{k=1}^n \bbm{1}_{t_k}}$. Since $\mbf{V}_{t_k}$ is stationary and ergodic, it follows from the strong law of large numbers that $n^{-1}\sum_{k=1}^n \bbm{1}_{t_k}\convas \bb{E}_{\theta^*} [\bbm{1}_{t_k}]<\varepsilon_0$. Hence,
\begin{align*}
      \bb{P}_{\theta^*}\Big(\liminf_{n\ra\infty}\Big\{\prod_{k=1}^n \big( 1-\sigma_-(\mbf{V}_{t_k}) \big)\leq \rho^{(1-2\varepsilon_0)n}\Big\}\Big)
      \geq \bb{P}_{\theta^*}\big(\liminf_{n\ra\infty}\{a_n\leq \rho^{-2\varepsilon_0 n}\}\big)=1,
\end{align*}
and (\ref{mix1}) follows.

Next, we show (\ref{mix2}). Let $\rho$ and $\varepsilon_0$ be defined as in (\ref{rho}) and (\ref{Itk}), respectively. For each $t\geq 2r$, set $\xi_t$ such that it satisfies $\rho^{\varepsilon_0\lf(t-r)/r\rf}=e^{-\alpha_1\xi_t}$. The existence of $\xi_t$ is guaranteed, since $\rho^{\varepsilon_0\lf(t-r)/r\rf}\in(0,1)$, and $e^{-\alpha_1\xi_t}$ is monotone decreasing in $\xi_t$ from 1 to 0. Then,
\begin{align*}
      \sum_{t=1}^\infty\bb{P}_{\theta^*}\big(\sigma_-(\mbf{V}_t)\leq C_1\rho^{\varepsilon_0\lf(t-r)/r\rf}\big)
      \leq(2r-1) +\sum_{t=2r}^\infty\bb{P}_{\theta^*}\big(\sigma_-(\mbf{V}_t)\leq C_1e^{-\alpha_1 \xi_t}\big)<\infty.
\end{align*}
By the Borel--Cantelli lemma, $\bb{P}_{\theta^*}\big(\sigma_-(\mbf{V}_t) \leq C_1\rho^{\varepsilon_0\lf(t-r)/r\rf}\ \textrm{i.o.}\big)=0$.
\hfill$\square$

\subsection{Proof of propositions and theorems in Section 4}
\textbf{Proof of Lemma \ref{perioddistance}: }
Write
\begin{align}
    &p_\theta(Y_t|\oline{\mbf{Y}}_{-m}^{t-1},S_{-m}=s_i,\mbf{X}_{-m+1}^t)\nonumber\\
    &=\sum_{s_t,s_{t-r},s_{-m}}
    g_\theta(Y_t|\oline{\mbf{Y}}_{t-1},s_{t},X_t)
    \bb{P}_\theta(S_t=s_t| S_{t-r}=s_{t-r},\allowbreak\oline{\mbf{Y}}_{t-r}^{t-1},\mbf{x}_{t-r+1}^t)\nonumber\\
    &\hspace{2cm}\times\bb{P}_\theta(S_{t-r}=s_{t-r}|S_{-m}=s_{-m},\oline{\mbf{Y}}_{-m}^{t-1},\mbf{X}_{-m+1}^{t-1})\delta_{s_i}(s_{-m})\label{equapyt}
\end{align}
and similarly for $p_\theta(Y_t|\oline{\mbf{Y}}_{-m'}^{t-1},S_{-m'}=s_j,\mbf{X}_{-m'+1}^t)$. It follows that
\begin{align*}
    &\Big|p_\theta(Y_t|\oline{\mbf{Y}}_{-m}^{t-1},S_{-m}=s_i,\mbf{X}_{-m+1}^t)-p_\theta(Y_t|\oline{\mbf{Y}}_{-m'}^{t-1},S_{-m'}=s_{j},\mbf{X}_{-m'+1}^t)\Big|\\
    =&\bigg|\sum_{s_t,s_{t-r},s_{-m}}
    \Big[g_\theta(Y_t|\oline{\mbf{Y}}_{t-1},s_{t},X_t)
    \bb{P}_\theta(S_t=s_{t}|S_{t-r}=s_{t-r},\oline{\mbf{Y}}_{t-r}^{t-1},\mbf{X}_{t-r+1}^t)\\
    &\hspace{2cm}\times\bb{P}_\theta(S_{t-r}=s_{t-r}|S_{-m}=s_{-m},\oline{\mbf{Y}}_{-m}^{t-1},\mbf{X}_{-m+1}^{t-1})\\
    &\hspace{2cm}\times\big(\delta_{s_i}(s_{-m})-\bb{P}_\theta(S_{-m}=s_{-m}|S_{-m'}=s_{j},\oline{\mbf{Y}}_{-m'}^{t-1},\mbf{X}_{-m'+1}^{t-1}) \big) \Big]\bigg|\\
    \leq&\prod_{i=1}^{\lf (t-r+m)/r\rf} \big(1-\sigma_-(\mbf{V}_{-m+ri})\big)
    \sum_{s_{t}}g_\theta(Y_t|\oline{\mbf{Y}}_{t-1},s_{t},X_t).
\end{align*}
Moreover, we rewrite the period predictive density as
\begin{align*}
    &p_\theta(Y_t|\oline{\mbf{Y}}_{-m}^{t-1},S_{-m}=s_i,\mbf{X}_{-m+1}^t)\\
    &=\sum_{s_t,s_{t-r},s_{-m}}
    \big(g_\theta(Y_t|\oline{\mbf{Y}}_{t-1},s_{t},X_t)
    \bb{P}_\theta(S_t=s_{t}|S_{t-r}=s_{t-r},\oline{\mbf{Y}}_{t-r}^{t-1},\mbf{X}_{t-r+1}^t)\\
    &\hspace{3cm}\times\bb{P}_\theta(S_{t-r}=s_{t-r}|S_{-m}=s_{-m},\oline{\mbf{Y}}_{-m}^{t-1},\mbf{X}_{-m+1}^{t-1})\delta_{s_i}(s_{-m})\big)
\end{align*}
and similarly for $p_\theta(Y_t|\oline{\mbf{Y}}_{-m'}^{t-1},S_{-m'}=s_j,\mbf{X}_{-m'+1}^t)$. It follows that
\begin{align*}
    &p_\theta(Y_t|\oline{\mbf{Y}}_{-m}^{t-1},S_{-m}=s_{i},\mbf{X}_{-m+1}^t)\wedge p_\theta(Y_t|\oline{\mbf{Y}}_{-m}^{t-1},S_{-m'}=s_j,\mbf{X}_{-m'+1}^t)\\
    \geq&\min_{s_t,s_{t-r}}
    \bb{P}_\theta(S_t=s_{t}|S_{t-r}=s_{t-r},\oline{\mbf{Y}}_{t-r}^{t-1},\mbf{X}_{t-r+1}^t)
    \times\sum_{s_{t}}g_\theta(Y_t|\oline{\mbf{Y}}_{t-1},s_{t},X_t)\\
    =&\sigma_-( \mbf{V}_t )\sum_{s_{t}}g_\theta(Y_t|\oline{\mbf{Y}}_{t-1},s_{t},X_t).
\end{align*}
By $|\log x-\log y|\leq|x-y|/(x\wedge y)$,
\begin{align*}
    |\log p_\theta(Y_t|\oline{\mbf{Y}}_{-m}^{t-1},S_{-m}=s_{i},\mbf{X}_{-m+1}^t)-\log p_\theta(Y_t|\oline{\mbf{Y}}_{-m'}^{t-1},S_{-m'}=s_{j},\mbf{X}_{-m'+1}^t)|\\
     \leq \frac{\prod_{i=1}^{\lf (t-r+m)/r\rf}\big(1-\sigma_-( \mbf{V}_{-m+ri} )\big)}{\sigma_-(\mbf{V}_{t})}.
\end{align*}
By (\ref{mix1}) and (\ref{mix2}) in Lemma \ref{mixingbound},
\begin{align*}
    \bb{P}_{\theta^*}\big(|\Delta_{t,m,s_i}(\theta)-\Delta_{t,m',s_j}(\theta)|\geq \frac{1}{C_1}\rho^{(1-3\varepsilon_0){\lf (t-r+m)/r\rf}}\ \mathrm{i.o.}\big)=0.
\end{align*}
Then (\ref{1.1}) follows because $(1-3\varepsilon_0){\lf (t-r+m)/r\rf}\geq \lf(t-r+m)/r\rf/2\geq\lf(t-r+m)/2r\rf\geq\lf (t+m)/3r\rf$. 
(\ref{1.2}) follows by replacing
$\bb{P}_\theta(S_{-m}=s_{-m}|S_{-m'}=s_{j},\oline{\mbf{Y}}_{-m'}^{t-1},\mbf{X}_{-m'+1}^t)$
in the proof of (\ref{1.1}) with $\bb{P}_\theta(S_{-m}=s_{-m}|\oline{\mbf{Y}}_{-m}^{t-1},\mbf{X}_{-m+1}^t)$. Finally, in light of (\ref{equapyt}),
      \begin{align*}
            &\sup_{\theta\in\Theta}\sup_{m\geq0}\max_{s\in\bb{S}}
            |\Delta_{t,m,s}(\theta)|\leq
            \max\big\{\big|\log b_+\big|,
            \big|\log(b_-(\mbf{Y}_{t-r}^t,X_t))\big|\big\}
      \end{align*}
follows from
\begin{align*}
    b_-(\mbf{Y}_{t-r}^t,X_t)
    \leq p_\theta(Y_t|\oline{\mbf{Y}}_{-m}^{t-1},S_{-m}=s,\mbf{X}_{-m+1}^t)
    \leq b_+.
\end{align*}
Hence, $\{\Delta_{t,m,s}(\theta)\}_{m\geq0}$ is uniformly bounded in $L^1(\bb{P}_{\theta^*})$ by Assumption \ref{bound1}.
\hfill$\square$

\noindent\textbf{Proof of Proposition \ref{uconvergence}: }
By the compactness of $\Theta$, it suffices to show $\forall \theta\in\Theta$,
\begin{align*}
    \limsup_{\delta\ra0}\limsup_{n\ra\infty}\sup_{\theta':|\theta'-\theta|\leq\delta}|n^{-1}\ell_n(\theta',s_0)-\ell(\theta)|=0,\quad \bb{P}_{\theta^*}\text{-a.s.}
\end{align*}
We decompose the difference as
\begin{align*}
    &\limsup_{\delta\ra0}\limsup_{n\ra\infty}\sup_{\theta':|\theta'-\theta|\leq\delta}|n^{-1}\ell_n(\theta',s_0)-\ell(\theta)|\\
    &=\limsup_{\delta\ra0}\limsup_{n\ra\infty}\sup_{\theta':|\theta'-\theta|\leq\delta}|n^{-1}\ell_n(\theta',s_0)-\ell(\theta')|
    +\limsup_{\delta\ra0}\sup_{\theta':|\theta'-\theta|\leq \delta}|\ell(\theta')-\ell(\theta)|.
\end{align*}
On the one hand, the first term
\begin{align*}
    &\limsup_{\delta\ra0}\limsup_{n\ra\infty}\sup_{\theta':|\theta'-\theta|\leq\delta}|n^{-1}\ell_n(\theta',s_0)-\ell(\theta')|\\
    &\leq\limsup_{n\ra\infty}\frac{1}{n}\sum_{t=1}^n\sup_{\theta'\in\Theta}|\Delta_{t,0,s_0}(\theta')-\Delta_{t,\infty}(\theta')|
    +\limsup_{n\ra\infty}\sup_{\theta'\in\Theta}|\frac{1}{n}\sum_{t=1}^n\Delta_{t,\infty}(\theta')-\ell(\theta')|\\&= 0,\quad \bb{P}_{\theta^*}\text{-a.s.}
\end{align*}
The last step follows from (\ref{1}) and (\ref{2}).
On the other hand, note that
\begin{align*}
    &\Delta_{0,m,s}(\theta)=p_\theta(Y_0|\oline{\mbf{Y}}_{-m}^{-1},S_{-m}=s_{-m},\mbf{X}_{-m+1}^0)
    =\frac{p_\theta(\mbf{Y}_{-m+1}^0|\oline{\mbf{Y}}_{-m},S_{-m}=s_{-m},\mbf{X}_{-m+1}^0)}{p_\theta(\mbf{Y}_{-m+1}^{-1}|\oline{\mbf{Y}}_{-m},S_{-m}=s_{-m},\mbf{X}_{-m+1}^{-1})},\\
    &\text{where } p_\theta(\mbf{Y}_{-m+1}^j|\oline{\mbf{Y}}_{-m},S_{-m}=s_{-m},\mbf{X}_{-m+1}^j)\\
    &\qquad\quad=\sum_{\mbf{s}_{-m+1}^j}\prod_{\ell=-m+1}^jq_\theta(s_\ell|s_{\ell-1},\oline{\mbf{Y}}_{\ell-1},X_\ell)\prod_{\ell=-m+1}^j g_\theta(Y_\ell|\oline{\mbf{Y}}_{\ell-1},s_\ell,X_\ell)\quad \text{for }j=0,-1.
\end{align*}
Therefore, $\Delta_{0,m,s}(\theta)$ is continuous with respect to $\theta$ from Assumption \ref{continuityass}. It follows that $\Delta_{0,\infty}(\theta)$ is continuous by uniform convergence.
Then, the second term is bounded by
\begin{align*}
    \limsup_{\delta\ra0}\sup_{\theta':|\theta'-\theta|\leq\delta}|\ell(\theta')-\ell(\theta)|
    &\leq\limsup_{\delta\ra0}\sup_{\theta':|\theta'-\theta|\leq\delta}\bb{E}_{\theta^*}[|\Delta_{0,\infty}(\theta')-\Delta_{0,\infty}(\theta)|]\\
    &\leq\limsup_{\delta\ra0}\bb{E}_{\theta^*}\Big[\sup_{\theta':|\theta'-\theta|\leq\delta}|\Delta_{0,\infty}(\theta')-\Delta_{0,\infty}(\theta)|\Big]\\
    & =\bb{E}_{\theta^*}\Big[\limsup_{\delta\ra0}\sup_{\theta':|\theta'-\theta|\leq\delta}|\Delta_{0,\infty}(\theta')-\Delta_{0,\infty}(\theta)|\Big]=0,
\end{align*}
where the first equality follows by the dominated convergence theorem.
\hfill$\square$

\begin{proposition}[Identification]
Under \ref{ergodic} and \ref{identifiability}, $\ell(\theta)\leq\ell(\theta^*)$ and $\ell(\theta)=\ell(\theta^*)$ if and only if $\theta=\theta^*$.
      \label{identification}
\end{proposition}
\noindent\textbf{Proof: }
\begin{align}
    \ell(\theta)&=\bb{E}_{\theta^*}\Big[ \lim_{m\ra\infty}\log p_\theta(Y_1|\oline{\mbf{Y}}_{-m}^0,S_{-m},\mbf{X}_{-m+1}^1) \Big]\notag\\
    & = \lim_{m\ra\infty} \bb{E}_{\theta^*}\Big[\log p_\theta(Y_1|\oline{\mbf{Y}}_{-m}^0,S_{-m},\mbf{X}_{-m+1}^1) \Big]\label{id1}\\
    &=\lim_{m\ra\infty}\bb{E}_{\theta^*}\big[\bb{E}_{\theta^*}[\log  p_\theta(Y_1|\oline{\mbf{Y}}_{-m}^0,S_{-m},\mbf{X}_{-m+1}^1)|\oline{\mbf{Y}}_{-m}^0,S_{-m},\mbf{X}_{-m+1}^1]\big].\notag
\end{align}
It follows that
\begin{align*}
    \ell(\theta^*)&-\ell(\theta)
    &=\lim_{m\ra\infty}\bb{E}_{\theta^*}\bigg[\bb{E}_{\theta^*}\Big[\log\frac{p_{\theta^*}(Y_1|\oline{\mbf{Y}}_{-m}^0,S_{-m},\mbf{X}_{-m+1}^1)}{p_\theta(Y_1|\oline{\mbf{Y}}_{-m}^0,S_{-m},\mbf{X}_{-m+1}^1)} \Big|\oline{\mbf{Y}}_{-m}^0,S_{-m},\mbf{X}_{-m+1}^1\Big] \bigg]\geq0.
\end{align*}
The nonnegativity follows because Kullback--Leibler divergence is nonnegative, and thus, the limit of its expectation is nonnegative. Next, we show that $\theta^*$ is the unique maximizer. The proof closely follows \citet[pp. 2269--2270]{douc2004asymptotic}. Assume $\ell(\theta)=\ell(\theta^*)$. For any $t\geq 1$ and $m\geq0$,
\begin{align*}
    \bb{E}_{\theta^*}[\log p_\theta(\mbf{Y}_1^t|\oline{\mbf{Y}}_{-m}^0,S_{-m},\mbf{X}_{-m+1}^t)]
    =\sum_{k=1}^t \bb{E}_{\theta^*} [\log p_\theta(Y_k|\oline{\mbf{Y}}_{-m}^{k-1},S_{-m},\mbf{X}_{-m+1}^k)].
\end{align*}
By (\ref{id1}) and stationarity, $\lim_{m\ra\infty} \bb{E}_{\theta^*}[\log p_\theta(\mbf{Y}_1^t|\oline{\mbf{Y}}_{-m}^0,S_{-m},\mbf{X}_{-m+1}^t)]=t\ell(\theta)$.
For $1\leq k\leq t-p+1$,
\begin{align}
    0&=t(\ell(\theta^*)-\ell(\theta)) =\lim_{m\ra\infty}\bb{E}_{\theta^*}\bigg[ \log\frac{p_{\theta^*}(\mbf{Y}_1^t|\oline{\mbf{Y}}_{-m}^0,S_{-m},\mbf{X}_{-m+1}^t)}{p_\theta(\mbf{Y}_1^t|\oline{\mbf{Y}}_{-m}^0,S_{-m},\mbf{X}_{-m+1}^t)} \bigg] \nonumber\\
    =&\limsup_{m\ra\infty}\bb{E}_{\theta^*}\bigg[ \log\frac{p_{\theta^*}(\mbf{Y}_{t-k+1}^t|\oline{\mbf{Y}}_{t-k},\oline{\mbf{Y}}_{-m}^{0},S_{-m},\mbf{X}_{-m+1}^t)}{p_{\theta}(\mbf{Y}_{t-k+1}^t|\oline{\mbf{Y}}_{t-k},\mbf{Y}_{-m}^{0},S_{-m},\mbf{X}_{-m+1}^t)} \nonumber \\
    & +\log\frac{p_{\theta^*}(\oline{\mbf{Y}}_{t-k}|\oline{\mbf{Y}}^{0}_{-m},S_{-m},\mbf{X}_{-m+1}^{t})}{p_{\theta}(\oline{\mbf{Y}}_{t-k}|\oline{\mbf{Y}}^{0}_{-m},S_{-m},\mbf{X}_{-m+1}^{t})}
    +\log\frac{p_{\theta^*}(\mbf{Y}_{1}^{t-k-p}|\mbf{Y}_{t-k-p+1}^{t},\oline{\mbf{Y}}_{-m}^{0},S_{-m},\mbf{X}_{-m+1}^t)}{p_{\theta}(\mbf{Y}_{1}^{t-k-p}|\mbf{Y}_{t-k-p+1}^{t},\oline{\mbf{Y}}_{-m}^{0},S_{-m},\mbf{X}_{-m+1}^t)} \bigg]\nonumber\\
    \geq&\limsup_{m\ra\infty} \bb{E}_{\theta^*}\bigg[ \log\frac{p_{\theta^*}(\mbf{Y}_{t-k+1}^t|\oline{\mbf{Y}}_{t-k},\oline{\mbf{Y}}_{-m}^{0},S_{-m},\mbf{X}_{-m+1}^t)}{p_{\theta}(\mbf{Y}_{t-k+1}^t|\oline{\mbf{Y}}_{t-k},\mbf{Y}_{-m}^{0},S_{-m},\mbf{X}_{-m+1}^t)} \bigg]\nonumber\\
    =& \limsup_{m\ra\infty} \bb{E}_{\theta^*}\bigg[ \log\frac{p_{\theta^*}(\mbf{Y}_{1}^k|\oline{\mbf{Y}}_{0},\oline{\mbf{Y}}_{-m-t+k}^{-t+k},S_{-m-t+k},\mbf{X}_{-m+1-t+k}^k)}{p_{\theta}(\mbf{Y}_{1}^k|\oline{\mbf{Y}}_{0},\oline{\mbf{Y}}_{-m-t+k}^{-t+k},S_{-m-t+k},\mbf{X}_{-m+1-t+k}^k)} \bigg] \label{ineq0t}.
\end{align}
This holds when we let $t\ra\infty$. It suffices to show that for all $t\geq0$ and all $\theta\in\Theta$,
\begin{align}
    \sup_{m\geq k} \Bigg|\bb{E}_{\theta^*}\bigg[\log \frac{p_{\theta^*}(\mbf{Y}_1^t|\oline{\mbf{Y}}_{0},\oline{\mbf{Y}}_{-m}^{-k},S_{-m},\mbf{X}_{-m}^t)}{p_{\theta}(\mbf{Y}_1^t|\oline{\mbf{Y}}_{0},\oline{\mbf{Y}}_{-m}^{-k},S_{-m},\mbf{X}_{-m}^t)}\bigg]-\bb{E}_{\theta^*}\bigg[\log \frac{p_{\theta^*}(\mbf{Y}_1^t|\oline{\mbf{Y}}_{0},\mbf{X}_{-p+1}^t)}{p_{\theta}(\mbf{Y}_1^t|\oline{\mbf{Y}}_{0},\mbf{X}_{-p+1}^t)}\bigg]\Bigg|\ra0,\label{KL1}
\end{align}
as $k\ra\infty$. From (\ref{ineq0t}) and (\ref{KL1}), if $\ell(\theta^*)=\ell(\theta)$, then $\bb{E}_{\theta^*}\big[\log \big( p_{\theta^*}(\mbf{Y}_1^t|\oline{\mbf{Y}}_{0},\mbf{X}_{-p+1}^t)/\allowbreak p_{\theta}(\mbf{Y}_1^t|\oline{\mbf{Y}}_{0},\mbf{X}_{-p+1}^t)\big)\big]=0$. The laws $\bb{P}_{\theta^*}(\mbf{Y}_1^t\in \cdot|\oline{\mbf{Y}}_{0},\mbf{X}_{-p+1}^t)$ and $\bb{P}_{\theta}(\mbf{Y}_1^t\in \cdot|\oline{\mbf{Y}}_{0},\mbf{X}_{-p+1}^t)$ agree. From Assumption \ref{identifiability}, $\theta=\theta^*$.

Next we show (\ref{KL1}). Define $U_{k,m}(\theta)\tq\log p_\theta(\mbf{Y}_1^t|\oline{\mbf{Y}}_{0},\oline{\mbf{Y}}_{-m}^{-k},S_{-m},\mbf{X}_{-m}^t)$, and $U(\theta)\tq\log p_\theta(\mbf{Y}_1^t|\oline{\mbf{Y}}_{0},\mbf{X}_{-p+1}^t)$. It is enough to show that for all $\theta\in\Theta$,
\begin{align*}
    \bb{E}_{\theta^*}\Big[ \sup_{m\geq k} |U_{k,m}(\theta)-U(\theta)| \Big]\ra0, \text{ as }k\ra\infty.
\end{align*}
Put
\begin{align*}
    &A_{k,m}=p_\theta(\oline{\mbf{Y}}_{0}^t|\oline{\mbf{Y}}_{-m}^{-k},S_{-m},\mbf{X}_{-m}^t),\quad
    A = p_\theta(\oline{\mbf{Y}}_{0}^t|\mbf{X}_{-p+1}^t),\\
    &B_{k,m}=p_\theta(\oline{\mbf{Y}}_{0}|\oline{\mbf{Y}}_{-m}^{-k},S_{-m},\mbf{X}_{-m}^0),\quad
    B = p_\theta(\oline{\mbf{Y}}_{0}|\mbf{X}_{-p+1}^0).
\end{align*}
Then
\begin{align}
    |p_\theta(\mbf{Y}_1^t|\oline{\mbf{Y}}_{0},\oline{\mbf{Y}}_{-m}^{-k},S_{-m},\mbf{X}_{-m}^t)&-p_\theta(\mbf{Y}_1^t|\oline{\mbf{Y}}_{0},\mbf{X}_{-p+1}^t)|=\Big|\frac{A_{k,m}}{B_{k,m}}-\frac{A}{B}\Big|\notag\\
    &\leq \frac{B|A_{k,m}-A|+A|B_{k,m}-B|}{B B_{k,m}}.\label{ab}
\end{align}
For all $t\geq 0$ and $k\geq p$, write
\begin{align*}
    &p_\theta(\oline{\mbf{Y}}_{0}^t|\oline{\mbf{Y}}_{-m}^{-k},S_{-m},\mbf{X}_{-m}^t)\\
    =&\int\int p_\theta(\oline{\mbf{Y}}_{0}^t|Z_{-p}=z_{-p},\mbf{X}_{-p+1}^t)\bb{P}_\theta(dz_{-p}|Z_{-k+1}=z_{-k+1},\mbf{X}_{-k-p+2}^{-p})\\
    &\hspace{7cm}\times\bb{P}_\theta(dz_{-k+1}|\oline{\mbf{Y}}_{-m}^{-k},S_{-m},\mbf{X}_{-m}^{-k+1})\\
    &p_\theta(\oline{\mbf{Y}}_{0}^t|\mbf{X}_{-p+1}^t)\\
    =&\int\int p_\theta(\oline{\mbf{Y}}_{0}^t|Z_{-p}=z_{-p},\mbf{X}_{-p+1}^t)\bb{P}_\theta(dz_{-p}|\mbf{X}_{-2p+1}^{-p})\\
    &\hspace{7cm}\times\bb{P}_\theta(dz_{-k+1}|\oline{\mbf{Y}}_{-m}^{-k},S_{-m},\mbf{X}_{-m}^{-k+1}),
\end{align*}
where $Z_t\tq(\oline{\mbf{Y}}_t,S_t)'$. The second expression holds because conditionally on $X_t$, $\{X_k\}_{k\geq t+1}$ is independent of $\{Z_k\}_{k\leq t}$, and conditionally on $X_{t}$, $\{X_k\}_{k\leq t-1}$ is independent of $\{Z_k\}_{k\geq t}$. An upper bound of their difference is
\begin{align*}
    &|p_\theta(\oline{\mbf{Y}}_{0}^t|\oline{\mbf{Y}}_{-m}^{-k},S_{-m},\mbf{X}_{-m}^t)
    -p_\theta(\oline{\mbf{Y}}_{0}^t|\mbf{X}_{-p+1}^t)|\\
    \leq& \int\int p_\theta(\oline{\mbf{Y}}_{0}^t|Z_{-p}=z_{-p},\mbf{X}_{-p+1}^t)\\
    \times&\big| \bb{P}_\theta(dz_{-p}|Z_{-k+1}=z_{-k+1},\mbf{X}_{-k-p+2}^{-p})-\bb{P}_\theta(dz_{-p}|\mbf{X}_{-2p+1}^{-p})\big|
    \bb{P}_\theta(dz_{-k+1}|\oline{\mbf{Y}}_{-m}^{-k},S_{-m},\mbf{X}_{-m}^t)\\
    \leq& b_+^{t+p} \int \|\bb{P}_\theta(z_{-p}\in \cdot |Z_{-k+1}=z_{-k+1},\mbf{X}_{-k-p+2}^{-p})-\bb{P}_\theta(z_{-p}\in \cdot|\mbf{X}_{-2p+1}^{-p})\|_{TV}\\
    &\hspace{8cm}\times\bb{P}_\theta(dz_{-k+1}|\oline{\mbf{Y}}_{-m}^{-k},S_{-m},\mbf{X}_{-m}^t).
\end{align*}
$\|\bb{P}_\theta(z_{-p}\in \cdot |Z_{-k+1}=z_{-k+1},\mbf{X}_{-k-p+2}^{-p})-\bb{P}_\theta(z_{-p}\in \cdot|\mbf{X}_{-2p+1}^{-p})\|_{TV}$ goes to zero as $k$ goes to infinity, owing to the Markovian property of $Z_t$ conditional on $\mbf{X}_{-\infty}^{+\infty}$ and Assumption \ref{ergodic}. Thus,            \begin{align}\label{abdiff}
    \sup_{m\geq k} |p_\theta(\oline{\mbf{Y}}_{0}^t|\oline{\mbf{Y}}_{-m}^{-k},S_{-m},\mbf{X}_{-m}^t)-p_\theta(\oline{\mbf{Y}}_{0}^t|\mbf{X}_{-p+1}^t)|\convas 0, \text{ as } k\ra\infty.
\end{align}
Using Assumption \ref{bound1},
\begin{align*}
    B=p_\theta(\oline{\mbf{Y}}_{0}|\mbf{X}_{-p+1}^0)
    =&\int p_\theta(\oline{\mbf{Y}}_{0}|Z_{-p}=z_{-p},\mbf{X}_{-p+1}^0)\bb{P}_\theta(dz_{-p}|\mbf{X}_{-2p+1}^{-p})\\
    \geq& \int \prod_{s=-p+1}^0 b_-(Y_s,\oline{\mbf{Y}}_{s-1},X_s) \bb{P}_\theta(dz_{-p}|\mbf{X}_{-2p+1}^{-p})>0.
\end{align*}
By (\ref{abdiff}), with $\bb{P}_{\theta^*}$-probability arbitrarily close to 1, $B_{k,m}$ is uniformly bounded away from zero for $m\geq k$ and $k$ sufficiently large. By (\ref{ab}) and (\ref{abdiff}),
\begin{align*}
    \sup_{m\geq k}|p_\theta(\mbf{Y}_1^t|\oline{\mbf{Y}}_{0},\oline{\mbf{Y}}_{-m}^{-k},S_{-m},\mbf{X}_{-m}^t)-p_\theta(\mbf{Y}_1^t|\oline{\mbf{Y}}_{0},\mbf{X}_{-p+1}^t)|\convp 0, \text{ as }k\ra\infty.
\end{align*}
Since 
$p_\theta(\mbf{Y}_1^t|\oline{\mbf{Y}}_0,\oline{\mbf{Y}}_{-m}^{-k},S_{-m},\mbf{X}_{-m}^t)
=\sum_{s}p_\theta(\mbf{Y}_1^t|\oline{\mbf{Y}}_0,S_{0}=s,\mbf{X}_{1}^t)
\bb{P}_{\theta}(S_{0}=s|\oline{\mbf{Y}}_0,S_{-m},\allowbreak\mbf{X}_{-m}^t)$, it is bounded below by $\prod_{s=1}^tb_-(Y_s,\oline{\mbf{Y}}_{s-1},X_s)$ and bounded above by $b_+^t$. The same lower bound can be attained for $p_\theta(\mbf{Y}_1^t|\oline{\mbf{Y}}_{0},\mbf{X}_{-p+1}^t)$. Use the inequality $|\log x-\log y|\leq |x-y|/(x\wedge y)$,
\begin{align*}
    \sup_{m\geq k} |U_{k,m}(\theta)-U(\theta)|\convp0, \text{ as } k\ra\infty.
\end{align*}
Using the bounds of $p_\theta(\mbf{Y}_1^t|\oline{\mbf{Y}}_0,\oline{\mbf{Y}}_{-m}^{-k},S_{-m},\mbf{X}_{-m}^t)$,
\begin{align*}
    \bb{E}_{\theta^*}\Big[ \sup_k \sup_{m\geq k} |U_{k,m}(\theta)| \Big]<\infty.
\end{align*}
The proof is completed by applying the bounded convergence theorem.
\hfill$\square$

\noindent\textbf{Proof of Theorem \ref{consistency}: }
As $\theta^*$ is a well-separated maximum of $\ell(\theta)$, we have $\forall \varepsilon>0$, $\sup_{\theta:|\theta-\theta^*|\geq\varepsilon} \ell(\theta)<\ell(\theta^*)$. For $\varepsilon$, $\delta>0$ exists such that $|\theta-\theta^*|>\varepsilon$ implies $\ell(\theta)<\ell(\theta^*)-\delta$, and thus,
\begin{align*}
    \bb{P}_{\theta^*}(|\hat{\theta}_{n,s_0}-\theta^*|>\varepsilon)
    \leq \bb{P}_{\theta^*}(\ell(\hat{\theta}_{n,s_0})<\ell(\theta^*)-\delta)
    = \bb{P}_{\theta^*}(\ell(\theta^*)-\ell(\hat{\theta}_{n,s_0})>\delta).
\end{align*}
The proof is completed by Proposition \ref{uconvergence} and
\begin{align*}
    &\ell(\theta^*)-\ell(\hat{\theta}_{n,s_0})
    =n^{-1}\ell_n(\theta^*,s_0)-\ell(\hat{\theta}_{n,s_0})+\ell(\theta^*)-n^{-1}\ell_n(\theta^*,s_0)\\
    \leq&n^{-1}\ell_n(\hat{\theta}_{n,s_0},s_0)-\ell(\hat{\theta}_{n,s_0})+\ell(\theta^*)-n^{-1}\ell_n(\theta^*,s_0)\\
    \leq& 2\sup_{\theta\in\Theta}|n^{-1}\ell_n(\theta,s_0)-\ell(\theta)|.
\end{align*}
\hfill$\square$

\subsection{Proof of lemmas and theorems in Subsection 5.1}
We first show that $\dot{\Delta}_{t,\infty}(\theta^*)$ is well-defined in $L^2(\bb{P}_{\theta^*})$. Since $\{\bb{E}_{\theta^*}[\phi_{\theta^*,k}|\oline{\mbf{Y}}_{-m}^{t},\mbf{X}_{-m}^{t}]\}_{m\geq0}$ is a martingale, by Jensen's inequality, $\{\|\bb{E}_{\theta^*}[\phi_{\theta^*,k}|\oline{\mbf{Y}}_{-m}^{t},\mbf{X}_{-m}^{t}]\|^2\}_{m\geq0}$ is a submartingale. Moreover, for any $m$,
\begin{align*}
\bb{E}_{\theta^*}\big[\|\bb{E}_{\theta^*}[\phi_{\theta^*,k}|\oline{\mbf{Y}}_{-m}^{t},\mbf{X}_{-m}^{t}]\|^2\big]&\leq\bb{E}_{\theta^*}\big[\bb{E}_{\theta^*}[\|\phi_{\theta^*,k}\|^2|\oline{\mbf{Y}}_{-m}^{t},\mbf{X}_{-m}^{t}]\big]
=\bb{E}_{\theta^*}\big[\|\phi_{\theta^*,k}\|^2\big]<\infty
\end{align*}
under Assumption \ref{finitemoment}. Then by the martingale convergence theorem (see, e.g., \citealp[p.508]{shiryaev1996}),
\begin{align}\label{appdefine2}
      \|\bb{E}_{\theta^*}[\phi_{\theta^*,k}|\oline{\mbf{Y}}_{-m}^t,\mbf{X}_{-m}^t]\|^2
      \convas\|\bb{E}_{\theta^*}[\phi_{\theta^*,k}|\oline{\mbf{Y}}_{-\infty}^t,\mbf{X}_{-\infty}^t]\|^2,
\end{align}
as $m\ra\infty$ and
\begin{align}
\bb{E}_{\theta^*}\big[\|\bb{E}_{\theta^*}[\phi_{\theta^*,k}|\oline{\mbf{Y}}_{-\infty}^t,\mbf{X}_{-\infty}^t]\|^2\big]<\infty.
      \label{appdefine3}
\end{align}

On the other hand, by setting $m=\infty$ in (\ref{mix23}),
\begin{align}
      \bb{E}_{\theta^*}\Big[\sum_{k=-\infty}^{t-1}\|\bb{E}_{\theta^*}[\phi_{\theta^*,k}|\oline{\mbf{Y}}_{-\infty}^{t},\mbf{X}_{-\infty}^{t}]-\bb{E}_{\theta^*}[\phi_{\theta^*,k}|\oline{\mbf{Y}}_{-\infty}^{t-1},\mbf{X}_{-\infty}^{t-1}]\|^2\Big]<\infty.\label{appdefine1}
\end{align}

Combining (\ref{appdefine3}) and (\ref{appdefine1}), $\dot{\Delta}_{t,\infty}(\theta^*)$ is well defined in $L^2(\bb{P}_{\theta^*})$.

To see that $\{\dot{\Delta}_{t,\infty}(\theta^*)\}_{k=-\infty}^{\infty}$ is an $(\mcal{F},\bb{P}_{\theta^*})$-adapted martingale increment sequence, note that $\bb{E}_{\theta^*}[\dot{\Delta}_{t,\infty}|\mcal{F}_{t-1}]=0$ follows from
\begin{align}
    &\bb{E}_{\theta^*}\Bigg[ \sum_{k=-\infty}^{t-1}
    \big(\bb{E}_{\theta^*}[\phi_{\theta^*,k}|\oline{\mbf{Y}}_{-\infty}^t,\mbf{X}_{-\infty}^{t}]
    -\bb{E}_{\theta^*}[\phi_{\theta^*,k}|\oline{\mbf{Y}}_{-\infty}^{t-1},\mbf{X}_{-\infty}^{t-1}]\big)\Big|\oline{\mbf{Y}}_{-\infty}^{t-1},\mbf{X}_{-\infty}^{t-1}\Bigg]=0,
    \label{martingale1}\\
    &\bb{E}_{\theta^*}[\phi_{\theta^*,t}|\oline{\mbf{Y}}_{-\infty}^{t-1},\mbf{X}_{-\infty}^{t-1}]
    =\bb{E}_{\theta^*}\big[\bb{E}_{\theta^*}[\phi_{\theta^*,t}|\oline{\mbf{Y}}_{-\infty}^{t-1},S_{t-1},\mbf{X}_{-\infty}^{t-1}]\big|\oline{\mbf{Y}}_{-\infty}^{t-1},\mbf{X}_{-\infty}^{t-1}]\big]=0.
    \label{martingale2}
\end{align}

\noindent\textbf{Proof of Lemma \ref{scoreapproximation}: }
      It suffices to show
      $\lim_{n\ra\infty}\bb{E}_{\theta^*}\big[\| \frac{1}{\sqrt{n}} \sum_{t=1}^n (\dot{\Delta}_{t,0,s}(\theta^*)-\dot{\Delta}_{t,0}(\theta^*))\|^2\big]=0$
      and $\lim_{n\ra\infty}\bb{E}_{\theta^*}\big[\| \frac{1}{\sqrt{n}} \sum_{t=1}^n (\dot{\Delta}_{t,0}(\theta^*)-\dot{\Delta}_{t,\infty}(\theta^*))\|^2\big]=0$.

      First, we show the first term.
      Using the Minkowski inequality and (\ref{mix21}),
      \begin{align*}
            &\big(\bb{E}_{\theta^*} \big[\|\frac{1}{\sqrt{n}}\sum_{t=1}^n (\dot{\Delta}_{t,0,s}(\theta^*)-\dot{\Delta}_{t,0}(\theta^*))\|^2\big]\big)^\frac{1}{2}\\
            =&\big( \bb{E}_{\theta^*}\big[\|\frac{1}{\sqrt{n}}\sum_{t=1}^n( \bb{E}_{\theta^*}[ \phi_{\theta^*,t}|\oline{\mbf{Y}}_{0}^t,S_0=s,\mbf{X}_0^t ]- \bb{E}_{\theta^*}[ \phi_{\theta^*,t}|\oline{\mbf{Y}}_{0}^t,\mbf{X}_0^t ])\|^2\big]\big)^{\frac{1}{2}}\\
            \leq&\frac{1}{\sqrt{n}}\sum_{t=1}^n\big( \bb{E}_{\theta^*}\big[\| \bb{E}_{\theta^*}[ \phi_{\theta^*,t}|\oline{\mbf{Y}}_{0}^t,S_0=s,\mbf{X}_0^t ]- \bb{E}_{\theta^*}[ \phi_{\theta^*,t}|\oline{\mbf{Y}}_{0}^t,\mbf{X}_0^t ]\|^2\big]\big)^{\frac{1}{2}}\\
            \leq&\frac{1}{\sqrt{n}}\sum_{t=1}^n\big(8(\bb{E}_{\theta^*}[\|\phi_{\theta^*,0}\|_\infty^4])^{\frac{1}{2}} \rho^{\lf (t-1)/2r\rf}\big)^{\frac{1}{2}}
            \leq\frac{1}{\sqrt{n}}4(\bb{E}_{\theta^*}[\|\phi_{\theta^*,0}\|_\infty^4])^{\frac{1}{4}}\frac{1}{\rho(1-\rho^{\frac{1}{4r}})}
      \end{align*}
which converges to zero as $n\ra\infty$.

      Next, consider the second term. By Lemma \ref{mixingbound2},
      \begin{align*}
            &\big(\bb{E}_{\theta^*}[\| \dot{\Delta}_{t,0}(\theta^*)-\dot{\Delta}_{t,\infty}(\theta^*)\|^2]\big)^{\frac{1}{2}}\\
            \leq& \big(\bb{E}_{\theta^*}[\|\bb{E}_{\theta^*}[\phi_{\theta^*,t} |\oline{\mbf{Y}}_{-\infty}^t,\mbf{X}_{-\infty}^t]
            -\bb{E}_{\theta^*}[\phi_{\theta^*,t} |\oline{\mbf{Y}}_0^t,\mbf{X}_0^t]\|^2]\big)^\frac{1}{2}\\
            &+\sum_{k=1}^{t-1}\big(\bb{E}_{\theta^*}[\|\bb{E}_{\theta^*}[\phi_{\theta^*,k} |\oline{\mbf{Y}}_{-\infty}^t,\mbf{X}_{-\infty}^t]
            -\bb{E}_{\theta^*}[\phi_{\theta^*,k} |\oline{\mbf{Y}}_{-\infty}^{t-1},\mbf{X}_{-\infty}^{t-1}]\\
            &\qquad\qquad-\bb{E}_{\theta^*}[\phi_{\theta^*,k} |\oline{\mbf{Y}}_{0}^t,\mbf{X}_{0}^t]
            -\bb{E}_{\theta^*}[\phi_{\theta^*,k} |\oline{\mbf{Y}}_0^{t-1},\mbf{X}_0^{t-1}]\|^2]\big)^\frac{1}{2}\\
            &+\sum_{k=-\infty}^{0} \big(\bb{E}_{\theta^*}[\|\bb{E}_{\theta^*}[\phi_{\theta^*,k} |\oline{\mbf{Y}}_{-\infty}^t,\mbf{X}_{-\infty}^t]
            -\bb{E}_{\theta^*}[\phi_{\theta^*,k} |\oline{\mbf{Y}}_{-\infty}^{t-1},\mbf{X}_{-\infty}^{t-1}]\|^2]\big)^\frac{1}{2}\\
            \leq & 4(\bb{E}_{\theta^*}[\|\phi_{\theta^*,0}\|_\infty^4])^{\frac{1}{4}}
            \big(\rho^{\lf(t-1)/4r\rf}+2\sum_{k=1}^{t-1}\rho^{\lf(k-1)/4r\rf}\wedge \rho^{\lf(t-1-k)/4r\rf}+\sum_{k=-\infty}^{0}\rho^{\lf(t-1-k)/4r\rf}\big)\\
            \leq &4(\bb{E}_{\theta^*}[\|\phi_{\theta^*,0}\|_\infty^4])^{\frac{1}{4}}\frac{6-\rho^{\frac{1}{4r}}}{1-\rho^\frac{1}{4r}}\rho^\frac{\frac{t}{2}-4r-1}{4r}.
      \end{align*}
      Using Minkowski inequality,
      \begin{align*}
            \big(\bb{E}_{\theta^*}[\| \frac{1}{\sqrt{n}} \sum_{t=1}^n (\dot{\Delta}_{t,0}(\theta^*)-\dot{\Delta}_{t,\infty}(\theta^*))\|^2]\big)^{\frac{1}{2}}
            \leq \frac{1}{\sqrt{n}}\sum_{t=1}^n \big(\bb{E}_{\theta^*}[\| \dot{\Delta}_{t,0}(\theta^*)-\dot{\Delta}_{t,\infty}(\theta^*)\|^2]\big)^{\frac{1}{2}}\\
            \leq \frac{1}{\sqrt{n}}\sum_{t=1}^n 4(\bb{E}_{\theta^*}[\|\phi_{\theta^*,0}\|_\infty^4])^{\frac{1}{4}}\frac{6-\rho^{\frac{1}{4r}}}{1-\rho^\frac{1}{4r}}\rho^\frac{\frac{t}{2}-4r-1}{4r}
      \end{align*}
      which converges to $0$ as $n\ra\infty$.
\hfill$\square$

\subsection{Proof of theorems in Subsection 5.2}
We define
\begin{align*}
      \Gamma_{t,m,s}(\theta)\tq\bb{E}_\theta\Big[\sum_{k=-m+1}^t\dot{\phi}_{\theta,k}|\oline{\mbf{Y}}_{-m}^t,S_{-m}=s,\mbf{X}_{-m}^t\Big]\\
      -\bb{E}_\theta\Big[\sum_{k=-m+1}^{t-1}\dot{\phi}_{\theta,k}|\oline{\mbf{Y}}_{-m}^{t-1},S_{-m}=s,\mbf{X}_{-m}^{t-1}\Big]\\
      \Phi_{t,m,s}(\theta)\tq \Var_\theta\Big[\sum_{k=-m+1}^t\phi_{\theta,k}|\oline{\mbf{Y}}_{-m}^t,S_{-m}=s,\mbf{X}_{-m}^t\Big]\\
      -\Var_\theta\Big[\sum_{k=-m+1}^{t-1}\phi_{\theta,k}|\oline{\mbf{Y}}_{-m}^{t-1},S_{-m}=s,\mbf{X}_{-m}^{t-1}\Big]
\end{align*}
so that $\nabla^2_\theta \ell_n(\theta,s_0)=\sum_{t=1}^n(\Gamma_{t,0,s_0}(\theta)+\Phi_{t,0,s_0}(\theta))$.
Similarly, we define
\begin{align*}
      \Gamma_{t,m}(\theta)&\tq\bb{E}_\theta\Big[\sum_{k=-m+1}^t\dot{\phi}_{\theta,k}|\oline{\mbf{Y}}_{-m}^t,\mbf{X}_{-m}^t\Big]
      -\bb{E}_\theta\Big[\sum_{k=-m+1}^{t-1}\dot{\phi}_{\theta,k}|\oline{\mbf{Y}}_{-m}^{t-1},\mbf{X}_{-m}^{t-1}\Big],\\
      \Phi_{k,m}(\theta)&\tq\Var_\theta\Big[\sum_{k=-m+1}^t \phi_{\theta,k}|\oline{\mbf{Y}}_{-m}^t,\mbf{X}_{-m}^t\Big]
      -\Var_\theta\Big[\sum_{k=-m+1}^{t-1}\phi_{\theta,k}|\oline{\mbf{Y}}_{-m}^{t-1},\mbf{X}_{-m}^{t-1}\Big].
\end{align*}
Lemmas \ref{gammap} and \ref{gammae} show that $\{\Gamma_{t,m,s}(\theta)\}_{m\geq0}$ converges uniformly with respect to $\theta\in G$ $\bb{P}_{\theta^*}$-a.s. and in $L^1(\bb{P}_{\theta^*})$ to a random variable that we denote by $\Gamma_{t,\infty}(\theta)$, and the limit does not depend on $s$. Lemmas \ref{phip} and \ref{phie} show similar results for $\{\Phi_{t,m,s}(\theta)\}_{m\geq0}$. We construct the stationary ergodic sequence by conditioning the observed Hessian on the entire history of $(\oline{\mbf{Y}}_t,X_t)$ from past infinity: $\Gamma_{t,\infty}(\theta)\tq\lim_{m\to\infty}\Gamma_{t,m,s}(\theta)$ and $\Phi_{t,\infty}(\theta)\tq\lim_{m\to\infty}\Phi_{t,m,s}(\theta)$.
Using the Louis missing information principle and Assumption \ref{dominatedbounded},
\begin{align}\label{I}
\bb{E}_{\theta^*}[\Gamma_{0,\infty}(\theta^*)+\Phi_{0,\infty}(\theta^*)]=-\bb{E}_{\theta^*}[\dot{\Delta}_{0,\infty}(\theta^*)\dot{\Delta}_{0,\infty}(\theta^*)']=-I(\theta^*).
\end{align}
Propositions \ref{econverge} and \ref{varconverge} and (\ref{I}) together yield Theorem \ref{HessianLLN}.
\begin{proposition}\label{econverge}
      Assume \ref{ergodic}--\ref{continuityass} and \ref{differentiable}--\ref{finitemoment}. Then, for all $s\in\bb{S}$ and $\theta\in G$,
      \begin{align*}
            \lim_{\delta\ra0}\lim_{n\ra\infty} \sup_{\theta':|\theta'-\theta|<\delta}\Big|\frac{1}{n}\sum_{t=1}^n\Gamma_{t,0,s}(\theta')-\bb{E}_{\theta^*}[\Gamma_{0,\infty}(\theta)]\Big|=0,\qquad \bb{P}_{\theta^*}\text{-a.s.}
      \end{align*}
\end{proposition}
\noindent\textbf{Proof: }
Write
\begin{align*}
    &\lim_{\delta\ra0}\lim_{n\ra\infty} \sup_{\theta':|\theta'-\theta|<\delta}\Big|\frac{1}{n}\sum_{t=1}^n\Gamma_{t,0,s}(\theta')-\bb{E}_{\theta^*}[\Gamma_{0,\infty}(\theta)]\Big|\\
    \leq & \lim_{\delta\ra0}\lim_{n\ra\infty} \sup_{\theta':|\theta'-\theta|<\delta}\frac{1}{n}\sum_{t=1}^n\Big|\Gamma_{t,0,s}(\theta')-\Gamma_{t,\infty}(\theta')\Big|\\
    &+\lim_{\delta\ra0}\lim_{n\ra\infty} \sup_{\theta':|\theta'-\theta|<\delta}\Big|\frac{1}{n}\sum_{t=1}^n\Gamma_{t,\infty}(\theta')-\bb{E}_{\theta^*}[\Gamma_{0,\infty}(\theta')]\Big|\\
    &+\lim_{\delta\ra0}\lim_{n\ra\infty} \sup_{\theta':|\theta'-\theta|<\delta}\bb{E}_{\theta^*}\big|\Gamma_{0,\infty}(\theta')-\Gamma_{0,\infty}(\theta)\big|.
\end{align*}
The first term on the right-hand side is zero $\bb{P}_{\theta^*}$-a.s. by Lemma \ref{gammap}. The second term is zero $\bb{P}_{\theta^*}$-a.s., owing to the ergodic theorem. The third term is zero from the continuity of $\Gamma_{0,\infty}(\theta)$ in $L^1(\bb{P}_{\theta^*})$, the proof of which follows a similar argument to the proof of \citet[Lemma 14]{douc2004asymptotic} and is omitted here.
\hfill$\square$

\begin{proposition}\label{varconverge}
       Assume \ref{ergodic}--\ref{continuityass} and \ref{differentiable}--\ref{finitemoment}. Then, for all $s\in\bb{S}$ and $\theta\in G$,
      \begin{align*}
            \lim_{\delta\ra0}\lim_{n\ra\infty} \sup_{\theta':|\theta'-\theta|<\delta}
            \Big|\frac{1}{n}\sum_{t=1}^n\Phi_{t,0,s}(\theta')-\bb{E}_{\theta^*}[\Phi_{0,\infty}(\theta)]\Big|=0,\qquad \bb{P}_{\theta^*}\text{-a.s.}
      \end{align*}
\end{proposition}
\noindent\textbf{Proof: }By Lemmas \ref{phip} and \ref{phie}, $\{\Phi_{t,m,s}(\theta)\}_{m\geq0}$ is a uniform Cauchy sequence with respect to $\theta$ $\bb{P}_{\theta^*}$-a.s. and in $L^1(\bb{P}_{\theta^*})$. The rest of the proof of Proposition \ref{varconverge} follows along the same line as the proof of Proposition \ref{econverge}.
\hfill$\square$

\subsection{Proof of lemmas and theorems in Section \ref{secdiscussion}}
\noindent\textbf{Proof of Theorem \ref{thmergodic}: }
First consider the case where the autoregressive coefficient depends on $\tilde{S}_t$. Define $H_{t,k}\tq A(\tilde{S}_t)A(\tilde{S}_{t-1})\ldots A(\tilde{S}_{t-k+1})$ with the convention $H_{t,0}=I_p$. We have for $k>1$,
\begin{align}
    &\bb{E}_{\theta^*}[\mathrm{vec}(H_{t,k}H_{t,k}')|\mbf{X}_{t-k+2}^t]\nonumber\\
    =&\bb{E}_{\theta^*}[(A(\tilde{S}_t)\otimes A(\tilde{S}_t))\ldots (A(\tilde{S}_{t-k+1})\otimes A(\tilde{S}_{t-k+1}))|\mbf{X}_{t-k+2}^t]\mathrm{vec}(\mathrm{I}_p)\nonumber\\
    =&\sum_{\tilde{s}_t,\ldots,\tilde{s}_{t-k+1}}\big[(A(\tilde{s}_t)\otimes A(\tilde{s}_t))\ldots(A(\tilde{s}_{t-k+1})\otimes A(\tilde{s}_{t-k+1}))
    \nonumber\\
    &\hspace{1cm}\times
    \prod_{\ell=t-k+2}^{t}
    \tilde{q}_{\theta^*}(\tilde{s}_\ell|\tilde{s}_{\ell-1},X_\ell)\bb{P}_{\theta^*}(\tilde{S}_{t-k+1}=\tilde{s}_{t-k+1}|\mbf{X}_{t-k+2}^t)\big]\mathrm{vec}(I_p).\label{vech}
\end{align}
We can write (\ref{vech}) in matrix form as 
\begin{align*}
    \bb{E}_{\theta^*}[\mathrm{vec}(H_{t,k}H_{t,k}')|\mbf{X}_{t-k+2}^t]
    =\mathbb{I}\prod_{\ell=0}^{k-2}M(X_{t-j})N_{t-k+1}\mathrm{vec}(\mathrm{I}_{p}),
\end{align*}
where $\mathbb{I}=(\mathrm{I}_{p^2},\ldots,\mathrm{I}_{p^2})$ consists of $J$ column partitions and
\begin{align*}
    N_{t-k+1}=
    \begin{pmatrix}
        (A(1)\otimes A(1))\bb{P}_{\theta^*}(\tilde{S}_{t-k+1}=1|\mbf{X}_{t-k+2}^t)\\
        \vdots\\
        (A(J)\otimes A(J))\bb{P}_{\theta^*}(\tilde{S}_{t-k+1}=J|\mbf{X}_{t-k+2}^t)
    \end{pmatrix}.
\end{align*}
It follows that
\begin{align*}
    \bb{E}_{\theta^*}[\|H_{t,k}\|^2|\mbf{X}_{t-k+2}^t]
    &\leq p^2\Big\|\mathbb{I}\prod_{\ell=0}^{k-2}M(X_{t-\ell})N_{t-k+1}\mathrm{vec}(\mathrm{I}_{p})\Big\|\\
    &\leq Jp^5\sum_{s=1}^J\|A(s)\|^2\prod_{\ell=0}^{k-2}\|M(X_{t-\ell})\|
    \leq C\prod_{\ell=0}^{k-2}\|M(X_{t-\ell})\|
\end{align*}
for some constant $C>0$.
Then 
$\bb{E}_{\theta^*}[\log\|H_{t,k}\||\mbf{X}_{t-k+2}^t]
=\frac{1}{2}\bb{E}_{\theta^*}[\log\|H_{t,k}\|^2|\mbf{X}_{t-k+2}^t]
\leq\frac{1}{2}\log\bb{E}_{\theta^*}[\|H_{t,k}\|^2|\mbf{X}_{t-k+2}^t]
\leq\frac{1}{2}\log C+\frac{1}{2}\sum_{\ell=0}^{k-2}\log \|M(X_{t-\ell})\|$ by Jensen's inequality. By the law of iterated expectations and the stationarity of $X_t$ and $M(X_t)$,
\begin{align*}
    \bb{E}_{\theta^*}[\log\|H_{t,k}\|]
    \leq \frac{1}{2}\log C+\frac{1}{2}k\bb{E}_{\theta^*}[\log\|M(X_t)\|]
    \leq \frac{1}{2}\log C+\frac{1}{2}k\log\bb{E}_{\theta^*}[\|M(X_t)\|].
\end{align*}
Now we can verify that the top Lyapounov exponent defined by 
\begin{align*}
    \gamma=\inf_{t\geq 1}\frac{1}{t}\bb{E}_{\theta^*}
    [\log \|A(\tilde{S}_t)A(\tilde{S}_{t-1})\ldots A(\tilde{S}_1)\|]
\end{align*}
is strictly negative. By
\begin{align*}
    \frac{1}{t}\bb{E}_{\theta^*}[\log \|A(\tilde{S}_t)A(\tilde{S}_{t-1})\ldots A(\tilde{S}_1)\|]
=\frac{1}{t}\bb{E}_{\theta^*}[\log \|H_{t,t}\|]\leq \frac{\log C}{2t}+\frac{1}{2}\log \bb{E}_{\theta^*}[\|M(X_t)\|]
\end{align*}
for all $t$, and $\bb{E}_{\theta^*}[\|M(X_t)\|]<1$, the top Lyapounov exponent
\begin{align*}
    \gamma=\inf_{t\geq 1}\frac{1}{t}\bb{E}_{\theta^*}
    [\log \|A(\tilde{S}_t)A(\tilde{S}_{t-1})\ldots A(\tilde{S}_1)\|]\leq \frac{1}{2}\log \bb{E}_{\theta^*}[\|M(X_t)\|]<0.
\end{align*}
By Theorem 1 of \citet{brandt1986} (See also Theorem 1 of \citealp{francq2001stationarity}), the process
\begin{align*}
\oline{\mbf{Y}}_t&=B(S_t,X_t)+A(\tilde{S}_t)B(S_{t-1},X_{t-1})+A(\tilde{S}_t)A(\tilde{S}_{t-1})B(S_{t-2},X_{t-2})+\dots\\
&=\sum_{k=0}^\infty H_{t,k}B(S_{t-k},X_{t-k})
\end{align*}
converges almost surely for each $t$ and is the unique strictly stationary solution of (\ref{vectorial}). Furthermore, since $\{(A(\tilde{S}_t),B(S_t,X_t))\}$ is strictly stationary ergodic, by Theorem 7.1.3 in \citet{durrett2013probability}, $\{\oline{\mbf{Y}}_t\}_t$ is ergodic.

Next consider the case where the autoregressive coefficient does not depend on $\tilde{S}_t$, i.e., $A(\tilde{S}_t)=A$. Choose $\epsilon$ such that $\rho(A)+\epsilon<1$. Then there exists $C>0$ such that $|(A^t)_{ij}|\leq C(\rho(A)+\epsilon)^t$ for all $t=1,2,\ldots$ and $i,j=1,\ldots,Jp$. Then $\|A^t\|\leq p^2C(\rho(A)+\epsilon)^t$ for all $t=1,2,\ldots$. 
The top Lyapounov exponent
\begin{align*}
    \gamma
    =\inf_{t\geq 1}\frac{1}{t}\log\|A^t\|
    \leq \inf_{t\geq 1}
    [\frac{1}{t}\log p^2C+\log(\rho(A)+\epsilon)]<1.
\end{align*}
Again by Theorem 1 of \citet{brandt1986}, $\{\oline{\mbf{Y}}_t\}_t$ is ergodic.
\hfill$\square$

\noindent\textbf{Proof of Lemma \ref{suffsigma}:}
Setting $C_1=1$,
\begin{align*}
    \bb{P}_{\theta^*}(\sigma_-(\oline{\mbf{Y}}_0,X_1)\leq e^{-\alpha_1\xi})
    =\bb{P}_{\theta^*}(|\log\sigma_-(\oline{\mbf{Y}}_0,X_1)|^{1+\delta}\geq (\alpha_1 \xi)^{1+\delta})\\
    \leq \bb{E}_{\theta^*}\big[|\log \sigma_-(\oline{\mbf{Y}}_0,X_1)|^{1+\delta}\big]/(\alpha_1 \xi)^{1+\delta}=C_2 \xi^{-(1+\delta)},
\end{align*}
where $C_2=\bb{E}_{\theta^*}[|\log \sigma_-(\oline{\mbf{Y}}_0,X_1))|^{1+\delta}]/\alpha_1^{1+\delta}$, and the second inequality follows from Markov's inequality.
\hfill$\square$

\noindent\textbf{Proof of Lemma \ref{lemlogistic}:}
We write
\begin{align*}
    \min_{\tilde{s}_0,\tilde{s}_1}\tilde{q}_\theta(\tilde{s}_1|\tilde{s}_0,X_1)
    =\min\Big\{\frac{\exp(X_t'\beta_1)}{1+\exp(X_t'\beta_1)}\bbm{1}\{\exp(X_t'\beta_1)<1\}
    +\frac{1}{1+\exp(X_t'\beta_1)}\bbm{1}\{\exp(X_t'\beta_1)\geq 1\},\\
    \frac{\exp(X_t'\beta_2)}{1+\exp(X_t'\beta_2)}\bbm{1}\{\exp(X_t'\beta_2)<1\}
    +\frac{1}{1+\exp(X_t'\beta_2)}\bbm{1}\{\exp(X_t'\beta_2)\geq 1\}\Big\}.
\end{align*}
Since $\Theta$ is compact, it suffices to show for all $\theta\in\Theta$ and $i=1,2$,
\begin{align}
    \bb{E}_{\theta^*}\big[\big|\log\big(\frac{\exp(X_t'\beta_i)}{1+\exp(X_t'\beta_i)}\bbm{1}\{\exp(X_t'\beta_i)<1\}
    +\frac{1}{1+\exp(X_t'\beta_i)}\bbm{1}\{\exp(X_t'\beta_i)\geq 1\}\big)\big|^2\big]<\infty.\label{loglogistic}
\end{align}
Note that
\begin{align*}
    &\Big|\log\Big(\frac{\exp(X_t'\beta_i)}{1+\exp(X_t'\beta_i)}\bbm{1}\{\exp(X_t'\beta_i)<1\}
    +\frac{1}{1+\exp(X_t'\beta_i)}\bbm{1}\{\exp(X_t'\beta_i)\geq 1\}\Big)\Big|\\
    =&-\log\Big(\frac{\exp(X_t'\beta_i)}{1+\exp(X_t'\beta_i)}\Big)\bbm{1}\{\exp(X_t'\beta_i)<1\}
    -\log\Big(\frac{1}{1+\exp(X_t'\beta_i)}\Big)\bbm{1}\{\exp(X_t'\beta_i)\geq 1\}\\
    \leq & -\log\Big(\frac{1}{2}\exp(X_t'\beta_i)\Big)\bbm{1}\{\exp(X_t'\beta_i)<1\}
    -\log \Big(\frac{1}{2\exp(X_t'\beta_i)}\Big)\bbm{1}\{\exp(X_t'\beta_i)\geq 1\}\\
    = & \log2 +|X_t'\beta_i|.
\end{align*}
Then the left-hand side of (\ref{loglogistic}) is bounded by $2(\log^2 2+\bb{E}_{\theta^*}[|X_t'\beta_i|^2])\leq 2\log^2 2+2\bb{E}_{\theta^*}[\|X_t\|^2]\|\beta_i\|^2<\infty$ if $\bb{E}_{\theta^*}[\|X_t\|^2]<\infty$. \hfill$\square$

\noindent\textbf{Proof of Lemma \ref{lemprobit}:}
We write
\begin{align*}
    \min_{\tilde{s}_0,\tilde{s}_1}\tilde{q}_\theta(\tilde{s}_1|\tilde{s}_0,X_1)
    =\min\{&\Phi(X_t'\beta_1)\bbm{1}\{X_t'\beta_1\leq0\}+\Phi^c(X_t'\beta_1)\bbm{1}\{X_t'\beta_1>0\},\\
    &\Phi(X_t'\beta_2)\bbm{1}\{X_t'\beta_2\leq0\}+\Phi^c(X_t'\beta_2)\bbm{1}\{X_t'\beta_2>0\}\},
\end{align*}
where $\Phi^c(t)\tq1-\Phi(t)$. 
Since $\Theta$ is compact, it suffices to show for all $\theta\in\Theta$ and $i=1,2$,
\begin{align}\label{logprobit}
    \bb{E}_{\theta^*}[|\log(\Phi(X_t'\beta_i)\bbm{1}\{X_t'\beta_i\leq0\}+\Phi^c(X_t'\beta_i)\bbm{1}\{X_t'\beta_i>0\})|^2]<\infty.
\end{align}
Using $\Phi^c(t)>\sqrt{\frac{2}{\pi}}\frac{1}{t+\sqrt{t^2+4}}\exp(-\frac{t^2}{2})$ for $t\geq0$ and $\log(x)\leq x-1$ for $x>0$,
\begin{align*}
    &|\log(\Phi(X_t'\beta_i)\bbm{1}\{X_t'\beta_i\leq 0\}+\Phi^c(X_t'\beta_i)\bbm{1}\{X_t'\beta_i>0\}|\\
    =&-\log\Phi^c(-X_t'\beta_i)\bbm{1}\{X_t'\beta_i\leq 0\}
    -\log\Phi^c(X_t'\beta_i)\bbm{1}\{X_t'\beta_i>0\}\\
    \leq & -\frac{1}{2}\log\frac{2}{\pi}
    +\log\Big(|X_t'\beta_i|+\sqrt{(X_t'\beta_i)^2+4}\Big)+\frac{1}{2}(X_t'\beta_i)^2\\
    \leq & -\frac{1}{2}\log \frac{2}{\pi}+\Big(|X_t'\beta_i|+\sqrt{(X_t'\beta_i)^2+4}-1\Big)+\frac{1}{2}(X_t'\beta_i)^2.
\end{align*}
The left-hand side of (\ref{logprobit}) is bounded by $\frac{3}{4}\bb{E}_{\theta^*}[(X_t'\beta_i)^4]+18\bb{E}_{\theta^*}[(X_t'\beta_i)^2]+45+\frac{3}{4}\log^2\frac{2}{\pi}<\infty$ if $\bb{E}_{\theta^*}[\|X_t\|^4]<\infty$. 
\hfill $\square$

\noindent\textbf{Proof of Lemma \ref{lemccp}:}
It suffices to show that $\bb{E}_{\theta^*}[|\log \tilde{q}_\theta(\tilde{s}_1|\tilde{s}_{0},\ldots,\tilde{s}_{-d+1},\oline{\mbf{Y}}_{t-1})|^2]<\infty$ for all $\tilde{s}_{-d+1},\ldots,\tilde{s}_1$.
First, for the case of $\tilde{s}_1=0$ and $\tilde{s}_0=0$,
\begin{align*}
    \tilde{q}_{\theta}(\tilde{s}_1=0|\tilde{s}_0=0,\tilde{s}_{-1},\dots,\tilde{s}_{-d+1},\oline{\mbf{Y}}_0)
    =\frac{\int_{-\infty}^{\tau\sqrt{1-\alpha^2}} \Phi\Big(\frac{\tau-\rho U_{0}}{\sqrt{1-\rho^2}}-\frac{\alpha x}{\sqrt{1-\alpha^2}\sqrt{1-\rho^2}}\Big)\varphi(x)dx}{\Phi(\tau\sqrt{1-\alpha^2})}.
\end{align*}

The following result holds for the numerator:
\begin{align*}
      &\int_{-\infty}^{\tau\sqrt{1-\alpha^2}} \Phi\Big(\frac{\tau-\rho U_0}{\sqrt{1-\rho^2}}-\frac{\alpha x}{\sqrt{1-\alpha^2}\sqrt{1-\rho^2}}\Big)\varphi(x)dx
      \\
      \geq& \int_{-|\tau\sqrt{1-\alpha^2}|-1}^{-|\tau\sqrt{1-\alpha^2}|}\Phi\Big(\frac{\tau-\rho U_0}{\sqrt{1-\rho^2}}+\frac{|\alpha| x}{\sqrt{1-\alpha^2}\sqrt{1-\rho^2}}\Big)\varphi(x)dx\\
      \geq&  \Phi\Big(\frac{\tau-\rho U_0}{\sqrt{1-\rho^2}}-\frac{|\alpha\tau\sqrt{1-\alpha^2}|+|\alpha|}{\sqrt{1-\alpha^2}\sqrt{1-\rho^2}}\Big)
      \Big( \Phi(-|\tau\sqrt{1-\alpha^2}|)-\Phi(-|\tau\sqrt{1-\alpha^2}|-1) \Big)\\
      \geq& \Phi(-D)\Big( \Phi(-|\tau\sqrt{1-\alpha^2}|)-\Phi(-|\tau\sqrt{1-\alpha^2}|-1) \Big)\\
      =& \Phi^c(D)\Big( \Phi(-|\tau\sqrt{1-\alpha^2}|)-\Phi(-|\tau\sqrt{1-\alpha^2}|-1) \Big)\\
      \geq & \sqrt{\frac{2}{\pi}} e^{-\frac{D^2}{2}}\frac{1}{D+\sqrt{D^2+4}}\Big( \Phi(-|\tau\sqrt{1-\alpha^2}|)-\Phi(-|\tau\sqrt{1-\alpha^2}|-1) \Big),
\end{align*}
where $D\tq \Big|\frac{\tau-\rho U_{0}}{\sqrt{1-\rho^2}}-\frac{|\alpha\tau\sqrt{1-\alpha^2}|+|\alpha|}{\sqrt{1-\alpha^2}\sqrt{1-\rho^2}}\Big|$. The last inequality follows from $\Phi^c(x)\geq \sqrt{\frac{2}{\pi}}e^{-x^2/2}\frac{1}{x+\sqrt{x^2+4}}$. Consequently,
\begin{align*}
      &\bb{E}_{\theta^*}\big[|\log \tilde{q}_{\theta}(\tilde{s}_1=0|\tilde{s}_0=0,\tilde{s}_{-1},\dots,\tilde{s}_{-d+1},\oline{\mbf{Y}}_0)|^2\big]\\
      \leq & \bb{E}_{\theta^*}\Big[\Big|\log\Big( e^{-\frac{D^2}{2}}\frac{1}{D+\sqrt{D^2+4}} \Big)\\
      &+\log\sqrt{\frac{2}{\pi}}+\log\Big(\Phi(-|\tau\sqrt{1-\alpha^2}|)-\Phi(-|\tau\sqrt{1-\alpha^2}|-1)\Big)-\log\big(\Phi(\tau\sqrt{1-\alpha^2})\big)\Big|^2\Big].
\end{align*}
We need show only that $\bb{E}_{\theta^*}\Big[\big( \log\big( e^{-\frac{D^2}{2}}\frac{1}{D+\sqrt{D^2+4}} \big) \big)^2\Big]<\infty$. Note that $D$ is folded normal distributed and has finite moments. We use $\log(x)\leq x-1$ for $x>0$ to obtain
\begin{align}
\begin{split}
      &\bb{E}_{\theta^*}\Big[\big( \log\big( e^{-\frac{D^2}{2}}\frac{1}{D+\sqrt{D^2+4}} \big) \big)^2\Big] = \bb{E}_{\theta^*}\Big[\big(-\frac{D^2}{2}-\log(D+\sqrt{D^2+4}) \big)^2\Big]\\
      =&\bb{E}_{\theta^*}\Big[ \frac{D^4}{4}+\big(\log(D+\sqrt{D^2+4})\big)^2+D^2\log(D+\sqrt{D^2+4}) \Big]\\
      \leq&\bb{E}_{\theta^*}\Big[ \frac{D^4}{4} +(D+\sqrt{D^2+4}-1)^2+D^2(D+\sqrt{D^2+4}-1) \Big]\\
      \leq&\bb{E}_{\theta^*}\Big[ \frac{D^4}{4} + (2D+1)^2+D^2(2D+1) \Big]<\infty.
\end{split}\label{D1}
\end{align}
Next, for the case of $\tilde{s}_1=1$ and $\tilde{s}_0=0$,
\begin{align*}
    \tilde{q}_{\theta}(\tilde{s}_1=1|\tilde{s}_0=0,\tilde{s}_{-1},\dots,\tilde{s}_{-d+1},\oline{\mbf{Y}}_0)
    =\frac{\int_{-\infty}^{\tau\sqrt{1-\alpha^2}} \Phi^c\Big(\frac{\tau-\rho U_{0}}{\sqrt{1-\rho^2}}-\frac{\alpha x}{\sqrt{1-\alpha^2}\sqrt{1-\rho^2}}\Big)\varphi(x)dx}{\Phi(\tau\sqrt{1-\alpha^2})}.
\end{align*}

The following result holds for the numerator
\begin{align*}
      &\int_{-\infty}^{\tau\sqrt{1-\alpha^2}} \Phi^c\Big(\frac{\tau-\rho U_0}{\sqrt{1-\rho^2}}-\frac{\alpha x}{\sqrt{1-\alpha^2}\sqrt{1-\rho^2}}\Big)\varphi(x)dx
      \\
      \geq& \int_{-|\tau\sqrt{1-\alpha^2}|-1}^{-|\tau\sqrt{1-\alpha^2}|}\Phi^c\Big(\frac{\tau-\rho U_0}{\sqrt{1-\rho^2}}-\frac{|\alpha| x}{\sqrt{1-\alpha^2}\sqrt{1-\rho^2}}\Big)\varphi(x)dx\\
      \geq&  \Phi^c\Big(\frac{\tau-\rho U_0}{\sqrt{1-\rho^2}}+\frac{|\alpha\tau\sqrt{1-\alpha^2}|+|\alpha|}{\sqrt{1-\alpha^2}\sqrt{1-\rho^2}}\Big)
      \Big( \Phi\big(-|\tau\sqrt{1-\alpha^2}|\big)-\Phi\big(-|\tau\sqrt{1-\alpha^2}|-1\big) \Big)\\
      \geq& \Phi^c(D_2)( \Phi(-|\tau\sqrt{1-\alpha^2}|)-\Phi(-|\tau\sqrt{1-\alpha^2}|-1) )\\
      \geq & \sqrt{\frac{2}{\pi}} e^{-\frac{D_2^2}{2}}\frac{1}{D_2+\sqrt{D_2^2+4}}\Big( \Phi\big(-|\tau\sqrt{1-\alpha^2}|\big)-\Phi\big(-|\tau\sqrt{1-\alpha^2}|-1\big) \Big)
\end{align*}
where $D_2\tq \Big|\frac{\tau-\rho U_{0}}{\sqrt{1-\rho^2}}+\frac{|\alpha\tau\sqrt{1-\alpha^2}|+|\alpha|}{\sqrt{1-\alpha^2}\sqrt{1-\rho^2}}\Big|$. Then, we can show $\bb{E}_{\theta^*}[|\log \tilde{q}_{\theta}(\tilde{s}_1=1|\tilde{s}_0=0,\tilde{s}_{-1},\dots,\tilde{s}_{-d+1},\oline{\mbf{Y}}_0)|^2]<\infty$ by showing that $\bb{E}_{\theta^*}\Big[\Big( \log\big( e^{-\frac{D_2^2}{2}}\frac{1}{D_2+\sqrt{D_2^2+4}} \big) \Big)^2\Big]<\infty$. The proof follows a similar procedure to that in (\ref{D1}).

Finally, for the case of $\tilde{s}_1=0,\tilde{s}_0=1$,
\begin{align*}
     &\int^{-\infty}_{\tau\sqrt{1-\alpha^2}} \Phi\Big(\frac{\tau-\rho U_0}{\sqrt{1-\rho^2}}-\frac{\alpha x}{\sqrt{1-\alpha^2}\sqrt{1-\rho^2}}\Big)\varphi(x)dx\\
     \geq& \sqrt{\frac{2}{\pi}} e^{-\frac{D^2}{2}}\frac{1}{D+\sqrt{D^2+4}}
     \Big( \Phi\big(|\tau\sqrt{1-\alpha^2}+1|\big)-\Phi\big(|\tau\sqrt{1-\alpha^2}|\big) \Big),
\end{align*}
and for the case of $\tilde{s}_1=1$, $\tilde{s}_0=1$,
\begin{align*}
      &\int^{-\infty}_{\tau\sqrt{1-\alpha^2}} \Phi^c\Big(\frac{\tau-\rho U_0}{\sqrt{1-\rho^2}}-\frac{\alpha x}{\sqrt{1-\alpha^2}\sqrt{1-\rho^2}}\Big)\varphi(x)dx\\
      \geq&\sqrt{\frac{2}{\pi}} e^{-\frac{D_2^2}{2}}\frac{1}{D_2+\sqrt{D_2^2+4}}
     \Big( \Phi\big(|\tau\sqrt{1-\alpha^2}+1|\big)-\Phi\big(|\tau\sqrt{1-\alpha^2}|\big) \Big).
\end{align*}
Therefore, the result follows.
\hfill$\square$

\section*{Appendix B: Auxiliary results}
\renewcommand{\theequation}{B.\arabic{equation}}
\renewcommand{\thesection}{B}
\setcounter{equation}{0}

\begin{lemma}[Minorization condition] \label{minorization}
    Let $m,n\in\bb{Z}$ with $-m\leq n$ and $\theta\in\Theta$. Conditionally on $\oline{\mbf{Y}}_{-m}^n$ and $\mbf{X}_{-m}^n$, $\{S_k\}_{-m\leq k\leq n}$ satisfies the Markov property. Assume \ref{bound1}. Then, for all $-m+r\leq k\leq n$, a function $\mu_k(\mbf{Y}_{k-r}^n,\mbf{X}_k^n,A)$ exists, such that:
    (a) for any $A\in\mcal{P}(\bb{S})$, ($\mbf{y}_{k-r}^n,\mbf{x}_k^n)\ra\mu_k(\mbf{y}_{k-r}^n,\mbf{x}_k^n,A)$ is a Borel function; and
    (b) for any $\mbf{y}_{k-r}^n$ and $\mbf{x}_k^n$, $\mu_k(\mbf{y}_{k-r}^n,\mbf{x}_{k}^n,\cdot)$ is a probability measure on $\mcal{P}(\bb{S})$. Moreover, for $A\in\mcal{P}(\bb{S})$, the following holds:
    \begin{align}
        \min_{s_{k-r}\in \bb{S} }
        \bb{P}_\theta(S_k\in A|S_{k-r}=s_{k-r},\oline{\mbf{Y}}_{-m}^n,\mbf{X}_{-m}^n)
        \geq \sigma_-(\oline{\mbf{Y}}_{k-r}^{k-1},\mbf{X}_{k-r+1}^k)\cdot\mu_k(\mbf{Y}_{k-r}^n,\mbf{X}_k^n,A).\label{ineqmin}
    \end{align}
\end{lemma}
\noindent\textbf{Proof:}
For $-m\leq k\leq n$, $\{S_k\}$ is a Markov chain conditionally, since
\begin{align*}
    &p_\theta(S_k|\mbf{S}_{-m}^{k-1},\oline{\mbf{Y}}_{-m}^n,\mbf{X}_{-m}^n)
    =\frac{p_\theta(S_k,\oline{\mbf{Y}}_k^n|\mbf{S}_{-m}^{k-1},\oline{\mbf{Y}}_{-m}^{k-1},\mbf{X}_{-m}^n)}
    {p_\theta(\oline{\mbf{Y}}_k^n|\mbf{S}_{-m}^{k-1},\oline{\mbf{Y}}_{-m}^{k-1},\mbf{X}_{-m}^n)}\\
    =&\frac{p_\theta(S_k,\oline{\mbf{Y}}_k^n|S_{k-1},\oline{\mbf{Y}}_{k-1},\mbf{X}_k^n)}
    {p_\theta(\oline{\mbf{Y}}_k^n|S_{k-1},\oline{\mbf{Y}}_{k-1},\mbf{X}_k^n)}
    =p_\theta(S_k|S_{k-1},\oline{\mbf{Y}}_{k-1}^n,\mbf{X}_k^n).
\end{align*}
To see the minorization condition, observe that
\begin{align*}
    &\bb{P}_\theta(S_k\in A|S_{k-r},\oline{\mbf{Y}}_{-m}^n,\mbf{X}_{-m}^n)
    =\bb{P}_\theta(S_k\in A|S_{k-r},\oline{\mbf{Y}}_{k-r}^n,\mbf{X}_{k-r+1}^n)\\
    =&\frac{\sum_{s_k\in A}p_\theta(\mbf{Y}_{k}^n|S_k=s_k,\oline{\mbf{Y}}_{k-1},\mbf{X}_{k}^n)
    \bb{P}_\theta(S_k=s_k|S_{k-r},\oline{\mbf{Y}}^{k-1}_{k-r},\mbf{X}_{k-r+1}^k)}
    {\sum_{s_k\in\bb{S}}p_\theta(\mbf{Y}_{k}^n|S_k=s_k,\oline{\mbf{Y}}_{k-1},\mbf{X}_{k}^n)
    \bb{P}_\theta(S_k=s_k|S_{k-r},\oline{\mbf{Y}}^{k-1}_{k-r},\mbf{X}_{k-r+1}^k)}.
\end{align*}
Since $\sigma_-(\oline{\mbf{Y}}_{k-r}^{k-1},\mbf{X}_{k-r+1}^k)\leq \bb{P}_\theta(S_k=s_k|S_{k-r},\oline{\mbf{Y}}_{k-r}^{k-1},\mbf{X}_{k-r+1}^k)\leq 1$,
it readily follows that
\begin{align*}
    \bb{P}_\theta(S_k\in A|S_{k-r},\oline{\mbf{Y}}_{-m}^n,\mbf{X}_{-m}^n)
    \geq \sigma_-(\oline{\mbf{Y}}_{k-r}^{k-1},\mbf{X}_{k-r+1}^k)\mu_k(\mbf{Y}_{k-r}^n,\mbf{X}_k^n,A)
\end{align*}
with
\begin{align}
    \mu_k(\mbf{Y}_{k-r}^n,\mbf{X}_k^n,A)\tq
    \frac{\sum_{s_k\in A}p_\theta(\mbf{Y}_k^n|S_k=s_k,\oline{\mbf{Y}}_{k-1},\mbf{X}_{k}^n)}
    {\sum_{s_k\in \bb{S}}p_\theta(\mbf{Y}_k^n|S_k=s_k,\oline{\mbf{Y}}_{k-1},\mbf{X}_{k}^n)}.\label{Mmeasure}
\end{align}
In order for $\mu_k(\mbf{Y}_{k-r}^n,\mbf{X}_k^n,A)$ to be a well-defined probability measure, we need show only that the denominator is strictly positive. The summand term in the denominator of (\ref{Mmeasure}) is
\begin{align*}
    p_\theta(\mbf{Y}_k^n|S_{k}=s_k,\oline{\mbf{Y}}_{k-1},\mbf{X}_k^n)
    &=\sum_{\mbf{s}_{k+1}^n} \prod_{\ell=k}^n g_\theta(Y_\ell|\oline{\mbf{Y}}_{\ell-1},s_{\ell},X_\ell)
    \prod_{\ell=k+1}^nq_\theta(s_\ell|s_{\ell-1},\oline{\mbf{Y}}_{\ell-1},X_\ell)\\
    &\geq 
    \prod_{\ell=k}^n b_-(Y_\ell,\oline{\mbf{Y}}_{\ell-1},X_\ell)
    \sum_{\mbf{s}_{k+1}^n}\prod_{\ell=k+1}^nq_\theta(s_\ell|s_{\ell-1},\oline{\mbf{Y}}_{\ell-1},X_\ell)>0.
\end{align*}
\hfill$\square$

\begin{lemma}[Minorization condition of the time-reversed Markov process]
      Let $m,n\in\bb{Z}$ with $-m\leq n$ and $\theta\in\Theta$. Conditionally on $\oline{\mbf{Y}}_{-m}^n$ and $\mbf{X}_{-m}^n$, $\{S_{n-k}\}_{0\leq k\leq n+m}$ satisfies the Markov property. Assume \ref{bound1}. Then, for all $r\leq k\leq n+m$, a function $\tilde{\mu}_k(\oline{\mbf{Y}}_{-m}^{n-k+r},\mbf{X}_{-m}^{n-k+r},A)$ exists such that:
      (a) for any $A\in\mcal{P}(\bb{S})$, $(\oline{\mbf{y}}_{-m}^{n-k+r},\mbf{x}_{-m}^{n-k+r})\ra\tilde{\mu}_k(\oline{\mbf{y}}_{-m}^{n-k+r},\mbf{x}_{-m}^{n-k+r},A)$ is a Borel function; and
      (b) for any $\oline{\mbf{y}}_{-m}^{n-k+r}$ and $\mbf{x}_{-m}^{n-k+r}$, $\tilde{\mu}_k(\oline{\mbf{y}}_{-m}^{n-k+r},\allowbreak \mbf{x}_{-m}^{n-k+r},\cdot)$ is a probability measure on $\mcal{P}(\bb{S})$. Moreover, for $A\in\mcal{P}(\bb{S})$, the following holds:
      \begin{align*}
            &\min_{s_{n-k+r}\in \bb{S} }
            \bb{P}_\theta(S_{n-k}\in A|S_{n-k+r}=s_{n-k+r},\oline{\mbf{Y}}_{-m}^n,\mbf{X}_{-m}^n)\\
            &\hspace{3cm}\geq \sigma_-(\oline{\mbf{Y}}^{n-k+r-1}_{n-k},\mbf{X}_{n-k+1}^{n-k+r})\cdot
            \tilde{\mu}_k(\oline{\mbf{Y}}_{-m}^{n-k+r-1},\mbf{X}_{-m}^{n-k+r},A).
      \end{align*}
\end{lemma}
\noindent\textbf{Proof: }
      To observe the Markovian property, for $2\leq k\leq m+n$,
      \begin{align*}
            &p_\theta(S_{n-k}|\mbf{S}_{n-k+1}^n,\oline{\mbf{Y}}_{-m}^n,\mbf{X}_{-m}^n)\\
            =&
            \frac{p_\theta(\mbf{S}_{n-k+2}^n,\oline{\mbf{Y}}_{n-k+1}^n|S_{n-k+1},\oline{\mbf{Y}}_{n-k},\mbf{X}_{-m}^n)
            p_\theta(\mbf{S}_{n-k}^{n-k+1},\oline{\mbf{Y}}_{-m}^{n-k}|\mbf{X}_{-m}^n)}
            {p_\theta(\mbf{S}_{n-k+2}^n,\oline{\mbf{Y}}_{n-k+1}^n|S_{n-k+1},\oline{\mbf{Y}}_{n-k},\mbf{X}_{-m}^n)
            p_\theta(S_{n-k+1},\oline{\mbf{Y}}_{-m}^{n-k}|\mbf{X}_{-m}^n)}\\
            =&p_\theta(S_{n-k}|S_{n-k+1},\oline{\mbf{Y}}_{-m}^{n-k},\mbf{X}_{-m}^{n-k+1}).
      \end{align*}
To observe the minorization condition, note that
\begin{align*}
    &\bb{P}_\theta(S_{n-k}\in A|S_{n-k+r},\oline{\mbf{Y}}_{-m}^n,\mbf{X}_{-m}^n)
    =\bb{P}_\theta(S_{n-k}\in A|S_{n-k+r},\oline{\mbf{Y}}_{-m}^{n-k+r-1},\mbf{X}_{-m}^{n-k+r})\\
    =&\frac{\sum_{s_{n-k}\in A}p_\theta(S_{n-k}=s_{n-k},\oline{\mbf{Y}}_{-m}^{n-k+r-1},\mbf{X}_{-m}^{n-k+r})
    \bb{P}_\theta(S_{n-k+r}|S_{n-k}=s_{n-k},\oline{\mbf{Y}}_{n-k}^{n-k+r-1},\mbf{X}_{n-k+1}^{n-k+r})}
    {\sum_{s_{n-k}\in\bb{S}}p_\theta(S_{n-k}=s_{n-k},\oline{\mbf{Y}}_{-m}^{n-k+r-1},\mbf{X}_{-m}^{n-k+r})
    \bb{P}_\theta(S_{n-k+r}|S_{n-k}=s_{n-k},\oline{\mbf{Y}}_{n-k}^{n-k+r-1},\mbf{X}_{n-k+1}^{n-k+r})}.
\end{align*}
Since $\sigma_-(\oline{\mbf{Y}}_{n-k}^{n-k+r-1},\mbf{X}_{n-k+1}^{n-k+r})\leq \bb{P}_\theta(S_{n-k+r}|S_{n-k}=s_{n-k},\oline{\mbf{Y}}_{n-k}^{n-k+r-1},\mbf{X}_{n-k+1}^{n-k+r})\leq 1$,
it readily follows that
\begin{align*}
    \bb{P}_\theta(S_{n-k}\in A|S_{n-k+r},\oline{\mbf{Y}}_{-m}^n,\mbf{X}_{-m}^n)\geq \sigma_-(\oline{\mbf{Y}}_{n-k}^{n-k+r-1},\mbf{X}_{n-k+1}^{n-k+r})\tilde{\mu}_k(\oline{\mbf{Y}}_{-m}^{n-k+r-1},\mbf{X}_{-m}^{n-k+r},A)
\end{align*}
with
\begin{align*}
    \tilde{\mu}_k(\oline{\mbf{Y}}_{-m}^{n-k+r-1},\mbf{X}_{-m}^{n-k+r},A)\tq
    p_\theta(S_{n-k}\in A|\oline{\mbf{Y}}_{-m}^{n-k+r-1},\mbf{X}_{-m}^{n-k+r}).
\end{align*}
\hfill$\square$

\begin{lemma}[Uniform ergodicity of the time-reversed Markov process]
\label{mixingreverse}
      Assume \ref{bound1}. Let $m,n\in\bb{Z}$, $-m\leq n$, and $\theta\in \Theta$. Then, for $-m\leq k\leq n$, all probability measures $\mu_1$ and $\mu_2$ defined on $\mcal{P}(\bb{S})$, and all $\oline{\mbf{Y}}_{-m}^n$,
      \begin{align*}
            \Big\|\sum_{s\in\bb{S}}\bb{P}_\theta(S_k\in\cdot|S_{n}=s,\oline{\mbf{Y}}_{-m}^n,\mbf{X}_{-m}^n)\mu_1(s)
            -\sum_{s\in\bb{S}}\bb{P}_\theta(S_k\in\cdot|S_{n}=s,\oline{\mbf{Y}}_{-m}^n,\mbf{X}_{-m}^n)\mu_2(s)\Big\|_{TV}\\
            \leq \prod_{i=1}^{\lf(n-k)/r\rf}\big( 1-\sigma_-( \oline{\mbf{Y}}_{n-ri}^{n-ri+r-1},\mbf{X}_{n-ri+1}^{n-ri+r} ) \big)
            =\prod_{i=1}^{\lf(n-k)/r\rf}\big(1-\sigma_-(\mbf{V}_{n+r-ri})\big).
      \end{align*}
Moreover,
\begin{align*}
      &\Big\|\sum_{s\in\bb{S}}\bb{P}_\theta(S_k\in\cdot|S_{n}=s,\oline{\mbf{Y}}_{-m}^n,\mbf{X}_{-m}^n,S_{-m})\mu_1(s)\\
            &\hspace{4cm}-\sum_{s\in\bb{S}}\bb{P}_\theta(S_k\in\cdot|S_{n}=s,\oline{\mbf{Y}}_{-m}^n,\mbf{X}_{-m}^n,S_{-m})\mu_2(s)\Big\|_{TV}\\
            &\hspace{1cm}\leq\prod_{i=1}^{\lf(n-k)/r\rf}\big( 1-\sigma_-( \oline{\mbf{Y}}_{n-ri}^{n-ri+r-1},\mbf{X}_{n-ri+1}^{n-ri+r}) \big)=\prod_{i=1}^{\lf(n-k)/r\rf}\big(1-\sigma_-(\mbf{V}_{n+r-ri})\big).
\end{align*}
\end{lemma}
\noindent The proof is along the same line as the proof of \citet[Lemma 10]{kasahara2019asymptotic} and is omitted.

\begin{lemma}
Assume \ref{ergodic}--\ref{bound2}. Then, $\varepsilon_0\in(0,\frac{1}{16r})$ and $\rho\in(0,1)$ exist such that for all $m,n\in\bb{Z}^+$,
      \begin{align}
            &\bb{E}_{\theta^*}\Big[\prod_{k_1=1}^{n_1}(1-\sigma_-(\mbf{V}_{t_{k_1}}))^m\wedge \prod_{k_2=1}^{n_2}(1-\sigma_-(\mbf{V}_{t_{k_2}}))^m\Big] \leq 2(\rho^{n_1} \wedge \rho^{n_2}),\label{mix4}\\
            &\bb{E}_{\theta^*}\Big[\prod_{k_1=1}^{n_1}(1-\sigma_-(\mbf{V}_{t_{k_1}}))^m\wedge \prod_{k_2=1}^{n_2}(1-\sigma_-(\mbf{V}_{t_{k_2}}))^m\wedge \prod_{k_3=1}^{n_3}(1-\sigma_-(\mbf{V}_{t_{k_3}}))^m\Big]\notag\\
            &\hspace{8cm} \leq 2(\rho^{n_1} \wedge \rho^{n_2}\wedge \rho^{n_3}).\label{mix42}
      \end{align}
\end{lemma}
\noindent\textbf{Proof: }
(\ref{mix4}) follows from (\ref{mix3}) given
$\bb{E}_{\theta^*}\big[\prod_{k_1=1}^{n_1}(1-\sigma_-(\mbf{V}_{t_{k_1}}))^m \wedge
\prod_{k_2=1}^{n_2}(1-\sigma_-(\mbf{V}_{t_{k_2}}))^m]\leq \min \{\bb{E}_{\theta^*}[\prod_{k_1=1}^{n_1}(1-\sigma_-(\mbf{V}_{t_{k_1}}))^m], \bb{E}_{\theta^*}[\prod_{k_2=1}^{n_2}(1-\sigma_-(\mbf{V}_{t_{k_2}}))^m]\}$.  (\ref{mix42}) similarly follows.
\hfill$\square$

\begin{lemma}
Assume \ref{ergodic}--\ref{bound2}. For $1\leq m\leq n$, $\{t_{k_{i_1}}\}_{1\leq i_1\leq a_m}$ and $\{t_{h_{i_2}}\}_{1\leq i_2\leq b_m}$ are two sequences of integers satisfying (a) $t_{k_{i_1}}< t_{k_{i_1'}}$ for $1\leq i_1<i_1'\leq a_m$; (b) $t_{h_{i_2}}< t_{h_{i_2'}}$ for $1\leq i_2<i_2'\leq b_m$; and (c) $t_{k_{i_1}}\neq t_{h_{i_2}}$ for all $1\leq i_1\leq a_m$ and $1\leq i_2\leq b_m$. Then it holds that for the same $\varepsilon_0$ and $\rho$ defined as in Lemma \ref{mixingbound}, there exists a random sequence $\{A_{\underline{t},\oline{t}}\}$ such that
\begin{align}
    \sum_{m=1}^n \Big[\prod_{i_1=1}^{a_m} (1-\sigma_-(\mbf{V}_{t_{k_{i_1}}})) \wedge \prod_{i_2=1}^{b_m} (1-\sigma_-(\mbf{V}_{t_{h_{i_2}}}))\Big]
    \leq \rho^{-2\varepsilon_0(\oline{t}-\underline{t}+1)}A_{\underline{t},\oline{t}}\sum_{m=1}^n \rho^{a_m \vee b_m},\label{mix12}
\end{align}
where $\underline{t}=\min\{t_{k_1},t_{h_1}\}$,  $\oline{t}=\max\{t_{k_{a_m}},t_{h_{b_m}}\}$, and $\bb{P}_{\theta^*}(A_{\underline{t},\oline{t}}\geq M\ \mathrm{i.o.})=0$ for a constant $M<\infty$.

For $1\leq m\leq n$, $\{t_{k_{i_1}}\}_{1\leq i_1\leq a_m}$, $\{t_{h_{i_2}}\}_{1\leq i_2\leq b_m}$, and $\{t_{\ell_{i_3}}\}_{1\leq i_3\leq c_m}$ are three sequences of integers satisfying (a) $t_{k_{i_1}}< t_{k_{i_1'}}$ for $1\leq i_1<i_1'\leq a_m$; (b) $t_{h_{i_2}}< t_{h_{i_2'}}$ for $1\leq i_2<i_2'\leq b_m$;  (c) $t_{\ell_{i_3}}< t_{\ell_{i_3'}}$ for $1\leq i_3<i_3'\leq c_m$; and (d) $t_{k_{i_1}}\neq t_{h_{i_2}}$, $t_{k_{i_1}}\neq t_{\ell_{i_3}}$ and $t_{h_{i_2}}\neq t_{\ell_{i_3}}$ for all $1\leq i_1\leq a_m$, $1\leq i_2\leq b_m$, and $1\leq i_3\leq c_m$. Then it holds that for the same $\varepsilon_0$ and $\rho$ defined as in Lemma \ref{mixingbound}, there exists a random sequence $\{A_{\underline{t},\oline{t}}\}$ such that
\begin{align}
    \sum_{m=1}^n \bigg[\prod_{i_1=1}^{a_m} (1-\sigma_-(\mbf{V}_{t_{k_{i_1}}})) \wedge \prod_{i_2=1}^{b_m} (1-\sigma_-(\mbf{V}_{t_{h_{i_2}}})) \wedge \prod_{i_3=1}^{c_m} (1-\sigma_-(\mbf{V}_{t_{\ell_{i_3}}}))\bigg]\notag\\
    \leq \rho^{-2\varepsilon_0(\oline{t}-\underline{t}+1)}A_{\underline{t},\oline{t}}\sum_{m=1}^n \rho^{a_m \vee b_m \vee c_m},\label{mix13}
\end{align}
where $\underline{t}=\min\{t_{k_1},t_{h_1},t_{\ell_1}\}$,  $\oline{t}=\max\{t_{k_{a_m}},t_{h_{b_m}},t_{\ell_{c_m}}\}$, and $\bb{P}_{\theta^*}(A_{\underline{t},\oline{t}}\geq M\ \mathrm{i.o.})=0$ for a constant $M<\infty$.
\label{mixapp}
\end{lemma}
\noindent\textbf{Proof: }
      First, we show (\ref{mix12}). We define $\rho$, $\bbm{1}_t$, and $\varepsilon_0$ as in (\ref{rho}), (\ref{Itk}), and (\ref{epsilon0}), respectively. Notice that $\varepsilon_0\in(0,\frac{1}{16r})$. Using $1-\sigma_-(\mbf{V}_t)\leq \rho^{1-\bbm{1}_t}$, it follows that
\begin{align*}
      \prod_{i_1=1}^{a_m} (1-\sigma_-(\mbf{V}_{t_{k_{i_1}}}))
      \leq \rho^{a_m-\sum_{i_1=1}^{a_m} \bbm{1}_{t_{k_{i_1}}}}
      \leq \rho^{a_m-\sum_{t=\underline{t}}^{\oline{t}}\bbm{1}_t},\\
      \prod_{i_2=1}^{b_m} (1-\sigma_-(\mbf{V}_{t_{\ell_{i_2}}}))
      \leq \rho^{b_m-\sum_{i_2=1}^{b_m} \bbm{1}_{t_{\ell_{i_2}}}}
       \leq \rho^{b_m-\sum_{t=\underline{t}}^{\oline{t}}\bbm{1}_t},
\end{align*}
and
\begin{align}
      \sum_{m=1}^n \Big(\prod_{i_1=1}^{a_m} (1-\sigma_-(\mbf{V}_{t_{k_{i_1}}})) \wedge
      \prod_{i_2=1}^{b_m} (1-\sigma_-(\mbf{V}_{t_{\ell_{i_2}}}))\Big)
      &\leq \sum_{m=1}^n \big(\rho^{a_m-\sum_{t=\underline{t}}^{\oline{t}}\bbm{1}_t} \wedge \rho^{b_m-\sum_{t=\underline{t}}^{\oline{t}}\bbm{1}_t}\big)\notag\\
      &=\rho^{-\sum_{t=\underline{t}}^{\oline{t}}\bbm{1}_t} \sum_{m=1}^n \rho^{a_m \vee b_m}.\label{omega2bound}
\end{align}
Since $\mbf{V}_{t}$ forms a stationary and ergodic sequence for $\underline{t}\leq t\leq \overline{t}$, the strong law of large numbers yields $(\oline{t}-\underline{t}+1)^{-1}(\sum_{t=\underline{t}}^{\oline{t}}\bbm{1}_t)\convas \bb{E}_{\theta^*} [\bbm{1}_1] <\varepsilon_0$, as $\oline{t}-\underline{t}\ra\infty$. Hence, $\bb{P}_{\theta^*}\big(\rho^{-\sum_{t=\underline{t}}^{\oline{t}}\bbm{1}_t}\geq \rho^{-2\varepsilon_0(\oline{t}-\underline{t}+1)}\ \mathrm{i.o.}\big)=0.$ Let $\{A_{\underline{t},\oline{t}}\}$ denote a random sequence such that $\bb{P}_{\theta^*}(A_{\underline{t},\oline{t}}\geq M\ \mathrm{i.o.})=0$ for a constant $M<\infty$. Then (\ref{omega2bound}) is bounded by
$\rho^{-2\varepsilon_0(\oline{t}-\underline{t}+1)}A_{\underline{t},\oline{t}}\sum_{m=1}^n \rho^{a_m \vee b_m}$. (\ref{mix13}) similarly follows.
\hfill$\square$

\begin{lemma}\label{mixingbound2}
      Assume \ref{ergodic}--\ref{bound2} and \ref{differentiable}--\ref{finitemoment}. Then, for $-m'<-m<k\leq t$,
      \begin{align}
            \begin{split}
                  &\bb{E}_{\theta^*}\Big[\big\|\bb{E}_{\theta^*}[\phi_{\theta^*,k} |\oline{\mbf{Y}}_{-m}^t,S_{-m}=s,\mbf{X}_{-m}^t]
                  -\bb{E}_{\theta^*}[\phi_{\theta^*,k} |\oline{\mbf{Y}}_{-m}^t,\mbf{X}_{-m}^t]\big\|^2\Big]\\
                  &\hspace{6cm}\leq 8\big(\bb{E}_{\theta^*}\big[\|\phi_{\theta^*,0}\|_\infty^4\big]\big)^{\frac{1}{2}} \rho^{\lf(k+m-1)/2r\rf},
            \end{split}\label{mix21}\\
            \begin{split}
                  &\bb{E}_{\theta^*}\Big[\|\bb{E}_{\theta^*}[\phi_{\theta^*,k}|\oline{\mbf{Y}}_{-m}^t,\mbf{X}_{-m}^t]-\bb{E}_{\theta^*}[\phi_{\theta^*,k}|\oline{\mbf{Y}}_{-m'}^t,\mbf{X}_{-m'}^t]\|^2\Big]\\
                  &\hspace{6cm}\leq 8 \big(\bb{E}_{\theta^*}\big[\|\phi_{\theta^*,0}\|_\infty^4\big]\big)^{\frac{1}{2}} \rho^{\lf(k+m-1)/2r\rf},
            \end{split}\label{mix22}\\
            \begin{split}
                  &\bb{E}_{\theta^*}\Big[\|\bb{E}_{\theta^*}[\phi_{\theta^*,k}|\oline{\mbf{Y}}_{-m}^{t},\mbf{X}_{-m}^{t}]-\bb{E}_{\theta^*}[\phi_{\theta^*,k}|\oline{\mbf{Y}}_{-m}^{t-1},\mbf{X}_{-m}^{t-1}]\|^2\Big]\\
                  &\hspace{6cm}\leq 8 \big(\bb{E}_{\theta^*}\big[\|\phi_{\theta^*,0}\|_\infty^4\big]\big)^{\frac{1}{2}} \rho^{\lf(t-1-k)/2r\rf}.
            \end{split}\label{mix23}
      \end{align}
\end{lemma}
\noindent\textbf{Proof: }
      First, we show (\ref{mix21}). Note that
      \begin{align*}
            &\|\bb{E}_{\theta^*}[\phi_{\theta^*,k}|\oline{\mbf{Y}}_{-m}^{t},S_{-m}=s,\mbf{X}_{-m}^{t}]-\bb{E}_{\theta^*}[\phi_{\theta^*,k}|\oline{\mbf{Y}}_{-m}^{t},\mbf{X}_{-m}^{t}]\|\\
            \leq & 2\|\phi_{\theta^*,k}\|_\infty
            \Big\|\sum_{s_{-m}} \bb{P}_{\theta^*}(S_{k-1}\in\cdot|\oline{\mbf{Y}}_{-m}^{t},S_{-m}=s_{-m},\mbf{X}_{-m}^{t})\delta_{s}(s_{-m})\\
            &\qquad\qquad-\sum_{s_{-m}} \bb{P}_{\theta^*}(S_{k-1}\in\cdot|\oline{\mbf{Y}}_{-m}^{t},S_{-m}=s_{-m},\mbf{X}_{-m}^{t})\bb{P}_{\theta^*}(S_{-m}=s_{-m}|\oline{\mbf{Y}}_{-m}^t,\mbf{X}_{-m}^t)\Big\|\\
            \leq & 2\|\phi_{\theta^*,k}\|_\infty
            \prod_{i=1}^{\lf(k+m-1)/r\rf} (1-\sigma_-(\mbf{V}_{-m+ri})),
      \end{align*}
      where the second inequality follows from Lemma \ref{mixing}.
      Using Lemma \ref{mixingbound} and H\"{o}lder's inequality, the second moment is bounded by
      \begin{align*}
            &4\bb{E}_{\theta^*}\Big[\|\phi_{\theta^*,k}\|^2_\infty
            \prod_{i=1}^{\lf(k+m-1)/r\rf} (1-\sigma_-(\mbf{V}_{-m+ri}))^2\Big]\\
            \leq &4\big(\bb{E}_{\theta^*}\big[\|\phi_{\theta^*,0}\|_\infty^4\big]\big)^{\frac{1}{2}}
            \Big(\bb{E}_{\theta^*}\Big[\prod_{i=1}^{\lf(k+m-1)/r\rf} (1-\sigma_-(\mbf{V}_{-m+ri}))^4\Big]\Big)^{\frac{1}{2}}\\
            \leq& 8\big(\bb{E}_{\theta^*}\big[\|\phi_{\theta^*,0}\|_\infty^4\big]\big)^{\frac{1}{2}} \rho^{\lf(k+m-1)/2r\rf}.
      \end{align*}
      We can show (\ref{mix22}) by replacing $\bb{P}_{\theta^*}(S_{k-1}\in\cdot|\oline{\mbf{Y}}_{-m}^t,S_{-m}=s,\mbf{X}_{-m}^t)$ with $\bb{P}_{\theta^*}(S_{k-1}\in\cdot|\oline{\mbf{Y}}_{-m'}^t,\mbf{X}_{-m'}^t)$. For (\ref{mix23}),
      \begin{align*}
            &\|\bb{E}_{\theta^*}[\phi_{\theta^*,k}|\oline{\mbf{Y}}_{-m}^{t},\mbf{X}_{-m}^{t}]-\bb{E}_{\theta^*}[\phi_{\theta^*,k}|\oline{\mbf{Y}}_{-m}^{t-1},\mbf{X}_{-m}^{t-1}]\|\\
            \leq & 2 \|\phi_{\theta^*,k}\|_\infty
            \Big\|\sum_{s_{t-1}} \bb{P}_{\theta^*}(S_k\in\cdot|S_{t-1}=s_{t-1},\oline{\mbf{Y}}_{-m}^{t-1},\mbf{X}_{-m}^{t-1})\bb{P}_{\theta^*}(S_{t-1}=s_{t-1}|\oline{\mbf{Y}}_{-m}^{t},\mbf{X}_{-m}^{t})\\
            &\hspace{3cm}-\sum_{s_{t-1}} \bb{P}_{\theta^*}(S_k\in\cdot|S_{t-1}=s_{t-1},\oline{\mbf{Y}}_{-m}^{t-1},\mbf{X}_{-m}^{t-1})\bb{P}_{\theta^*}(S_{t-1}=s_{t-1}|\oline{\mbf{Y}}_{-m}^{t-1},\mbf{X}_{-m}^{t-1})\Big\|\\
            \leq &2 \|\phi_{\theta^*,k}\|_\infty
            \prod_{i=1}^{\lf(t-1-k)/r\rf} (1-\sigma_-(\mbf{V}_{t-1+r-ri})),
      \end{align*}
      where the second inequality follows from Lemma \ref{mixingreverse}.
      The bound for its second moment follows similarly to the proof for (\ref{mix21}).
\hfill$\square$

\begin{lemma}\label{gammap}
Assume \ref{ergodic}--\ref{bound2} and \ref{differentiable}--\ref{finitemoment}. There exist a random sequence $\{A_{t,m}\}$ and a random variable $K\in L^1(\bb{P}_{\theta^*})$ such that for all $t\geq1$ and $0\leq m\leq m'$,
\begin{align}
    &\max_{s_i,s_j}\sup_{\theta\in G}\|\Gamma_{t,m,s_i}(\theta)-\Gamma_{t,m',s_j}(\theta)\|\leq  K(t\vee m)^2 \rho^{(t+m)/4r}A_{t,m},\label{gammap1}\\
    &\max_{s}\sup_{\theta\in G}\|\Gamma_{t,m,s}(\theta)-\Gamma_{t,m}(\theta)\|\leq  K(t\vee m)^2 \rho^{(t+m)/4r}A_{t,m},\label{gammap2}
\end{align}
where $\bb{P}_{\theta^*}(A_{t,m}\geq M\ \mathrm{i.o.})=0$ for $M<\infty$.
\end{lemma}
\noindent\textbf{Proof: }
      Put $\|\dot{\phi}_k\|_\infty \tq \max_{s_k,s_{k-1}}\sup_{\theta\in G}\|\dot{\phi}_{\theta,k}\|$. For $-m'<-m<k\leq t$,
      \begin{align*}
            &\|\bb{E}_\theta[\dot{\phi}_{\theta,k}|\oline{\mbf{Y}}_{-m}^t,S_{-m}=s,\mbf{X}_{-m}^t]
            -\bb{E}_\theta[\dot{\phi}_{\theta,k}|\oline{\mbf{Y}}_{-m}^{t},\mbf{X}_{-m}^{t}]\|\\
            &\hspace{7cm}\leq 2 \|\dot{\phi}_k\|_{\infty}\prod_{i=1}^{\lf(k+m-1)/r\rf}(1-\sigma_-(\mbf{V}_{-m+ri})),\\
            &\|\bb{E}_\theta[\dot{\phi}_{\theta,k}|\oline{\mbf{Y}}_{-m}^t,S_{-m}=s_i,\mbf{X}_{-m}^t]
            -\bb{E}_\theta[\dot{\phi}_{\theta,k}|\oline{\mbf{Y}}_{-m'}^{t},S_{-m'}=s_j,\mbf{X}_{-m'}^{t}]\|\\
            &\hspace{7cm}\leq 2 \|\dot{\phi}_k\|_{\infty}\prod_{i=1}^{\lf(k+m-1)/r\rf}(1-\sigma_-(\mbf{V}_{-m+ri})),\\
            &\|\bb{E}_\theta[\dot{\phi}_{\theta,k}|\oline{\mbf{Y}}_{-m}^t,S_{-m}=s,\mbf{X}_{-m}^t]
            -\bb{E}_\theta[\dot{\phi}_{\theta,k}|\oline{\mbf{Y}}_{-m}^{t-1},S_{-m}=s,\mbf{X}_{-m}^{t-1}]\|\\
            &\hspace{7cm}\leq 2 \|\dot{\phi}_k\|_{\infty}\prod_{i=1}^{\lf(t-1-k)/r\rf}(1-\sigma_-(\mbf{V}_{t+r-ri})),\\
            &\|\bb{E}_\theta[\dot{\phi}_{\theta,k}|\oline{\mbf{Y}}_{-m}^t,\mbf{X}_{-m}^t]
            -\bb{E}_\theta[\dot{\phi}_{\theta,k}|\oline{\mbf{Y}}_{-m}^{t-1},\mbf{X}_{-m}^{t-1}]\|
            \leq 2 \|\dot{\phi}_k\|_{\infty}\prod_{i=1}^{\lf(t-1-k)/r\rf}(1-\sigma_-(\mbf{V}_{t+r-ri})).
      \end{align*}

      First, we show (\ref{gammap2}). Note that
      \begin{align}
            &\|\Gamma_{t,m,s}-\Gamma_{t,m}\|
            \leq 2\|\dot{\phi}_t\|_{\infty}
            \prod_{i=1}^{\lf(t+m-1)/r\rf} \big(1-\sigma_-(\mbf{V}_{-m+ri})\big) \notag \\
            &+4\sum_{k=-m+1}^{t-1}\|\dot{\phi}_k\|_\infty
            \Big(\prod_{i=1}^{\lf(k+m-1)/r\rf} \big(1-\sigma_-(\mbf{V}_{-m+ri})\big)\wedge
            \prod_{i=1}^{\lf(t-k-1)/r\rf} \big(1-\sigma_-(\mbf{V}_{t+r-ri})\big)\Big)\notag\\
            &\leq 4\max_{-m+1\leq k\leq t}\|\dot{\phi}_k\|_\infty \sum_{k=-m+1}^t\Big( \prod_{i=1}^{\lf(k+m-1)/r\rf} \big(1-\sigma_-(\mbf{V}_{-m+ri})\big)\wedge
            \prod_{i=1}^{\lf(t-k-1)/r\rf} \big(1-\sigma_-(\mbf{V}_{t+r-ri})\big)\Big). \label{gammaapp2}
      \end{align}
      The first part of (\ref{gammaapp2}) is bounded by
      \begin{align}
            4\max_{-m+1\leq k\leq t}\|\dot{\phi}_k\|_\infty
            &\leq 4\sum_{k=-m+1}^t(|k|\vee 1)^2\frac{1}{(|k|\vee 1)^2}\|\dot{\phi}_k\|_\infty\notag\\
            &\leq 4(t\vee m)^2\sum_{k=-\infty}^\infty \frac{1}{(|k|\vee1)^2} \|\dot{\phi}_k\|_\infty.
            \label{gammamax}
      \end{align}
      We proceed to bound the second part of (\ref{gammaapp2}). Since $-m+r\lf (k+m-1)/r \rf < t+r- r\lf (t-k-1)/r \rf $, we can apply Lemma \ref{mixapp}. Using $\rho^{\lf(k+m-1)/r\rf}\wedge\rho^{\lf(t-k-1)/r\rf}=\rho^{\lf(t-k-1)/r\rf}$ for $k\leq \frac{t-m}{2}$ and $\rho^{\lf(k+m-1)/r\rf}$ for $k\geq \frac{t-m}{2}$,
      \begin{align}
                  &\rho^{-2\varepsilon_0(t+m+r-1)}\sum_{k=-m+1}^{t}\big(\rho^{\lf\frac{k+m-1}{r}\rf}\wedge\rho^{\lf\frac{t-k-1}{r}\rf}\big)\notag \\
                  \leq &\rho^{-2\varepsilon_0(t+m+r-1)} \Big( \sum_{k\leq (t-m)/2}\rho^{\lf\frac{t-k-1}{r}\rf}+\sum_{k\geq(t-m)/2}\rho^{\lf\frac{k+m-1}{r}\rf} \Big)\notag \\
                  \leq & 2\frac{\rho^{\frac{t+m}{4r}}}{\rho^{r-1}(1-\rho^{\frac{1}{r}})}.
            \label{lemma101}
      \end{align}
      Because of (\ref{gammamax}), (\ref{lemma101}), and Lemma \ref{mixapp}, (\ref{gammap2}) follows.

      Next, we follow a similar procedure to show (\ref{gammap1}). Note that
     \begin{align}
            &\|\Gamma_{t,m,s_i}(\theta)-\Gamma_{t,m',s_j}(\theta)\|\leq
            2\|\dot{\phi}_t\|_\infty\prod_{i=1}^{\lf(t+m-1)/r\rf} \big(1-\sigma_-(\mbf{V}_{-m+ri})\big)\nonumber\\
            &+ 4 \sum_{k=-m+1}^{t-1}\|\dot{\phi}_k\|_\infty
            \Big(\prod_{i=1}^{\lf(k+m-1)/r\rf} \big(1-\sigma_-(\mbf{V}_{-m+ri})\big)\wedge
            \prod_{i=1}^{\lf(t-k-1)/r\rf} \big(1-\sigma_-(\mbf{V}_{t+r-ri})\big)\Big)\nonumber\\
            & +2\sum_{k=-m'+1}^{-m} \|\dot{\phi}_k\|_\infty \prod_{i=1}^{\lf(t-k-1)/r\rf}\big(1-\sigma_-(\mbf{V}_{t+r-ri})\big).
            \label{gammaapp1}
      \end{align}
      The first two terms on the right-hand side can be bounded as above. We proceed to show the bound for the third term. Define $\rho$, $\bbm{1}_t$, and $\varepsilon_0$ as in (\ref{rho}), (\ref{Itk}), and (\ref{epsilon0}), respectvely. Notice that the $\rho$ and $\varepsilon_0$ are the same as the ones in Lemma \ref{mixingbound}. Using $1-\mbf{V}_t\leq \rho^{1-\bbm{1}_t}$,
      \begin{align*}
            &\sum_{k=-m'+1}^{-m} \|\dot{\phi}_k\|_\infty \prod_{i=1}^{\lf(t-k-1)/r\rf}\big(1-\sigma_-(\mbf{V}_{t+r-ri})\big)
            \leq \sum_{k=-m'+1}^{-m} \|\dot{\phi}_k\|_\infty \rho^{\lf(t-k-1)/r\rf-\sum_{i=1}^{\lf(t-k-1)/r\rf} \bbm{1}_{t+r-ri}}\\
            &\leq \sum_{k=-m'+1}^{-m} \|\dot{\phi}_k\|_\infty \rho^{\lf(t-k-1)/r\rf-\sum_{i=-m'+r+1}^{t} \bbm{1}_i}
            \leq  \rho^{-\sum_{i=-m'+r+1}^{t} \bbm{1}_i} \sum_{k=-m'+1}^{-m} \|\dot{\phi}_k\|_\infty\rho^{\lf(t-k-1)/r\rf}.
      \end{align*}
      Using (\ref{lemma22}) and (\ref{lemma23}), $\bb{E}_{\theta^*}[\rho^{-\sum_{i=-m'+r+1}^{t} \bbm{1}_i}]\leq 2$ for all $m'$ and $t$. Then $\rho^{-\sum_{i=-m'+r+1}^{t} \bbm{1}_i}\leq \rho^{-\sum_{i=-\infty}^{+\infty} \bbm{1}_i}$, which is in $L^1(\bb{P}_{\theta^*})$.

Since $-\frac{k}{2r}\leq \frac{t-m-2k}{2r}$ for $k\leq m$, we can follow a similar procedure to \citet[p. 2295]{douc2004asymptotic} to obtain
      \begin{align*}
            2\sum_{k=-m'+1}^{-m} \|\dot{\phi}_k\|_\infty\rho^{\lf(t-k-1)/r\rf}&\leq2\rho^{\frac{t+m}{2r}}\sum_{k=-m'+1}^{-m} \|\dot{\phi}_k\|_\infty\rho^{\frac{t-m-2k}{2r}-1}\\
            \leq2\rho^{\frac{t+m}{2r}}\sum_{k=-\infty}^{\infty} \|\dot{\phi}_k\|_\infty\rho^{ \frac{|k|}{2r}-1}
            &\leq 2\rho^{\frac{t+m}{4r}}\sum_{k=-\infty}^{\infty} \|\dot{\phi}_k\|_\infty\rho^{ \frac{|k|}{2r}-1}.
      \end{align*}
      Then (\ref{gammap1}) follows.
\hfill$\square$

\begin{lemma}\label{gammae}
      Assume \ref{ergodic}--\ref{bound2} and \ref{differentiable}--\ref{finitemoment}. Then, for all $t\geq 1$ and $0\leq m\leq m'$,
      \begin{align}
            \lim_{m\ra\infty}\bb{E}_{\theta^*}[\max_{s_i,s_j}\sup_{\theta\in G}\|\Gamma_{t,m,s_i}(\theta)-\Gamma_{t,m',s_j}(\theta)\|]=0,\label{gammae1}\\
            \lim_{m\ra\infty}\bb{E}_{\theta^*}[\max_{s}\sup_{\theta\in G}\|\Gamma_{t,m,s}(\theta)-\Gamma_{t,m}(\theta)\|]=0.\label{gammae2}
      \end{align}
\end{lemma}
\noindent\textbf{Proof: }
      (\ref{gammae1}) follows from (\ref{gammaapp1}) and
      \begin{align*}
            &\bb{E}_{\theta^*}\big[\sup_{\theta\in G}\max_{s_i,s_j\in\bb{S}}\|\Gamma_{t,m,s_i}(\theta)-\Gamma_{t,m',s_j}(\theta)\|\big]\\
            \leq & 2 ( \bb{E}_{\theta^*} [\|\dot{\phi}_0\|_\infty^2] )^\frac{1}{2}
            \big(\big(\bb{E}_{\theta^*} \big[\prod_{i=1}^{\lf(t+m-1)/r\rf} \big(1-\sigma_-(\mbf{V}_{-m+ri})\big)^2\big]\big)^\frac{1}{2}\\
            &+ 2 \sum_{k=-m+1}^{t-1}
            \big(\bb{E}_{\theta^*}\big[ \prod_{i=1}^{\lf(k+m-1)/r\rf} \big(1-\sigma_-(\mbf{V}_{-m+ri})\big)^2\wedge
            \prod_{i=1}^{\lf(t-k-1)/r\rf} \big(1-\sigma_-(\mbf{V}_{t+r-ri})\big)^2\big]\big)^\frac{1}{2}\\
            & +\sum_{k=-m'+1}^{-m}  \big(\bb{E}_{\theta^*}\big[\prod_{i=1}^{\lf(t-k-1)/r\rf}\big(1-\sigma_-(\mbf{V}_{t+r-ri})\big)^2\big]\big)^\frac{1}{2}\big)
            \leq 8\big(\bb{E}_{\theta^*}[\|\dot{\phi}_0\|_\infty^2]\big)^\frac{1}{2} \frac{\rho^{\frac{t+m-2}{4r}-1}}{1-\rho^{\frac{1}{2r}}},
      \end{align*}
      which converges to 0. Similarly, we can obtain (\ref{gammae2}) by using (\ref{gammaapp2}).
\hfill$\square$

\begin{lemma}\label{phip}
      Assume \ref{ergodic}--\ref{bound2} and \ref{differentiable}--\ref{finitemoment}. Then, there exist  a random sequence $\{A_{t,m}\}$ and a random variable $K\in L^1(\bb{P}_{\theta^*})$ such that, for all $t\geq1$ and $0\leq m\leq m'$,
      \begin{align*}
            &\max_{s}\sup_{\theta\in G}\|\Phi_{t,m,s}(\theta)-\Phi_{t,m}(\theta)\|\leq  K(t\vee m)^3 \rho^{(t+m)/8r}A_{t,m},\\
            &\max_{s_i,s_j}\sup_{\theta\in G}\|\Phi_{t,m,s_i}(\theta)-\Phi_{t,m',s_j}(\theta)\|\leq  K(t\vee m)^3 \rho^{(t+m)/8r}A_{t,m},
      \end{align*}
      where $\bb{P}_{\theta^*}(A_{t,m}\geq M\ \mathrm{i.o.})=0$ for $M<\infty$.
\end{lemma}
\noindent\textbf{Proof: }
      Put $\|\phi_k\|_\infty \tq \max_{s_k,s_{k-1}}\sup_{\theta\in G}\|\phi_{\theta,k}\|$. For $m'\geq m\geq0$, all $-m<\ell\leq k\leq n$, all $\theta\in G$, and all $s_{-m}\in\bb{S}$,
      \begin{align*}
            &\|\Cov_\theta[\phi_{\theta,k},\phi_{\theta,\ell}|\oline{\mbf{Y}}_{-m}^n,\mbf{X}_{-m}^n]\|
            \leq 2\|\phi_{k}\|_{\infty} \|\phi_{\ell}\|_{\infty} \prod_{i=1}^{\lf(k-\ell-1)/r\rf}(1-\sigma_-(\mbf{V}_{\ell+ri})),\\
            &\|\Cov_\theta[\phi_{\theta,k},\phi_{\theta,\ell}|\oline{\mbf{Y}}_{-m}^n,S_{-m}=s_{-m},\mbf{X}_{-m}^n]\|
            \leq  2\|\phi_{k}\|_{\infty} \|\phi_{\ell}\|_{\infty}\prod_{i=1}^{\lf(k-\ell-1)/r\rf}(1-\sigma_-(\mbf{V}_{\ell+ri})),\\
            &\|\Cov_\theta[\phi_{\theta,k},\phi_{\theta,\ell}|\oline{\mbf{Y}}_{-m}^n,S_{-m}=s_{-m},\mbf{X}_{-m}^n]-\Cov_\theta[\phi_{\theta,k},\phi_{\theta,\ell}|\oline{\mbf{Y}}_{-m}^n,\mbf{X}_{-m}^n]\|\\
            &\hspace{6cm}\leq 6\|\phi_{k}\|_{\infty} \|\phi_{\ell}\|_{\infty}\prod_{i=1}^{\lf(\ell+m-1)/r\rf}(1-\sigma_-(\mbf{V}_{-m+ri})),\\
            &\|\Cov_\theta[\phi_{\theta,k},\phi_{\theta,\ell}|\oline{\mbf{Y}}_{-m}^n,\mbf{X}_{-m}^n]-\Cov_\theta[\phi_{\theta,k},\phi_{\theta,\ell}|\oline{\mbf{Y}}_{-m}^{n-1},\mbf{X}_{-m}^{n-1}]\|\\
            &\hspace{6cm}\leq  6 \|\phi_{k}\|_{\infty} \|\phi_{\ell}\|_{\infty}\prod_{i=1}^{\lf(n-k-1)/r\rf}(1-\sigma_-(\mbf{V}_{n+r-ri})),\\
            &\|\Cov_\theta[\phi_{\theta,k},\phi_{\theta,\ell}|\oline{\mbf{Y}}_{-m}^n,S_{-m}=s_{-m},\mbf{X}_{-m}^n]-\Cov_\theta[\phi_{\theta,k},\phi_{\theta,\ell}|\oline{\mbf{Y}}_{-m}^{n-1},S_{-m}=s_{-m},\mbf{X}_{-m}^{n-1}]\|\\
            &\hspace{6cm}\leq  6 \|\phi_{k}\|_{\infty} \|\phi_{\ell}\|_{\infty}\prod_{i=1}^{\lf(n-k-1)/r\rf}(1-\sigma_-(\mbf{V}_{n+r-ri})).
      \end{align*}
      We define $\Lambda_a^b\tq\sum_{i=a}^b \phi_{\theta,i}$. Then, $\Phi_{t,m,s}(\theta)-\Phi_{t,m',s}(\theta)$ may be decomposed as $A+2B+C$, where
      \begin{align*}
            A=&\Var_\theta[\Lambda_{-m+1}^{t-1}|\oline{\mbf{Y}}_{-m}^{t},S_{-m}=s,\mbf{X}_{-m}^{t}]
            -\Var_\theta[\Lambda_{-m+1}^{t-1}|\oline{\mbf{Y}}_{-m}^{t-1},S_{-m}=s,\mbf{X}_{-m}^{t-1}]\\
            &-\Var_\theta[\Lambda_{-m+1}^{t-1}|\oline{\mbf{Y}}_{-m}^{t},\mbf{X}_{-m}^{t}]
            +\Var_\theta[\Lambda_{-m+1}^{t-1}|\oline{\mbf{Y}}_{-m}^{t-1},\mbf{X}_{-m}^{t-1}],\\
            B=&\Cov_\theta[\Lambda_{-m+1}^{t-1},\phi_{\theta,t}|\oline{\mbf{Y}}_{-m}^t,S_{-m}=s,\mbf{X}_{-m}^t]
            -\Cov_\theta[\Lambda_{-m+1}^{t-1},\phi_{\theta,t}|\oline{\mbf{Y}}_{-m}^t,\mbf{X}_{-m}^t],\\
            C=&\Var_\theta[\phi_{\theta,t}|\oline{\mbf{Y}}_{-m}^{t},S_{-m}=s,\mbf{X}_{-m}^{t}]
            -\Var_\theta[\phi_{\theta,t}|\oline{\mbf{Y}}_{-m}^{t},\mbf{X}_{-m}^{t}].
      \end{align*}
      We have
      \begin{align*}
            &\|A\|\leq \max_{-m+1\leq \ell\leq k\leq t-1} \|\phi_{k}\|_\infty\|\phi_{\ell}\|_\infty \times
            2\sum_{-m+1\leq \ell\leq k\leq t-1} \Big(\big(2\times6\prod_{i=1}^{\lf(\ell+m-1)/r\rf}(1-\sigma_-(\mbf{V}_{-m+ri}))\big)\\
            &\hspace{3cm}\wedge \big(4\times2\prod_{i=1}^{\lf(k-\ell-1)/r\rf}(1-\sigma_-(\mbf{V}_{\ell+ri}))\big)
            \wedge \big(2\times 6\prod_{i=1}^{\lf(t-k-1)/r\rf}(1-\sigma_-(\mbf{V}_{t+r-ri}))\big)\Big),\\
            &\|B\|\leq \max_{-m+1\leq k\leq t-1} \|\phi_{k}\|_\infty\|\phi_{t}\|_\infty \times
            \sum_{-m+1\leq k\leq t-1} \Big(\big(6\prod_{i=1}^{\lf(k+m-1)/r\rf}(1-\sigma_-(\mbf{V}_{-m+ri}))\big)\\
            &\hspace{3cm}\wedge \big(2\times 2\prod_{i=1}^{\lf(t-k-1)/r\rf}(1-\sigma_-(\mbf{V}_{k+ri}))\big)\Big),\\
            &\|C\|\leq \|\phi_{t}\|_\infty^2 \times
            6\prod_{i=1}^{\lf(t+m-1)/r\rf}(1-\sigma_-(\mbf{V}_{-m+ri})).
      \end{align*}
      Similar to the calculation on p. 2299 of \citet{douc2004asymptotic}, we derive
      \begin{align*}
            \max_{-m+1\leq \ell\leq k\leq t-1} \|\phi_{k}\|_\infty\|\phi_{\ell}\|_\infty
            \leq (m^3+t^3)\sum_{k=-\infty}^\infty \frac{1}{(|k|\vee1)^2}\|\phi_k\|_\infty^2.
      \end{align*}
      In view of Lemma \ref{mixapp} and
      \begin{align*}
            \rho^{-2\varepsilon_0(t+m-r+1)}\sum_{-m+1\leq\ell\leq k\leq t-1} \Big( \rho^{\lf(\ell+m-1)/r\rf} \wedge \rho^{\lf(k-\ell-1)/r\rf}\wedge \rho^{\lf(t-k-1)/r\rf}\Big)\\
             \leq 4\frac{1}{\rho(1-\rho)(1-\rho^{\frac{1}{2}})}\rho^{\frac{t+m-4}{8r}},
      \end{align*}
      there exists  a random sequence $\{A_{t,m}\}$ such that
      \begin{align*}
            &\|A\| \leq (m^3+t^3)\sum_{k=-\infty}^\infty \frac{1}{(|k|\vee1)^2}\|\phi_k\|_\infty^2 \frac{96\rho^{\frac{t+m}{8r}-1}}{\rho^{\frac{1}{r}}(1-\rho^{\frac{1}{r}})(1-\rho^{\frac{1}{2r}})}A_{t,m},\\
            &\|B\| \leq (m^3+t^3)\sum_{k=-\infty}^\infty \frac{1}{(|k|\vee1)^2}\|\phi_k\|_\infty^2 \frac{12\rho^{\frac{t+m}{4r}-1}}{\rho^\frac{1}{r}(1-\rho^{\frac{1}{r}})}A_{t,m},\\
            &\|C\| \leq(m^3+t^3)\sum_{k=-\infty}^\infty \frac{1}{(|k|\vee1)^2}\|\phi_k\|_\infty^2 6 \rho^{\lf(t+m-1)/4r\rf}A_{t,m},
      \end{align*}
        where $\bb{P}_{\theta^*}(A_{t,m}\geq M\ \mathrm{i.o.})=0$ for $M<\infty$.
      The difference $\Phi_{t,m,s_i}(\theta)-\Phi_{t,m',s_j}(\theta)$ can be decomposed as $A+2B+C-D-2E-2F$, where
\begin{align*}
      A=&\Var_\theta[\Lambda_{-m+1}^{t-1}|\oline{\mbf{Y}}_{-m}^{t},S_{-m}=s_i,\mbf{X}_{-m}^{t}]
      -\Var_\theta[\Lambda_{-m+1}^{t-1}|\oline{\mbf{Y}}_{-m}^{t-1},S_{-m}=s_i,\mbf{X}_{-m}^{t-1}]\\
      &-\Var_\theta[\Lambda_{-m+1}^{t-1}|\oline{\mbf{Y}}_{-m'}^{t},S_{-m'}=s_j,\mbf{X}_{-m'}^{t}]
      +\Var_\theta[\Lambda_{-m+1}^{t-1}|\oline{\mbf{Y}}_{-m'}^{t-1},S_{-m'}=s_j,\mbf{X}_{-m'}^{t-1}],\\
      B=&\Cov_\theta[\Lambda_{-m+1}^{t-1},\phi_{\theta,t}|\oline{\mbf{Y}}_{-m}^t,S_{-m}=s_i,\mbf{X}_{-m}^t]\\
      &-\Cov_\theta[\Lambda_{-m+1}^{t-1},\phi_{\theta,t}|\oline{\mbf{Y}}_{-m'}^t,S_{-m'}=s_j,\mbf{X}_{-m'}^t],\\
      C=&\Var_\theta[\phi_{\theta,t}|\oline{\mbf{Y}}_{-m}^{t},S_{-m}=s_i,\mbf{X}_{-m}^{t}]
      -\Var_\theta[\phi_{\theta,t}|\oline{\mbf{Y}}_{-m'}^t,S_{-m'}=s_j,\mbf{X}_{-m'}^t],\\
      D=&\Var_\theta[\Lambda_{-m'+1}^{-m}|\oline{\mbf{Y}}_{-m'}^{t},S_{-m'}=s_j,\mbf{X}_{-m'}^{t}]
      -\Var_\theta[\Lambda_{-m'+1}^{-m}|\oline{\mbf{Y}}_{-m'}^{t-1},S_{-m'}=s_j,\mbf{X}_{-m'}^{t-1}],\\
      E=&\Cov_\theta[\Lambda_{-m+1}^{t-1},\Lambda_{-m'+1}^{-m}|\oline{\mbf{Y}}_{-m'}^t,S_{-m'}=s_j,\mbf{X}_{-m'}^t]\\
      &-\Cov_\theta[\Lambda_{-m+1}^{t-1},\Lambda_{-m'+1}^{-m}|\oline{\mbf{Y}}_{-m'}^{t-1},S_{-m'}=s_j,\mbf{X}_{-m'}^{t-1}],\\
      F=&\Cov_\theta[\Lambda_{-m'+1}^{-m},\phi_{\theta,t}|\oline{\mbf{Y}}_{-m'}^t,S_{-m'}=s_j,\mbf{X}_{-m'}^t].
\end{align*}
$\|A\|,\|B\|$, and $\|C\|$ are bounded as above. For the other terms, we have
\begin{align*}
      &\|D\|\leq 2\sum_{-m'+1\leq\ell\leq k\leq-m} \big(6\prod_{i=1}^{\lf(t-k-1)/r\rf}(1-\sigma_-(\mbf{V}_{t+r-ri}))\\
      &\hspace{4cm}\wedge 2\times 2\prod_{i=1}^{\lf(k-\ell-1)/r\rf}(1-\sigma_-(\mbf{V}_{\ell+ri}))\big)\|\phi_k\|_\infty\|\phi_\ell\|_\infty,\\
      &\|E\|\leq \sum_{k=-m+1}^{t-1}\sum_{\ell=-m'+1}^{-m} \big(6\prod_{i=1}^{\lf(t-k-1)/r\rf}(1-\sigma_-(\mbf{V}_{t+r-ri}))\\
      &\hspace{4cm}\wedge2\times 2\prod_{i=1}^{\lf(k-\ell-1)/r\rf}(1-\sigma_-(\mbf{V}_{\ell+ri}))\big)\|\phi_k\|_\infty\|\phi_\ell\|_\infty,\\
      &\|F\|\leq \sum_{-m'+1\leq\ell\leq-m} 2\prod_{i=1}^{\lf(t-\ell-1)/r\rf}(1-\sigma_-(\mbf{V}_{\ell+ri}))\|\phi_\ell\|_\infty\|\phi_t\|_\infty.
\end{align*}
Define $\rho$, $\bbm{1}_t$, and $\varepsilon_0$ as in (\ref{rho}), (\ref{Itk}), and (\ref{epsilon0}), respectively. Notice that the $\rho$ and $\varepsilon_0$ are the same as the ones in Lemma \ref{mixingbound}. Using $1-\mbf{V}_t\leq \rho^{1-\bbm{1}_t}$,
      \begin{align*}
            &\sum_{-m'+1\leq\ell\leq k\leq-m}\Big(6\prod_{i=1}^{\lf(t-k-1)/r\rf}(1-\sigma_-(\mbf{V}_{-m+ri}))
            \wedge2\times 2\prod_{i=1}^{\lf(k-\ell-1)/r\rf}(1-\sigma_-(\mbf{V}_{\ell+ri}))\Big)\|\phi_k\|_\infty\|\phi_\ell\|_\infty\\
            &\leq 12 \sum_{-m'+1\leq\ell\leq k\leq-m}\big(\rho^{\lf(t-k-1)/r\rf-\sum_{i=1}^{\lf(t-k-1)/r\rf} \bbm{1}_{-m+ri}}\wedge \rho^{\lf(k-\ell-1)/r\rf-\sum_{i=1}^{\lf(k-\ell-1)/r\rf} \bbm{1}_{\ell+ri}} \big)\|\phi_k\|_\infty\|\phi_\ell\|_\infty\\
            &\leq 12 \rho^{-\sum_{i=-m'+1}^t \bbm{1}_i}  \sum_{-m'+1\leq\ell\leq k\leq-m}\big( \rho^{\lf(t-k-1)/r\rf}\wedge \rho^{\lf(k-\ell-1)/r\rf}\big)\|\phi_k\|_\infty\|\phi_\ell\|_\infty.
      \end{align*}

      Using (\ref{lemma22}) and (\ref{lemma23}), $\bb{E}_{\theta^*}[\rho^{-\sum_{i=-m'}^{t} \bbm{1}_i}]\leq 2$ for all $m'$ and $t$. Then $\rho^{-\sum_{i=-m'}^{t} \bbm{1}_i}\leq \rho^{-\sum_{i=-\infty}^{+\infty} \bbm{1}_i}$, which is in $L^1(\bb{P}_{\theta^*})$.
Following a similar procedure to  \citet[pp. 2301]{douc2004asymptotic}, we obtain
      \begin{align*}
            &\sum_{-m'+1\leq\ell\leq k\leq-m} ((\rho^{\lf(t-k-1)/r\rf}\wedge \rho^{\lf(k-\ell-1)/r\rf})\|\phi_k\|_\infty\|\phi_\ell\|_\infty)\\
            &\hspace{3cm}\leq \rho^{\frac{t+m-2}{8r}-1}\sum_{\ell=-\infty}^\infty \rho^{\frac{|\ell|}{8r}}\|\phi_\ell\|_\infty \sum_{k=-\infty}^\infty \rho^{\frac{|k|}{4r}}\|\phi_k\|_\infty.
      \end{align*}
Thus,
      \begin{align*}
            \|D\| \leq 12  \rho^{\frac{t+m-2}{8r}-1}\rho^{-\sum_{i=-\infty}^{+\infty} I_i}\sum_{\ell=-\infty}^\infty \rho^{\frac{|\ell|}{8r}}\|\phi_\ell\|_\infty \sum_{k=-\infty}^\infty \rho^{\frac{|k|}{4r}}\|\phi_k\|_\infty.
      \end{align*}
      Similarly, we can derive
      \begin{align*}
            &\|E\| \leq 6 \rho^{\frac{t+m-2}{8r}-1}\rho^{-\sum_{i=-\infty}^{+\infty} I_i}\sum_{\ell=-\infty}^\infty \rho^{\frac{|\ell|}{8r}}\|\phi_\ell\|_\infty \sum_{k=-\infty}^\infty \rho^{\frac{|k|}{4r}}\|\phi_k\|_\infty,\\
            &\|F\| \leq 2\rho^{\frac{t+m}{4r}-1}\rho^{-\sum_{i=-\infty}^{+\infty} I_i} \sum_{\ell=-\infty}^\infty \rho^{\frac{3|\ell|}{4r}}\|\phi_\ell\|_\infty \sum_{k=-\infty}^\infty \rho^{\frac{|k|}{2r}}\|\phi_k\|_\infty.
      \end{align*}
      The proof is complete.
\hfill$\square$

\begin{lemma}\label{phie}
      Assume \ref{ergodic}--\ref{bound2} and \ref{differentiable}--\ref{finitemoment}. Then, for all $t\geq 1$ and $0\leq m\leq m'$,
      \begin{align*}
            \bb{E}_{\theta^*}[\max_{s_i,s_j}\sup_{\theta\in G}\|\Phi_{t,m,s_i}(\theta)-\Phi_{t,m',s_j}(\theta)\|]\ra0,\quad\text{as } m\ra\infty,\\
            \bb{E}_{\theta^*}[\max_{s}\sup_{\theta\in G}\|\Phi_{t,m,s}(\theta)-\Phi_{t,m}(\theta)\|]\ra0,\quad \text{as }m\ra\infty.
      \end{align*}
\end{lemma}
The proof of Lemma \ref{phie} follows from the inequalities in Lemma \ref{phip} using a similar argument to that of Lemma \ref{gammae}.

\section*{Appendix C: Tables and figures of Empirical Application}

\begin{table}[hbt]
\caption{Maximum Likelihood Estimation Result}
\begin{center}
\begin{tabular}{cccccc}
\hline
               & FTP                                                      & CLI                                                                     & BCI                                                      & CCI                                                      & CLI,BCI,CCI                                                             \\ \hline
$\mu(1)$       & \begin{tabular}[c]{@{}c@{}}-2.170\\ (0.196)\end{tabular} & \begin{tabular}[c]{@{}c@{}}-1.296\\ (0.099)\end{tabular}                & \begin{tabular}[c]{@{}c@{}}-2.327\\ (0.185)\end{tabular} & \begin{tabular}[c]{@{}c@{}}-2.139\\ (0.301)\end{tabular} & \begin{tabular}[c]{@{}c@{}}-1.757\\ (0.118)\end{tabular}                \\
$\mu(2)$       & \begin{tabular}[c]{@{}c@{}}0.164\\ (0.041)\end{tabular}  & \begin{tabular}[c]{@{}c@{}}0.189\\ (0.046)\end{tabular}                 & \begin{tabular}[c]{@{}c@{}}0.161\\ (0.042)\end{tabular}  & \begin{tabular}[c]{@{}c@{}}0.164\\ (0.044)\end{tabular}  & \begin{tabular}[c]{@{}c@{}}0.184\\ (0.041)\end{tabular}                 \\
$\gamma_1$     & \begin{tabular}[c]{@{}c@{}}0.048\\ (0.041)\end{tabular}  & \begin{tabular}[c]{@{}c@{}}-0.015\\ (0.046)\end{tabular}                & \begin{tabular}[c]{@{}c@{}}0.078\\ (0.044)\end{tabular}  & \begin{tabular}[c]{@{}c@{}}0.043\\ (0.043)\end{tabular}  & \begin{tabular}[c]{@{}c@{}}0.004\\ (0.045)\end{tabular}                 \\
$\gamma_2$     & \begin{tabular}[c]{@{}c@{}}0.201\\ (0.044)\end{tabular}  & \begin{tabular}[c]{@{}c@{}}0.146\\ (0.041)\end{tabular}                 & \begin{tabular}[c]{@{}c@{}}0.170\\ (0.051)\end{tabular}  & \begin{tabular}[c]{@{}c@{}}0.207\\ (0.045)\end{tabular}  & \begin{tabular}[c]{@{}c@{}}0.138\\ (0.044)\end{tabular}                 \\
$\gamma_3$     & \begin{tabular}[c]{@{}c@{}}0.139\\ (0.047)\end{tabular}  & \begin{tabular}[c]{@{}c@{}}0.196\\ (0.042)\end{tabular}                 & \begin{tabular}[c]{@{}c@{}}0.147\\ (0.049)\end{tabular}  & \begin{tabular}[c]{@{}c@{}}0.139\\ (0.048)\end{tabular}  & \begin{tabular}[c]{@{}c@{}}0.181\\ (0.046)\end{tabular}                 \\
$\gamma_4$     & \begin{tabular}[c]{@{}c@{}}0.006\\ (0.050)\end{tabular}  & \begin{tabular}[c]{@{}c@{}}0.113\\ (0.048)\end{tabular}                 & \begin{tabular}[c]{@{}c@{}}0.005\\ (0.049)\end{tabular}  & \begin{tabular}[c]{@{}c@{}}0.006\\ (0.050)\end{tabular}  & \begin{tabular}[c]{@{}c@{}}0.055\\ (0.047)\end{tabular}                 \\
$\sigma$       & \begin{tabular}[c]{@{}c@{}}0.496\\ (0.016)\end{tabular}  & \begin{tabular}[c]{@{}c@{}}0.501\\ (0.015)\end{tabular}                 & \begin{tabular}[c]{@{}c@{}}0.498\\ (0.016)\end{tabular}  & \begin{tabular}[c]{@{}c@{}}0.496\\ (0.018)\end{tabular}  & \begin{tabular}[c]{@{}c@{}}0.492\\ (0.017)\end{tabular}                 \\
$\beta_{10}$   & \begin{tabular}[c]{@{}c@{}}-0.733\\ (1.247)\end{tabular} & \begin{tabular}[c]{@{}c@{}}0.962\\ (0.800)\end{tabular}                 & \begin{tabular}[c]{@{}c@{}}-1.702\\ (2.574)\end{tabular} & \begin{tabular}[c]{@{}c@{}}-1.002\\ (1.191)\end{tabular} & \begin{tabular}[c]{@{}c@{}}-272.889\\ ($9.354\times10^7$)\end{tabular}  \\
$\beta_{11}$   &                                                          & \begin{tabular}[c]{@{}c@{}}-1.025\\ (1.425)\end{tabular}                & \begin{tabular}[c]{@{}c@{}}-1.610\\ (4.941)\end{tabular} & \begin{tabular}[c]{@{}c@{}}9.294\\ (24.429)\end{tabular} & \begin{tabular}[c]{@{}c@{}}-1242.074\\ ($3.734\times10^8$)\end{tabular} \\
$\beta_{12}$   & \multicolumn{1}{l}{}                                     & \multicolumn{1}{l}{}                                                    & \multicolumn{1}{l}{}                                     & \multicolumn{1}{l}{}                                     & \begin{tabular}[c]{@{}c@{}}954.478\\ ($3.170\times10^8$)\end{tabular}   \\
$\beta_{13}$   & \multicolumn{1}{l}{}                                     & \multicolumn{1}{l}{}                                                    & \multicolumn{1}{l}{}                                     & \multicolumn{1}{l}{}                                     & \begin{tabular}[c]{@{}c@{}}-88.191\\ ($5.591\times10^7$)\end{tabular}   \\
$\beta_{20}$   & \begin{tabular}[c]{@{}c@{}}-4.921\\ (0.693)\end{tabular} & \begin{tabular}[c]{@{}c@{}}-754.443\\ ($4.733\times10^4$)\end{tabular}  & \begin{tabular}[c]{@{}c@{}}-5.662\\ (0.905)\end{tabular} & \begin{tabular}[c]{@{}c@{}}-5.330\\ (1.576)\end{tabular} & \begin{tabular}[c]{@{}c@{}}-6.692\\ (4.052)\end{tabular}                \\
$\beta_{21}$   &                                                          & \begin{tabular}[c]{@{}c@{}}-1463.306\\ ($9.198\times10^4$)\end{tabular} & \begin{tabular}[c]{@{}c@{}}-4.849\\ (1.979)\end{tabular} & \begin{tabular}[c]{@{}c@{}}-3.628\\ (3.175)\end{tabular} & \begin{tabular}[c]{@{}c@{}}-12.361\\ (9.912)\end{tabular}               \\
$\beta_{22}$   &                                                          &                                                                         &                                                          &                                                          & \begin{tabular}[c]{@{}c@{}}4.837\\ (10.309)\end{tabular}                \\
$\beta_{23}$   & \multicolumn{1}{l}{}                                     & \multicolumn{1}{l}{}                                                    & \multicolumn{1}{l}{}                                     & \multicolumn{1}{l}{}                                     & \begin{tabular}[c]{@{}c@{}}-1.177\\ (4.383)\end{tabular}                \\
log-likelihood & -327.563                                                 & -318.012                                                                & -325.201                                                 & -325.035                                                 & -313.165                                                                \\
$p$-value      &                                                          & $7.113\times 10^{-5}$                                                   & $0.094$                                                  & $0.080$                                                  & $6.649\times10^{-5}$                                                    \\ \hline
\end{tabular}
\end{center}
\label{table:empirical2}
\footnotesize
\renewcommand{\baselineskip}{11pt}
\textbf{Note: }Standard errors are in parentheses. The last row reports the $p$-value for the test of not including predetermined variables in the transition probabilities.
\end{table}

\begin{figure}[!hbt]
\hspace{-2cm}
\includegraphics[width=18cm]{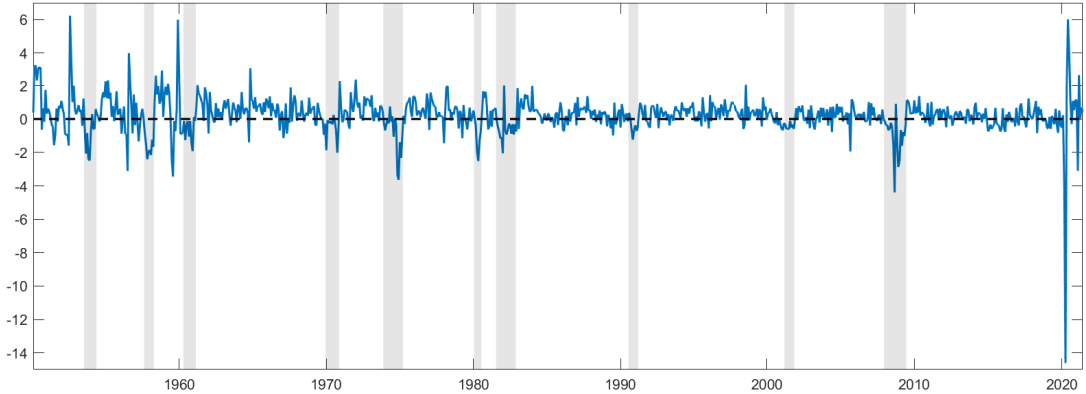}
\caption{Growth rate of U.S. monthly industrial production, Jan 1950--June 2021}
\end{figure}

\end{document}